\documentclass[preprint,secnumarabic,nobalancelastpage,nofootinbibt]{revtex4-1}

\usepackage{graphics}      % standard graphics specifications
\usepackage{graphicx}      % alternative graphics specifications
\usepackage{longtable}     % helps with long table options
\usepackage{url}           % for on-line citations
\usepackage{bm,color}            % special 'bold-math' package
\usepackage{amssymb, amsmath, enumerate, theorem, epsfig,subfigure,setspace}
\usepackage{amsfonts}
\usepackage{mathtools}
\usepackage{booktabs,tabularx}
\newcommand{\nb}[1]{\textcolor{black}{#1}}

\begin{document}

\title{Analog Photonics Computing for Information Processing, Inference and Optimisation}
  \author{Nikita Stroev$^1$ and Natalia G. Berloff$^{2}$ }
\email[correspondence address: ]{N.G.Berloff@damtp.cam.ac.uk}
\affiliation{$^1$Department of Physics of Complex Systems, Weizmann Institute of Science, Rehovot 76100, Israel}
\affiliation{$^2$Department of Applied Mathematics and Theoretical Physics, University of Cambridge, Cambridge CB3 0WA, United Kingdom}

\begin{abstract}

%Optical computing can be defined as the science of using photons and optics-related technologies for effective and efficient computing purposes. Since high-performance computational methods have become necessary for many scientific disciplines and the industry, it is clear that the demand for high-performance photonic components will inevitably increase. Photonic architectures can offer attractive properties such as practical computation scaling overcoming conventional computing hardware,  massive parallelisation, low energy consumption, broad bandwidth, and low latency. 
%Nevertheless, these platforms have specific drawbacks, which limit their applicability and put them in the special-purpose hardware niche. The first part of this review discusses physics of optical computing. It gives a brief history of optical computing development and describes modern analogue computing platforms and their architectures, focusing on neural network implementations. We discuss the special-purpose optimisers: physics-specific platforms suitable for this type of hardware. The second part of this review concerns the mathematical aspect of optical computing, such as a mathematical description of optical optimisers, various possible applications and the interconnections among them. Additionally, we cover the main directions of technological development in optical computing and some estimates of optical computing efficiency. Finally, we discuss future perspectives on the field and touch upon the domain of optical quantum computing.
\nb{This review  presents an overview of the current state-of-the-art in photonics computing, which leverages photons, photons coupled with matter, and optics-related technologies for effective and efficient computational purposes. It covers the history and development of photonics computing and modern analogue computing platforms and architectures, focusing on optimization tasks and neural network implementations. The authors examine special-purpose optimizers, mathematical descriptions of photonics optimizers, and their various  interconnections. Disparate applications are discussed, including direct encoding, logistics, finance, phase retrieval, machine learning, neural networks, probabilistic graphical models, and image processing, among many others.  The main directions of technological advancement and associated challenges in photonics computing are explored, along with an assessment of its efficiency. Finally, the paper discusses prospects and the field of optical quantum computing, providing insights into the potential applications of this technology.}
\end{abstract}

\maketitle

\section{Introduction}
%Modern problems of computers
In 1965 Intel co-founder Gordon Moore formulated an empirical observation that the number of transistors in a microprocessor will double nearly every two years, the statement which is known as Moore's law \cite{moore1965cramming,moore2000readings}. This prediction was followed by the forecast of reaching a saturation point by 2015. The progress of conventional computer architectures was very close to Moore's vision, \nb{ and } reaching the saturation point was just a matter of time. The miniaturization of silicon transistors recently managed to break the 7-nanometre barrier, which was believed to be the limit. Also, Moore's law usually comes with several essential indicators, such as the processor's thread performance and clock frequency, which reached the point of saturation much faster than the density of the transistors. All of these factors limit the scaling performance of modern computers. However, there are other reasons for the saturation of conventional computing power growth, which are the consequences of Moore's law. For example, increasing the number of transistors allows one to obtain more powerful systems. Still, the processing speed will inevitably decrease with the concomitant increase in heat production, while increased energy consumption is connected with the growth of the performance. Another critical issue is the so-called von Neumann bottleneck \cite{von1993first}, arising from the architecture design. It refers to the computer system throughput limitation due to the characteristic of bandwidth for incoming and outcoming data \cite{edwards2016eager,naylor2007reduceron}. All these issues pose severe problems to the future of conventional computer development. As a result, the alternatives to von Neumann systems started to emerge \cite{lent2016molecular,shin2019heterogeneous}.

%The promise of optical computer
One turns to alternative hardware architectures and purpose-built devices to keep up with the scaling performance. As such, universal quantum computing promises to decrease the algorithmic complexity of solving challenging tasks by exploiting the entangled states. However, in contrast to this high-risk and high-reward strategy (also discussed below  in the optical setting), there is an option to replace electrons with photons but remain in the scope of classical or classical with a transient quantum coherence regime of optical computing. The motivation for such transition is clear since photons move at the speed of light, have low heat production, have high density and can be efficiently coupled to matter to exploit nonlinear behaviours. Moreover, optical technologies have matured and entered our everyday lives, such as fibre optic channels that carry the global traffic of information or optical readers of compact disks. However,  the conversion of photons into electrons is required for compatibility with CMOS architectures. Such conversion takes a significant portion of energy, slows down the overall process of information processing, and presents a severe technological bottleneck in this type of hybrid technology. 

%Existing optical hardware
Despite these difficulties, optical hardware is exploited in computing devices. For example, different application-specific photonic hardware can operate on a reasonable scale in data centres for heavy machine learning (ML) applications and large-scale optimization. Moreover, neural network (NN) architectures are nearly ideally suited for optical hardware with the potential to achieve high efficiency, fast computing times, and low energy consumption due to the desired physical properties of the photonic systems.
%and because the element base can be realized by optical components, sharing the mathematical description between their software analogues. 
%We devote part of this review to exploring this connection more deeply. 
Nevertheless, at this point, optical computing can not be associated with mainstream technology. It is unlikely that optics will ever replace electronics as the universal platform in the foreseeable future. The additional reason is the technological inertia accumulated through the years by significant investments in CMOS technologies. Partially, the rapid development of what we call conventional computers in the early years led to an ever-increasing gap with computing using photonics, which will occupy its own place in the domain of application-specific hardware.

%Reviews references
%The material presented in this review is self-sufficient and contains essential key points that can be found across several other review articles aiming at covering optical computing technologies. We point out a few reviews that share many similarities in the covered material. 

There are many excellent reviews on the topic of optical computing. The challenges of modern computing and new opportunities for optics are discussed in \cite{li2021challenges}. This work presents the latest research progress of analogue optical computing, focusing on three main directions: vector/matrix manipulation, reservoir computing (RC) and photonic Ising machine. Moreover, it covers the topic of computing efficiencies, such as the ratio of performance and power dissipation and the error/precision interplay of such hardware. Another excellent review considers analogue optical computing in the context of artificial intelligence (AI) applications \cite{WU2021}. This work provides an overview of the latest accomplishments of optical computing, considering the realization of different AI models and NN paradigms. One can find additional information in other reviews \cite{genty2021machine,xiang2021review,chen2021highlighting,shastri2021photonics,wetzstein2020inference,kalinin2020nonlinear,dolev2013optical}, which appeared due to the recent interest in deep learning methods and their success in many domains.

 What differentiates our review from those listed above is that we treat analogue optical computing using the concept of universality of the underlying dynamical systems description. The advantage of optical computing comes from ultrafast emulation of the dynamics \cite{solli2015analog}.
We focus on physical optimisers that exploit bifurcation dynamics and threshold operation and aim at solving nonlinear problems, therefore, going beyond the speed-up of performing the linear operations that optics is so efficient at.

%Analog Optical Computing in a Broad Scence
%One can treat analog optical computing in a more general way because the dynamical processes there can be close to emulating real complicated systems. However, ultrafast optical systems have very rapid dynamics in comparison with the simulations on a digital computer \cite{solli2015analog}. Good example is the identification of optical rogue waves as rare solitons among ordinary events by optical means which is similar to the rare events that follow heavy-tailed statistics, like earthquakes, stock market crashes and other unusual behaviour \cite{solli2007optical}. The set of such phenomena is quite big, thus we limit the scope of this review to more practical, universal and direct types of computation.

%Content
We organised our review as follows. Section \ref{Analog optical computing} provides a short history of optical computing together with the modern analogue computing platforms focusing on NN implementation and other neuromorphic systems. Section \ref{Nonlinear optimization specific optical machines} discusses the special-purpose optimisers and several examples of such devices. This section connects the operational regimes of such machines with the complexity classes and addresses the scalability of this approach. 
Section \ref{Description of physical optical platforms for optimization} focuses on the physics of optical computing devices based on laser networks, optical parametric oscillators (OPOs) in fibre, photon or polariton networks, as well as their mathematical models. The second part of this review investigates the mathematical structures of different assignments and their emulation by the physical systems. The following Section \ref{Mathematical formulation of applications} lists a wide range of possible applications across different applied domains. The final part consists of our subjective perspective on the future technological development of optical computing field in Section \ref{Main directions of technological development in optical computing} and passing remarks about quantum optical devices in Section \ref{Optical quantum computing}. Finally, Section \ref{Final remarks} summarises the results.

\section{Analog optical computing}
\label{Analog optical computing}

%The demand for the optical type computation
Modern technologies demand vast data flows, creating various challenges for the development of the semiconductor industry and pushing classic electrical circuits to their physical limits. %Scientists, engineers and computer designers look to optics as the answer to the arising questions. 
The developments range from more mainstream such as optical components that can be integrated into traditional computers or play the role of specific hardware, dealing with computationally heavy tasks or supplementing such calculations, to ambitious ones, such as all-optical digital computer architecture. 
%The first approach offered the best short-term prospects with many attractive advantages and proved efficient.

\subsection{Brief prehistory of the optical computing}
\label{Brief prehistory of the optical computing}

Although optical computing is an emerging technology that has gained more momentum over time (especially considering the popularity and efficiency of the latest data-driven approaches), many significant advances have been made in previous decades. Therefore, before describing the particular systems, their advantages and their applications, we briefly discuss the  progress that enabled the future developments. More information and additional details can be found in \cite{ambs2010optical}. % with the particular emphasis on different time periods, their dominating perspective on the field, technological advances and special types of optical devices. 

%Principles of Optical Information Processing
The generic optical processor architecture comprises three plane parts: the input, the processing and the output planes. Early on, the input plane was a fixed image slide with its later change to a spatial light modulator (SLM), introduced to perform the input signal conversion. The processing plane can be composed of lenses, nonlinear components, or holograms, while the final output part is composed of photodetectors or a camera.

%First and second architecture of the correlator

\begin{figure}[!h]
	\begin{minipage}[c]{0.95\linewidth}
		\includegraphics[width=\linewidth]{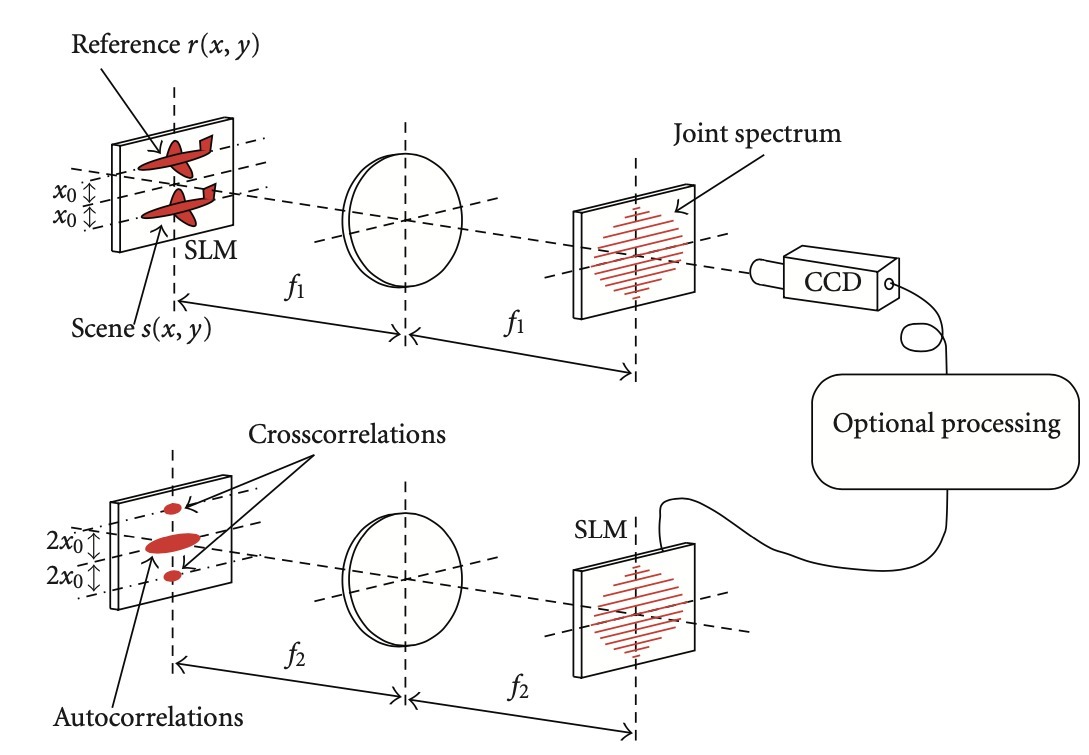}
		\caption{The optical setup of the joint transform correlator (JTC). The figure is taken from \cite{ambs2010optical}.}
		\label{1_old_optical}
	\end{minipage}
	\hfill
\end{figure}

The first promising applications for optical processors were pattern recognition tasks, which influenced the prototypes of optical correlators. The simple architecture called 4-f was based on the work on spatial filtering, see \cite{marechal1953filtre}. The Fourier transform property of a lens is the standard function of many optical computing schemes, taking advantage of the speed and parallelism of light. The second type of correlator architecture was presented in 1966 by Weaver and Goodman \cite{weaver1966technique}, which is called the joint transform correlator (JTC), see Fig.~\ref{1_old_optical}

%The Rise of Optical Computing (1945–1980)
Before 1950 there were significant steps in development of optical technologies such as the theory of image formation in the microscope \cite{abbe1873beitrage}, developed by Abbe, the development of phase contrast filter by Zernike \cite{zernike1934diffraction} and the appearance of the information optics after Elias Snitzer in 1952 \cite{elias1995optics,elias1952fourier}. Other major inventions of that period were holography (by Gabor,1948) \cite{gabor1948new} and the development of the laser in 1960 \cite{maiman1960optical,maiman1960stimulated}. The consequent introduction of the off-axis hologram allowed the separation of the different terms of the reconstruction \nb{solving  3D reconstruction tasks} \cite{leith1963wavefront,leith1964wavefront} in 1962 by Leith and Upatnieks, which basically led to practical holography and was further enhanced by Lohmann, creating the first computer-generated hologram \cite{brown1966complex,lohmann1967binary} in 1966. Early SLMs were based on the Pockels effect with few prospective devices \cite{dumont1974phototitus,iwasa1976optical,oliver1978real}. Liquid crystal technology is the most commonly used technology for SLMs today. Another significant step was the invention of the first optical transistor \cite{jain1976optical}, the hope for small integrated circuits. 

%Optical Computing Golden Age (1980–2004)
The period from 1980 to 2004 was vibrant and productive.
Active progress was going in the field of holography, particularly new encoding methods and the point-oriented methods were developed to achieve high quality and high diffraction efficiency optical reconstructions of the CGHs \cite{hauck1984computer}. More than 50 types of SLM were introduced in the eighties and nineties \cite{poon1998spatial}. Optical transistors presented another active area of research with the appearance of the micro-electromechanical systems (MEMS) technology \cite{abraham1983optical}. Vertical-cavity surface-emitting lasers (VCSELs) and the self-electrooptic effect (SEED) devices entered the markets in 90s \cite{jackson1994photonic}. In general, many aspects of modern optical interconnections and their components were introduced and studied during this period. 

%Short Historical Conclusions
The optical technologies development provided the necessary experience in the capabilities of the optical devices and led to the maturation of the experimental element base. Optical computing received a second chance after the success of so-called deep NNs, which share many similarities with the previous neural-like optical architectures.

\subsection{Modern optical computing}
\label{Modern optical computing}

Today, numerous research topics benefit from the progress in optical computing; therefore, the field is no longer so well defined. For example, some of the algorithms initially developed for pattern recognition using optical processors are now used successfully in digital computers. Other fields, such as biophotonics, largely benefit from past optical processing research.

%Optical Computer Architecture and Optical on-chip Calculations
\nb{Transistor} is the fundamental building block of modern electronic computers. Therefore, one must find an equivalent optical transistor to replace electronic components with optical ones. To assemble such transistors into the higher-level components to create an all-optical computer's central processing unit (CPU), one has to design the optical processor and optical storage and organise the optical data transfer. However, such an approach faces many challenges, while the potential of optics in large architecture consisting of higher-level components can be seen as somewhat speculative \cite{tucker2010role}. Among persisting problems are the scalability of the optical logic devices due to the bad logic-level restoration, cascadability, fan-out and input-output isolation, energy consumption issues and non-miniature device footprint. Moreover, coupling these potential devices with the electrical components will require \nb{ conversion of information carried} from photons to electrons, which is relatively slow and energy-consuming.

%Linear Elements
However, the development of integrated photonics continued \cite{bogaerts2020programmable}. It led to attempts to create linear logic elements, such as all-optical logic gates \cite{singh2014all,jandieri2020functional}, improve the existing optical transistors and develop new ones in the context of the all-optical processing \cite{minzioni2019roadmap,woods2009optical}. One can use SLM, micro-lens array and holographic elements in free space to realize optical linear interconnection. Such linear elements are essential components in various optical computing devices.

%Nonlinear Elements
Nonlinearity is another essential component in optical schemes; however, its realisation meets specific difficulties as light beams pass through each other unperturbed in a pure vacuum. To force these beams to interact, one has to set up a high-energy experiment, which is challenging to realise in practice. There are two other ways to realise the nonlinearity: introduce the digital readout mechanism, implemented by the charged-coupled device (CCD), send it to a computer with further nonlinear processing before feeding it back to SLM, or develop fully optical nonlinear activation materials with high enough intensity of the beams (utilising absorption, refraction or scattering processes). Nonlinearities can be divided into local (as needed in neural architectures) and global systems (such as reservoir computing systems, see below). Combining the linear and nonlinear elements led to the developing of specialised isolated devices. As a result, optical computing research has seen a resurgence in activity, centring around new developments in photonic hardware accelerators and neuromorphic computing.

% Neuromorphic Computing
Neuromorphic computing usually denotes the use of integrated systems to mimic neuro-biological architectures. Although it is very close to the domain of AI, with the stress on the word ``artificial'', which deals with the intelligent designed machines or agents, we will use neuromorphic computing in the general sense to describe any neural systems, be it brain or nature-inspired or artificially designed. Modern key focus areas are \nb{trying to emulate the neural structure and functionality of the human brain, including probabilistic computing, which incorporates uncertainties}. Optics has required ingredients to emulate NNs \cite{wu2021artificial,shastri2021photonics}.

\subsection{Optical neural networks}
\label{Optical neural networks}

% Element base
Optics has long been \nb{considered as a promising technology for implementation} of matrix multiplication and interconnects. Artificial neural networks (ANN) have been widely exploited for industrial and fundamental applications, and this new technological demand created a renewed case for photonic NNs. Although most ANN hardware systems are electronic-based, their optical implementation is \nb{promising because of their built-in} parallelism and low energy consumption. Disparate ANNs vary by types of constituent elements, mathematical operations and the architecture used. In \nb{photonic  ANNs}, the mathematical operations are mapped to the \nb{characteristics} of optical propagation set by programmable linear optics and nonlinearity. A scalar synaptic weight describes pairwise connections between artificial neurons. At the same time, the layout of interconnections can be given as a matrix-vector operation, where the input to each neuron is the dot product of the output from connected neurons with assigned weights.

%Photonic realizations 1
Photonic realizations of ANNs fall into three categories. First, free-space systems rely on diffraction, Fourier transforms, etc. \cite{paek1987optical,lin2018all}. They have high scalability and can simultaneously process large numbers of neurons but suffer limited connectivity. One example is a reconfigurable, scalable two-layer NN for the classifying phases of a statistical Ising model \cite{zuo2019all}. Second, SLMs program linear operations and Fourier lenses implement the summation by collecting the light power encoded signal. However, in the case of free optics, the nonlinear optical activation functions are realized in a complicated manner, e.g. with the laser-cooled atoms with electromagnetically induced transparency \cite{zuo2019all}. Finally, on-chip \nb{designs} based on wavelength multiplexing \cite{tait2017neuromorphic} or beamsplitter meshes \cite{shen2017deep} can achieve programmable all-to-all coupling but need to scale better. One on-chip design was proposed in \cite{zhang2021optical}, where the optical platform takes advantage of encoding information in both phase and magnitude, thus making it possible to execute complex arithmetic by optical interference, which suits performing handwriting recognition tasks. Mach–Zehnder Interferometers (MZIs) can perform many functions, such as dividing and modulating the light signals, separating the reference and the main light beams, and implementing a complex-valued weight matrix.

\begin{figure}[htbp]
	\begin{minipage}[c]{0.95\linewidth}
		\includegraphics[width=\linewidth]{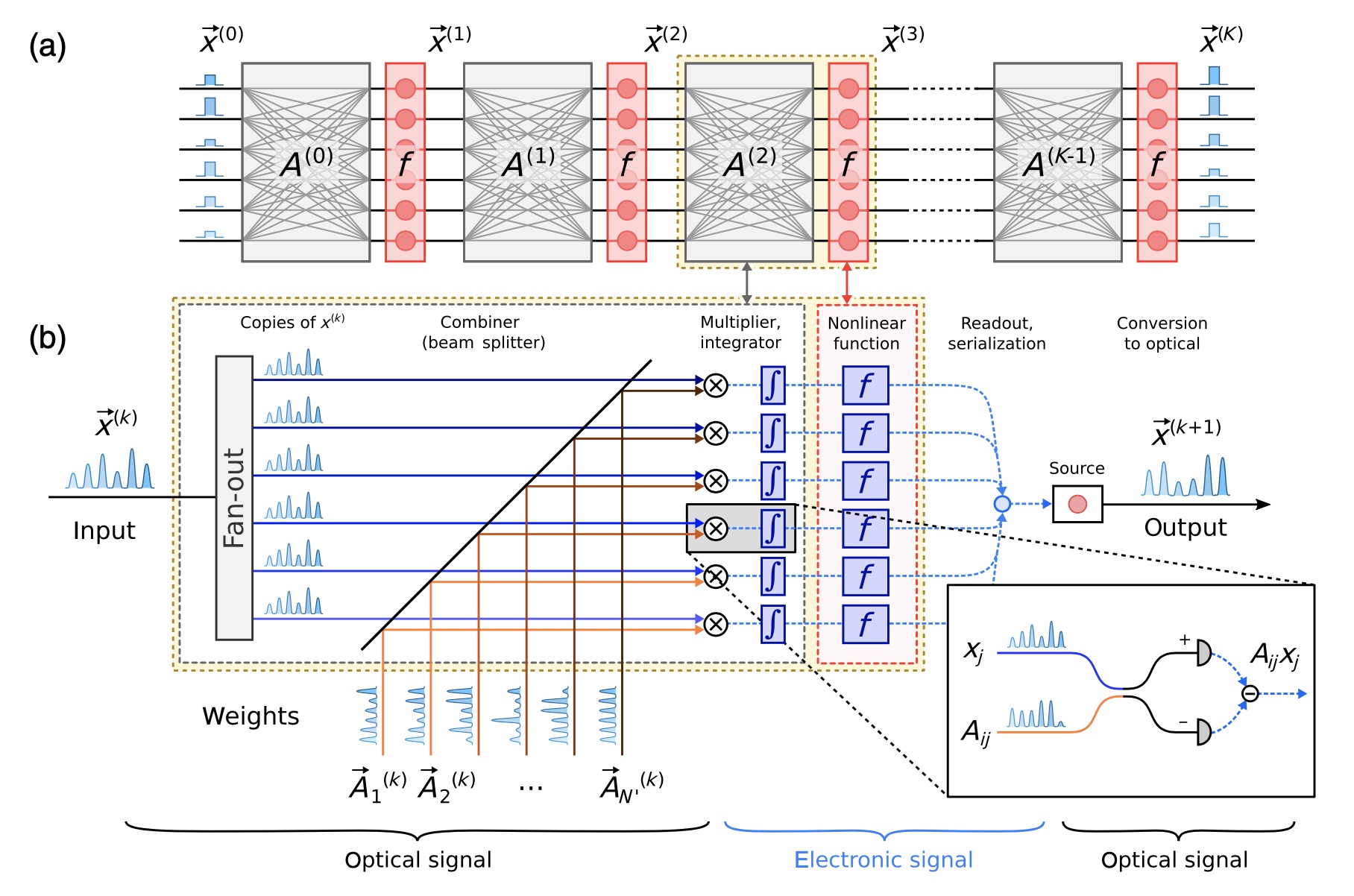}
		\caption{(a) One layer of an optical NN with $k$ layers consists of matrix-vector multiplication (grey) and non-linearities (red). (b) One-level implementation. Matrix multiplication is performed by combining the input and weight signals and performing balanced homodyne detection. The final signals are sent through a non-linear function (red), serialized, and sent to the following layer's input. Figure from \cite{hamerly2019large}}.
		\label{2_onn_1}
	\end{minipage}
	\hfill
\end{figure}

%Photonic realizations 2
Tunable waveguides can multiply optical signals, while wavelength-division multiplexing can add signals. Wavelength-division multiplexing can be achieved by the accumulation of carriers in semiconductors \cite{kravtsov2011ultrafast,tait2014broadcast}, electronic currents \cite{shainline2017superconducting,bangari2019digital}, or photon-induced changes of the material \cite{feldmann2019all}. To achieve the full potential in on-chip architectures, one must require long-range connections between neurons, assisted with photonic waveguides that outperform metal wires connections of conventional electronics but fall behind free-optics solutions. In particular, silicon photonic platforms demonstrated efficient neuromorphic architectures \cite{tait2017neuromorphic,shainline2017superconducting,shen2017deep}.  \nb{Coherent input light can be transformed using an array of beam splitters and phase shifters \cite{reck1994experimental}. This is done by assigning inputs to different waveguides and modulating their power. Another approach to weight configuration is modulating the effective refractive index of signal-carrying waveguides. Non-volatile synapse implementations, also known as all-optical, do not require electrical inputs for tuning. Instead, they use optically induced changes in chalcogenide materials to control light propagation in waveguides \cite{gholipour2015amorphous}. This could lead to improved heat dissipation.}

\begin{figure}[!h]
	\begin{minipage}[c]{0.95\linewidth}
		\includegraphics[width=\linewidth]{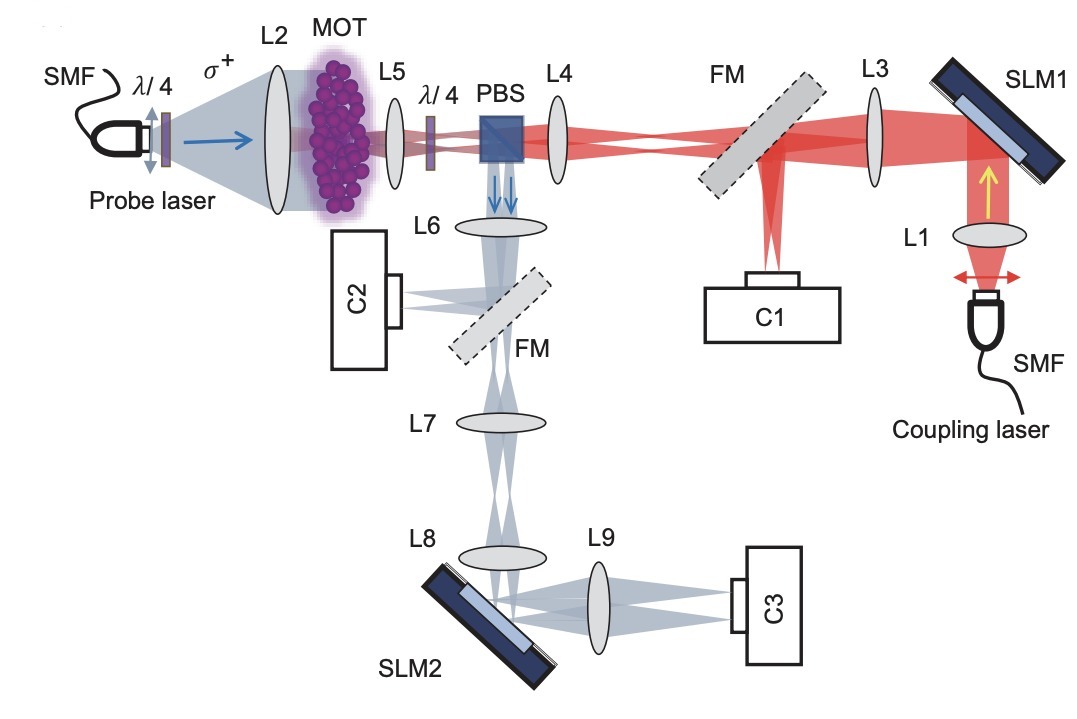}
		\caption{In fully functioned $2$-layer all optical NN the first layer comprises a linear operation by the first SLM (SLM1) which encodes a certain pattern and a nonlinear activation function based on the electromagnetic induced transparency at magneto-optical trap (MOT). The second layer contains the second SLM (SLM2), converting four beams into two output beams at camera C3. The collimated coupling laser beam passing lens L1 is incident on the SLM1, which generates four beams at the focal plane of L3, which is monitored by a flip mirror (FM) and camera C1. Four beams are imaged on the MOT through a 4-f system comprising L4 and L5. A probe laser is going opposite the coupling beam, which is imaged on camera C2 through L5 and L6. L7 and L8 achieve further amplification. Four beams are incident on SLM2, generating two beams and then focusing on camera C3. Figure and description is from \cite{zuo2019all}.}
		\label{3_onn_2}
	\end{minipage}
	\hfill
\end{figure}

%Photonic realizations 3
A scheme based on homodyne detection has several scaling advantages over on-chip approaches. \nb{It has linear  
chip-area scaling and constant circuit depth \cite{hamerly2019large}.  The input vector is encoded onto a pulse train and fanned out to an array of homodyne detectors. Each detector computes a product between the input and a row of a matrix encoded into optical pulse trains. The accumulated charge on the homodyne detector performs matrix-vector multiplication.} The output is sent through an electrical nonlinearity and converted back to optical signal using a modulator. The advantage of the homodyne detection scheme is that the matrix elements (weights) are encoded optically and can be dynamically reconfigured. This procedure requires a reduced number of photonic components: the number of modulators, detectors, and beamsplitter grows linearly with the number of neurons. The homodyne detection architecture can be parallelized to implement general matrix-matrix multiplication by routing the light out of plane \cite{hamerly2019large}. This is useful in practical NNs that reuse weights (either natively in convolutional layers or through batching).

\begin{figure}[!h]
	\begin{minipage}[c]{0.95\linewidth}
		\includegraphics[width=\linewidth]{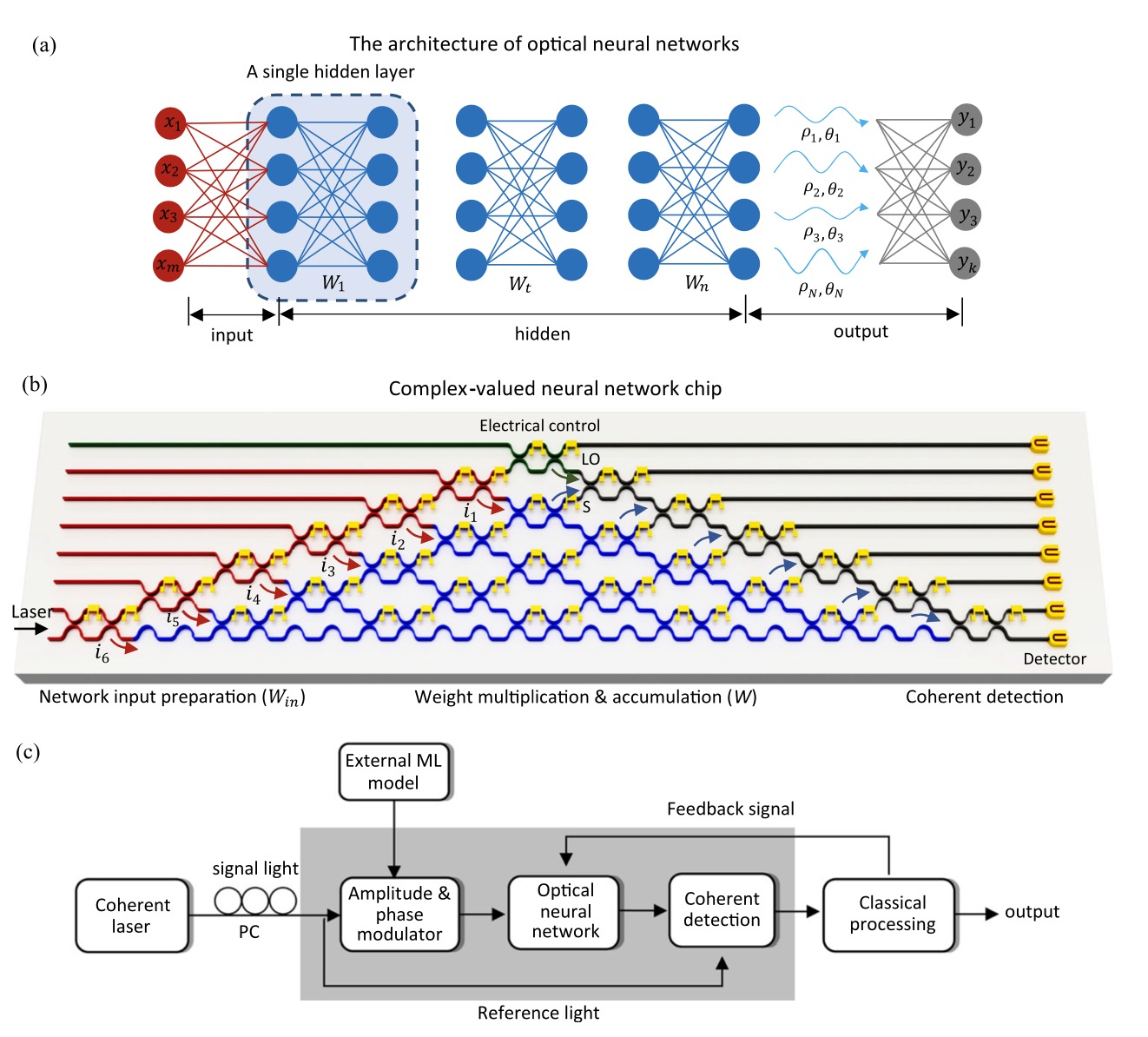}
		\caption{Complex-valued coherent optical NN. (a) Scheme with an input layer, multiple hidden layers, and an output layer. (b) The schematic of the optical neural chip in implementing complex-valued networks. The single chip performs all stages, such as the input preparation, weight multiplication and coherent detection. The division and modulation of the light signals ($i_{1}-i_{6}$) are realized by the MZIs (red). Green MZI separates the reference light. Blue MZIs are used to implement the $6\times 6$ complex-valued weight matrix. Grey MZIs are used for on-chip coherent detection. (c) The workflow of the ONC system. Figure from \cite{zhang2021optical}.}
		\label{4_onn_3}
	\end{minipage}
	\hfill
\end{figure}

%Nonlinearities
Nonlinearity in ANN is required to implement the thresholding effect of the neuron. Some photonic devices exhibit nonlinear neuron-like (gate-like) transfer functions. However, the challenge is to achieve cascadability. \nb{Photonic neurons need to be able to respond to multiple optical inputs, apply a nonlinearity, and produce an optical output that can drive other photonic neurons.} Integrated photonic solutions use either optical/electrical/optical (O/E/O) or all-optical design to achieve such cascadability. In the O/E/O approach, nonlinearities may be introduced during the E/O conversion stage by employing lasers, saturated modulators or photodetector–modulators \cite{williamson2019reprogrammable} or in the electronic domain only (e.g. the \nb{nonlinear behavior of spiking photonic neurons can be achieved using a superconducting electronic signal pathway} \cite{mccaughan2019superconducting}).

%Variants of NNs
NN architectures can take different forms: feed-forward and back-forward, layered and recurrent, spiking or continuous etc. \nb{Different neural models have unique signal representations, training methods, and network topologies.} Weight configurations can differ depending on the training type: supervised training, unsupervised or programmatic ‘compilation’. Topology describes the graph structure of neuron connectivity, and often it is advantageous to ANN operation to constrain the topology to guide weight configurations. Therefore, hardware implementation details may differ between different ANN, while the key technologies necessary for practical realization include active on-chip electronics and light sources. Many photonic architectures have already been demonstrated: recurrent ANN, continuous-time and programmed by compiler \cite{tait2017neuromorphic}; feed-forward, single-valued and externally trained ANN \cite{shen2017deep}; spiking, feed-forward ANN with both external and local training \cite{feldmann2019all}; \nb{a feed-forward multilayer artificial neural network  created using semiconducting few-photon light-emitting diodes and superconducting-nanowire single-photon detectors} \cite{shainline2017superconducting}; diffractive networks with a nonlinearity \cite{zuo2019all}. The computational tasks solved by these platforms cover the main functions attributed to ML and AI: image and audio recognition and classification, simulation of dynamical systems, combinatorial optimization and many other applications, which we will discuss in Section \ref{Mathematical formulation of applications}. Some of the architectures and their different experimental realizations are shown in Figs.~\ref{2_onn_1},~\ref{3_onn_2} and ~\ref{4_onn_3}.

%Short summary
The key merit of NN hardware is the level of energy consumption, which can be evaluated as petaMAC (multiply-accumulate operations) per second per mm${}^2$ processing speeds \cite{nahmias2019photonic} and attojoule per MAC energy efficiencies \cite{nozaki2019femtofarad}. In general, current optoelectronic hardware offers great advantages for implementing  ANN, but eliminating the electrical contribution will inevitably be beneficial. For practical applications of neuromorphic photonic systems, one needs to reduce heat dissipation during information
transfer between electrons and photons. \nb{Improving optical sources, high-efficiency modulators, and photonic analogue-to-digital interfaces can reduce the need for electronic processors. However, current photonic platforms lack functionality such as logic gates, high-level compilers and assemblers, analogue-digital-analogue conversion, and memory. While photonics has advantages in connectivity and linear operations over electronics, on-chip memory is challenging. In-memory computing, where processing is performed in situ with an array of memory devices called memristors, has been established} \cite{rios2015integrated,rios2019memory}; however, reading and writing at high frequencies is still challenging. The recent trends in the development of the ANN show the increasing demand to lower the power consumption of the devices. At the same time, the requirements for parallelism and scalability remain the same through the years \cite{schuman2017survey}. Thus, the optical domain offers a promising solution to future hardware requirements.

%Unconventional computing methods
\subsection{Reservoir and other neuromorphic computing systems}
\label{Reservoir and other neuromorphic computing systems}

%Intro
Reservoir computing (RC) is a recurrent NN-based framework for computation that maps input signals into a specific computational space of the fixed nonlinear system dynamics. This system is usually called a "reservoir", and its state is passed to a simple readout mechanism, specifically trained to get the final output \cite{tanaka2019recent}. The original concepts of RC can be traced to the liquid-state machines \cite{maass2002real} and echo-state networks \cite{jaeger2001echo}. Many physical systems can reproduce this computational framework, and the optical/photonics domain is no exception. The extension of RC to deep hierarchical RC allows one to create more efficient models and simultaneously investigate the inherent role of layered composition in recurrent structures. Another promising research direction is to combine RC with quantum physical systems to access larger computational space.

%Advantages and disadvantages
The idea of RC is to exploit the rich nonlinear dynamics of controllable nonlinear systems and simultaneously overcome the disadvantages of recurrent architectures with their challenging and time-consuming training for both hardware and software systems. The RC training is performed only at the readout stage, as the reservoir dynamics are fixed. This readout framework enjoys the benefits of a particular photonic physical system, such as speed or energy consumption, reducing learning costs. Another RC benefit is learning temporal (dynamic) dependencies compared to the feed-forward architectures used for static (non-temporal) data processing.
The simplicity of the training procedure in RC is attractive. However, accessing complex dynamics without rigorous understanding can lead to many problems. Operating within the RC framework usually needs extensive experiments and experimental verification due to the need for a unified theory of RC. Another disadvantage is the performance instability due to the noise present, typical for nearly chaotic dynamical systems.

\begin{figure}[!h]
	\begin{minipage}[c]{0.95\linewidth}
		\includegraphics[width=\linewidth]{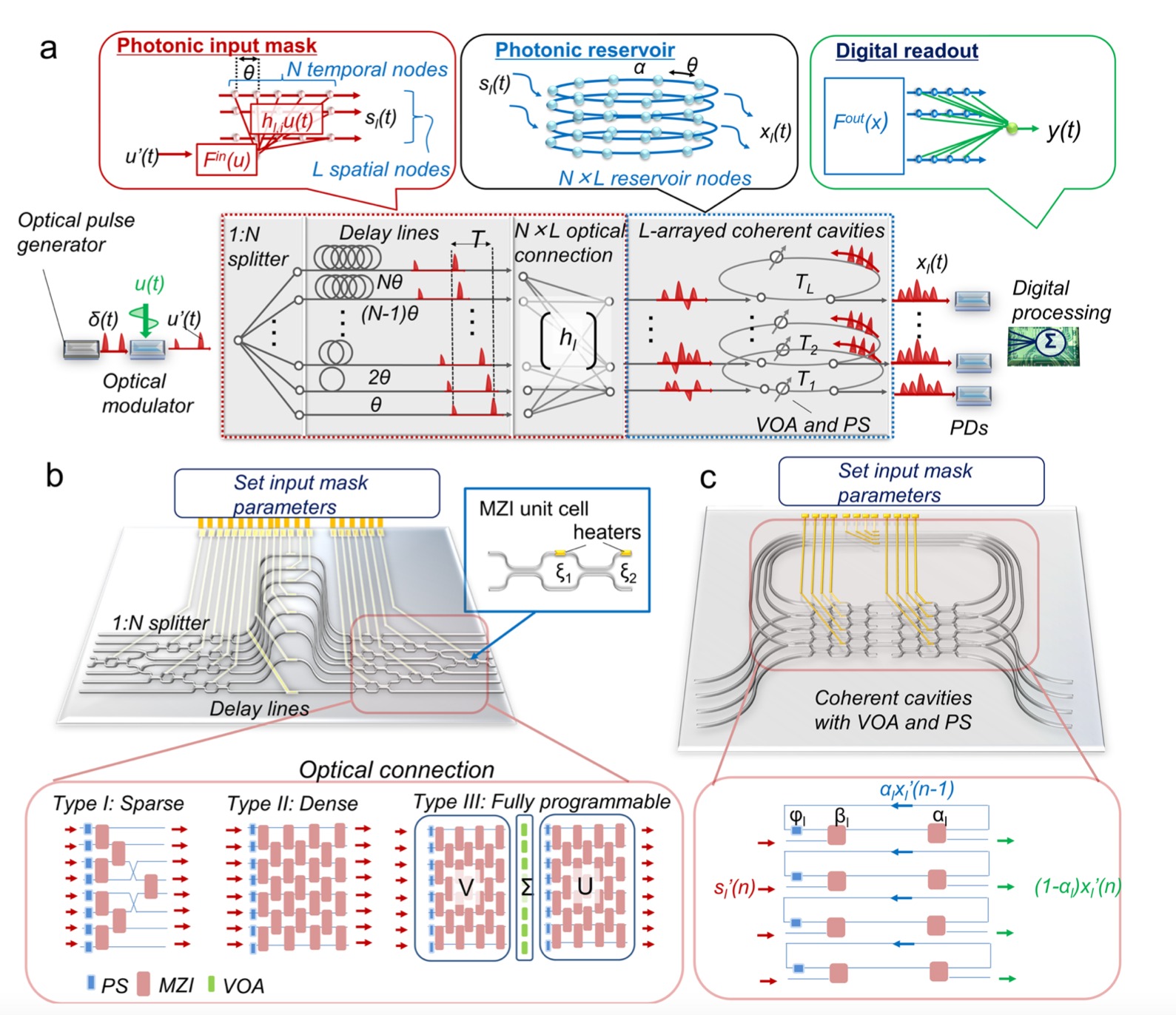}
		\caption{(a) Schematics of proposed reservoir computing (RC) architecture. The electric input signal $u(t)$ is coded on optical pulse, $\delta (t)$ is coded by optical modulator. The sinusoidal nonlinearity is achieved by the electro-optic conversion ($F^{in}$). RC system comprises the $L$-array of optical cavities with $N$ temporal nodes with $N \times L$ virtual nodes. Photodetectors (PDs) get the output signals from optical circuit and generate nonlinear conversion ($F^{out}$). The digital processing unit detects signals $|x_{l}|^{2}$ and weights and sums them to obtain the final output $y(t)$. (b)Schematic of the waveguide layout for input mask circuit with the three types of installations for optical connection, and (c) - input mask reservoir circuit. The input weights $h_{l}$ and reservoir parameters can be tuned by the phase shifter, MZI and variable optical attenuator setup.
		Figure and the description from \cite{nakajima2021scalable}.}
		\label{5_reservoir}
	\end{minipage}
	\hfill
\end{figure}

%Implementation cases (systems, parameters, tasks, efficiency)
Nevertheless, many successful cases of RC are being applied to practical problems, such as temporal pattern classification, time series forecasting, pattern generation, adaptive filtering and control, and system approximations. Moreover, RC can be used conventionally for static data processing. The first all-optical implementation of RC was demonstrated within a simple optical delayed feedback loop combined with the nonlinearity of an optical amplifier \cite{duport2012all}. Concerning the free-space optics principles, an image processing task was successfully solved using a predesigned configuration with a diffraction grating and Fourier imaging with randomly interconnected microring resonators \cite{mesaritakis2015high}. The reservoir consisting of a diffractive optical element was described based on an $8 \times 8$ laser array (of VCSELs) and an SLM. It showed rich dynamics with the potential for scaling up \cite{brunner2015reconfigurable}. Further modifications of this setup with a laser illumination field and digital micro-mirror device allowed one to realise the large-scale RC scheme with $2025$ diffractively coupled photonic nodes applied to a time series prediction task \cite{bueno2018reinforcement}. The recurrent 24-node silicon photonic NN, in which microring weight banks configure connections, was programmed using a “neural compiler” to solve a differential system emulation task with a 294-fold acceleration against a conventional benchmark \cite{tait2017neuromorphic}.

%Hybrid RC
Some hybrid architectures, such as opto-electronic devices, similarly benefit from the RC concept. For example, excellent performance has been obtained for speech recognition \cite{paquot2012optoelectronic,larger2012photonic,martinenghi2012photonic}, chaotic time series prediction \cite{larger2012photonic,soriano2013optoelectronic,ortin2015unified},  and radar signal forecasting \cite{duport2016fully}, with the operating speed in the megahertz range and the potential to increase it to gigahertz speed, at the same time preserving the state-of-the-art numerical accuracy. Additional cases of successful RC have been reported in literature \cite{tanaka2019recent,van2017advances,paquot2012optoelectronic}. We will consider the NN and RC architectures cases involving quantum effects in Section \ref{Quantum optical devices}. Another example is illustrated by Fig.~\ref{5_reservoir}.

\section{Nonlinear optimization with optical machines}
\label{Nonlinear optimization specific optical machines}

%Intro
\nb{Problems that can be solved by optical hardware cover a wide range of optimization  problems: linear and nonlinear in binary, integer, real or complex variables, with or without constraints}.  Such a general framework can include many applications across social sciences, finance, telecommunications, aerospace,  biological and chemical industries \cite{applications}.

%Nonlinear optimization
\nb{Nonlinear optimisation problems are quadratic to second order around the vicinity of the optimal solution so the usual simplification involves  Quadratic Programming (QP) that  minimizes quadratic functions of variables subject to linear constraints. }Such approximation can be successfully performed even outside the feasible solutions space. QPs can be met in the least squares regression or as a part of a bigger problem, such as support vector machine (SVM) training. The apparent correspondence between the QP objective function and 2-local spin Hamiltonians of various physical systems allows one to map the problem into the physical setup. Here, the \nb{variables} are associated with ``spins'' and the objective function with a ``Hamiltonian'' that encodes the interactions patterns and strengths between spins. 

%Quasi-equilibrium systems
%In this section, we consider the dynamical systems that are neither programmable strictly nor perform the prepared sequence of given operations (like NN with fixed weights). 
\nb{A system can find the optimal solution or ground state of a spin Hamiltonian using either quasi-equilibrium or non-equilibrium regimes. In thermodynamic equilibrium, the system may use quantum annealing with a time-dependent Hamiltonian. This involves adiabatic evolution from an initial trivial Hamiltonian to a final Hamiltonian encoding the original objective function. However, this process can become inefficient for larger systems and sophisticated Hamiltonians due to a shrinking spectral gap \cite{farhi}.}

%Non-equilibrium systems
Many open non-equilibrium gain-dissipative systems, such as lasers and photonic or polaritonic condensates are non-Hermitian systems and, therefore, do not have a ground state. Instead, they tend to minimise losses on the route to coherence. One can use geometric analogies to describe their operational principle as the approach of the surface of the optimisation cost function (loss landscape) from below. There are two main processes that lead to loss minimisation: bosonic stimulation below the threshold and the coherence of operations at the threshold responsible for the quality of the solution. After increasing the gain to the point where it overcomes the linear losses and is stabilised by the nonlinearity of the gain saturation, the emergent coherent state minimises the losses (equivalent to maximisation of the total number of particles). It hence achieves the loss minimisation mapped into the objective spin Hamiltonian. The system elements' resulting evolution closely resembles the Hopfield Networks' dynamics, proposed to be used to solve quadratic optimisation problems forty years ago \cite{Hopfield1982}. Despite the successes and a lot of excitement generated back then, the optimisers based on Hopfield networks were almost forgotten primarily due to the high connectivity required between neurons and the concomitant evolution time of the networks used by classical architecture. \nb{Hopfield networks have recently regained interest because they can now be emulated with physical systems such as electronic circuits or photonic neural networks. Photonic systems have an advantage over electronic systems because they operate on a much faster time scale and can carry many signals through a single optical waveguide. This means that a photonic implementation of Hopfield networks as optimizers can have large dimensionality, dense connectivity, and quickly converge to the optimal solution.}

\subsection{Spin Hamiltonians and hard optimization}
\label{Spin Hamiltonians}

%Spin Hamiltonians

\nb{Real-life decision or optimization problems can be challenging for conventional classical computers. These problems can often be reformulated into finding the ground state of a spin Hamiltonian, which can be emulated with a simulator quantum, classical  or hybrid. The overhead in the number of additional variables needed during this mapping is at most polynomial. However, such spin model Hamiltonians are experimentally challenging to implement and control. Two classes of spin Hamiltonians are generally considered: Ising and XY. The Ising model attracts the most attention because many challenging discrete combinatorial optimization problems can be mapped into it \cite{lucas2014ising}. The minimization of the Ising Hamiltonian with the coupling matrix $\bf{J}$ (the minimization of the quadratic unconstrained binary optimisation problem (QUBO)) is formulated as}
\begin{equation}
\min_{s_i} \quad -\,\sum_{i=1}^N \sum_{j=1,j<i}^N J_{ij} s_i s_j , \quad {\rm subject} \ {\rm to} \quad s_i \in \{-1,1\}.
\label{ising}
\end{equation}

\nb{Going  beyond quadratic  would lead to a $k$-local spin Hamiltonian: }
\begin{equation}
\min_{{ s_i}} \quad -\,\sum_{i_1,i_2,...i_k}^N Q_{i_1,i_2,...i_l,...,i_k} s_{i_1} s_{i_2}...s_{i_l}...s_{i_k} \quad {\rm subject} \ {\rm to} \quad s_{i_l} \in \{-1,1\}.
\label{hobo}
\end{equation}
\nb{This problem known a higher-order binary optimization problems (HOBO)  appears in many contexts including   including $k$-SAT \cite{maxsat},  number factorization \cite{FactorisationMicrosoft}, computing the partition function of a four-dimensional pure lattice gauge theory \cite{de2010mapping}, molecular similarity measurement \cite{hernandez2016novel}, and molecular conformational sampling \cite{marchand2019variable}.
 }
In the XY model ``spins" are continuous vectors ${\bf s_j} = (\cos \theta_j,\sin \theta_j)$ and the  quadratic continuous optimization problem (QCO) becomes
\begin{equation}
\min_{\bf s_i} \quad -\,\sum_{i,j<i} J_{ij} {\bf s}_i \cdot {\bf s}_j = \min_{\theta_i}\quad  -\sum_{i,j<i} J_{ij} \cos(\theta_i - \theta_j) \quad {\rm subject} \ {\rm to} \quad \theta_i \in [0,2\pi),
\label{xy}
\end{equation}
and directly applicable for phase retrieval problems \cite{PhaseRetrieval1,PhaseRetrieval2,PhaseRetrieval3,nir2018}. When phases $\theta_j$ are restricted to take on discrete values $2\pi /n$ with an integer $n>2$, Eq.(\ref{xy}) becomes the $n-$state Clock model with applications in chemical materials and protein folding research \cite{PottsProteins}.  

%Appearence of the phase
The appearance of continuous spins is a common feature in many optical systems because short photonic impulses can be characterized through amplitude and phase variables. Some of the optical hardware for ML take advantage of this feature. For example, complex-valued NNs \cite{zhang2021optical}, or more unusual concept of analogue transformations using a nonlinear set of functions were proposed \cite{stroev2021neural}.

%\subsection{P, NP, NP-complete problems}
%\label{P, NP, NP-complete problems}

% Intro to computational complexity
\nb{The computational complexity of a problem is determined by the time or number of operations required to solve it as the problem’s size increases. A problem belongs to the P class if a polynomial algorithm exists for solving it. If a polynomial algorithm exists for verifying a solution, the problem belongs to the non-deterministic polynomial-time (NP) class. Most difficult problems in NP are called NP-complete. They are equivalent to each other in that either all of them or none admit a polynomial-time algorithm. A problem is called NP-complete if the existence of an efficient algorithm for its solution implies the existence of such an algorithm for all NP-complete problems.
If a decision problem with a yes or no answer is NP-complete, then its corresponding optimization problem is said to be NP-hard. This means that NP-hard problems are not any easier to solve than the corresponding NP-complete decision problems. The computational complexity of the Ising model has been studied and shown to be NP-hard for certain cases  \cite{Barahona1982}. The existence of universal spin Hamiltonians has been established, meaning that all classical spin models can be reproduced within such a model \cite{Cubitt_universality}. The mapping of various NP problems, including Karp’s 21 NP-complete problems, to Ising models with polynomial overhead has been formulated \cite{lucas2014ising}.   }

\nb{The procedure for creating  ``hard" instances for spin Hamiltonians was developed based on  statistical physics, here the hardness of problems is  connected to a first-order phase transition in a system \cite{zdeborova2008statistical,mezard2009information,zdeborova2016statistical,Helmut_wishart}.} Indeed, even a medium size hard instance is difficult to solve on a classical computer due to the exponential growth of operations with size. Having a unified set of optimization problems with tunable hardness is an ongoing research direction. It will allow for an objective benchmark of classical and/or quantum simulators and algorithms. Otherwise, it would be hard to evaluate the performance of state-of-the-art platforms and methods.

%Easy and hard instances 3
Current research made a good starting point in developing a standardised procedure for performance evaluation. For example, the ``optimisation simplicity criterion" was recently proposed to identify computationally simple instances \cite{kalinin2020complexity}. Optical machines with their mode selection operation often follow the dominant eigenvalue of the coupling matrix and find minimisers that correspond to the signs of the principal eigenvector components. If the minimisers of a given problem have this property, the solution will be found easily in polynomial (at most quadratic) time. One such popular example is the Ising model on the M\"{o}bius ladder graph  \cite{kalinin2020complexity}. By rewiring the M\"{o}bius ladder graph to random 3-regular graphs, one can probe an intermediate computational complexity between P and NP-hard classes with several numerical methods. Another way to construct instances for testing involves planted ensemble technique \cite{Helmut_wishart,hamze2018near}.

\section{Description of physical optical platforms for optimization}
\label{Description of physical optical platforms for optimization}

%Intro to platforms
 \nb{The concept of using simulators or analogue processing devices  predated the modern computers \cite{hertz1991introduction}. In recent years, different physical platforms have been competing to solve classical optimization problems faster than conventional hardware. This competition has led to the emergence of a new field of coherent networks/Hamiltonian  simulators at the intersection of laser and condensed matter physics, engineering, and complexity theories. Various physical systems have emerged as promising platforms for solving computational problems. We shall overview some of these systems following the structure we used in \cite{kalinin2020nonlinear}. }

\subsection{Laser networks}
\label{Complex laser networks}

%Laser networks
\nb{A new generation of complex lasers, such as degenerate cavity lasers, multimode fiber amplifiers, large-aperture VCSELs, and random lasers, have many advantages over traditional laser resonators when used for computing \cite{controlcoh}. They have many spatial degrees of freedom and controllable  nonlinear interactions. The spatial coherence of emission can be tuned over a wide range and the output beams may have arbitrary profiles providing pairwise couplings. These properties allow complex lasers to be used for reservoir computing (RC) \cite{photonicRC}  or to represent the coupling matrix interactions.  In laser networks, coupling can be engineered by mutual light injection from one laser to another. This introduces losses that depend on the relative phases between the lasers and drives the system to phase-locking that minimizes losses \cite{lasers}.
Degenerate cavity lasers are particularly promising  as Hamiltonian simulators as all their transverse modes have nearly identical $Q$ which leads to simultaneous syncronization  of a large number of transverse modes \cite{controlcoh}.}

%Equations
The evolution of the $N$ single transverse and longitudinal modes class--B lasers are  described by \cite{ratelasers,nirprl2017} 
\begin{eqnarray}
\frac{d A_i}{d t} &=&(G_i-\alpha_i)\frac{A_i}{\tau_p}+\sum_{j}J_{ij}\frac{A_j}{\tau_p}\cos(\theta_i-\theta_j),\label{l1}\\
\frac{d \theta_i}{d t} &=&\Omega_i-\sum_{j}J_{ij}\frac{A_j}{\tau_p A_i}\sin(\theta_i-\theta_j),\label{l2}\\
\frac{d G_i}{d t} &=&\frac{1}{\tau_c}[P_i-G_i(1+|A_i|^2)],
\label{l3}
\end{eqnarray}
\nb{where  for laser $i$, $A_i$[$\theta_i$] is amplitude [phase],   $G_i$ is gain, $P_i$ is pump strength, $\alpha_i$ is loss, $ \Omega_i$  is frequency detuning, $\tau_p$ and $\tau_c$ are the cavity round trip time and   the carrier
lifetime, respectively. The coupling strengths between $i$-th and $j$-th lasers are represented by $J_{ij}$. }
If all  amplitudes  are equal, then Eq.~(\ref{l2}) reduces to the Kuramoto equation of coupled phase oscillators
\begin{equation}
\frac{d \theta_i}{d t} =\Omega_i-\frac{1}{\tau_p}\sum_{j}J_{ij}\sin(\theta_i-\theta_j).
\label{kuramoto}
\end{equation}
Equation (\ref{kuramoto})  is widely used to describe the emergence of coherent behaviour in complex systems \cite{kuramoto,kuramoto2}. %By LaSalle invariance principle \cite{khalil}, every trajectory of the Kuramoto model converges to a minimum of the XY Hamiltonian.

\nb{It has been demonstrated that the probability of finding the global minimum of the XY Hamiltonian agrees between the experimental realisation of the laser array and with numerical simulations. However, simulating the Kuramoto model Eq.~(\ref{kuramoto}) on the same matrix of coupling strength yields a much lower probability of finding the global minimum. This implies that amplitude dynamics provide a mechanism to reach lower energy states. Consequently, cavity lasers can be used as an efficient physical simulator for finding the global minimum of the XY Hamiltonian and solving phase retrieval problems. A particularly successful example was a digital degenerate cavity laser \cite{nir2018}. It is an all-optical system that uses a nonlinear lasing process to find a solution that best satisfies the constraint on the Fourier magnitudes of light scattered from an object.}

\subsection{Optical parametric oscillators}
\label{Coherent Ising machine}

%CIM intro
\nb{A network of coupled degenerate optical parametric oscillators (DOPOs)  has been used to minimize the Ising Hamiltonian \cite{CIM1,CIM3,CIM4,CIM5,CIM6,CIM_eqs2015,CIM_eqs2019}. DOPO is a  nonlinear oscillator with two possible phase states above the threshold that can be associated with the Ising spins. These artificial Ising spins are encoded by the optical phase of short laser pulses generated by a  nonlinear optical process, i.e. optical parametric amplification. The DOPO-based simulator, coherent Ising machine (CIM), is a gain-dissipative system in which the ground state of the Ising Hamiltonian corresponds to the lowest loss configuration. The optimal solution is found by driving the system close to the near-threshold regime, where other local energy minima are still unstable. }

\begin{figure}[!h]
	\begin{minipage}[c]{0.95\linewidth}
		\includegraphics[width=\linewidth]{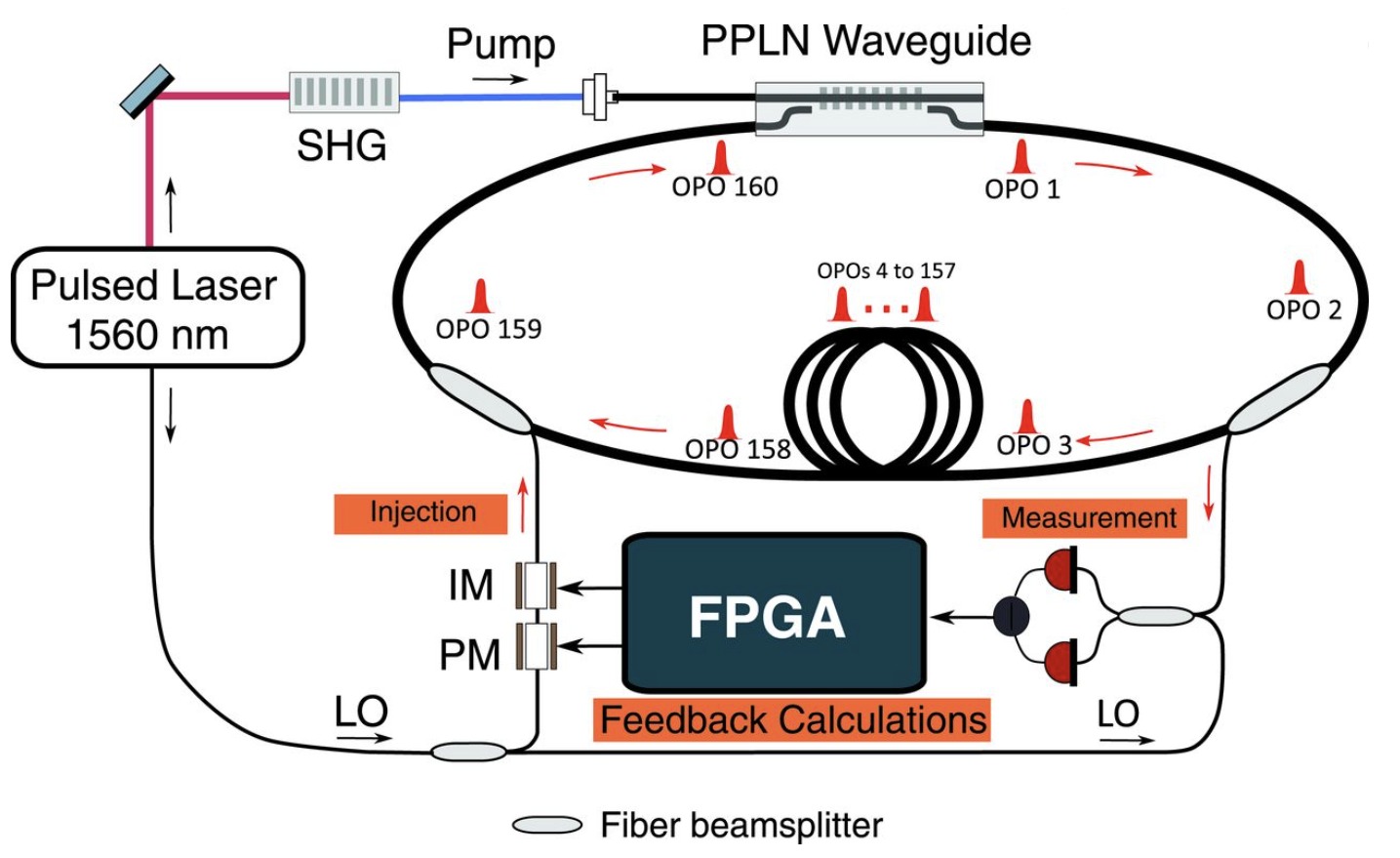}
		\caption{Schematics of the coherent Ising machine (CIM) with the feedback mechanism. The time-multiplexed pulse degenerate parametric oscillator is formed by a non-linear crystal (periodically polarized lithium niobate (PPLN)) in a fibre optic ring cavity containing $160$ pulses. The feedback signal couples the independent pulses in the cavity and is computed from the measurements from different pulse fractions.IM - intensity modulator; PM - phase modulator; LO - local oscillator; SHG - second-harmonic generation; FPGA - field-programmable gate array. The figure is taken from \cite{CIM1}.}
		\label{8_cim}
	\end{minipage}
	\hfill
\end{figure}

%CIM implementations
Currently, most successful implementations of CIMs use a fibre-based degenerate DOPOs  and a measurement-based feedback coupling, in which a matrix-vector multiplication is performed on the FPGA embedded in the feedback loop, see the scheme depicted in Fig.~\ref{8_cim}. The computational performance of such a scalable optical processor, bounded by the electronic feedback, was demonstrated for various large-scale Ising problems \cite{CIM1,CIM3,CIM4}. The comparison of a possible CIM's speedup over classical algorithms is an ongoing study \cite{CIM5,Dwave_vs_CIM}. Furthermore, the ability to implement arbitrary coupling connections \cite{CIM1} between any two spins implies better scalability than the solid-state based annealer, i.e. D-Wave machine \cite{CIM3}.

%CIM equations
In CIM, each Ising spin (a DOPO)  is described by 
\begin{equation}
	\frac{d \Psi_i}{dt} = p \Psi_i^* - \Psi_i - |\Psi_i|^2 \Psi_i + \sum_j J_{ij} \Psi_j,
	\label{CIM_ai}
\end{equation}
\nb{where $\Psi_i=A_i \exp[i \theta_i]$ is the complex amplitude, $\theta_i$ is the phase, $p$ is pump intensity, and  linear and nonlinear losses are normalized. A portion of the light is extracted from the cavity after each round trip, homodyned against a reference pulse to produce $\Psi_i$ that is fed to the FPGA, where the last term of Eq.~(\ref{CIM_ai}) is computed and   converted  back to light for the next round trip. }

CIM's essential elements are DOPOs with an unconventional operating mechanism called mode selection or gain-dissipative principle. 
Each DOPO is prepared in a linear superposition state of different excitations to implement a quantum parallel search.
The cost function is mapped to the effective loss, photon decay rate, of the given network by setting the coupling coefficient proportional to the $J_{ij}$, which encodes the information about the given task. The ground state of the Ising Hamiltonian corresponds to an oscillation mode with the minimum network loss. The system reaches the ground state with a minimum loss at the threshold pump rate.
It starts oscillating as a single stable mode, which triggers photons' stimulated emission and affects the saturation for all the other modes.
Detecting this single oscillation mode will give us the solution to the desired problem.

\nb{\subsection{Networks of light-matter particles}}
\label{Photon and polariton networks}

%Microcavity exciton-polaritons
Photons have both attractive and not properties concerning computational assignments. However, despite the commonly known optical platforms, such as free optical setups or systems of lasers, it is possible to bind the photons with the matter wave excitations. This gives rise to unique designs, combining the photons with matter, such as exciton-polaritons. \nb{Exciton-polaritons  (or simply polaritons) are quasi-particles that result from coupling of light confined inside semiconductor microcavities and bound electron-hole pairs. They are bosons and can form a condensed state above a critical density \cite{Kasprzak2006}.  The design and choice of material allow for control of the polariton mass and the realization of solid-state nonequilibrium condensates at room temperature in organic structures. In the limit of small gain, solid-state condensates resemble equilibrium Bose-Einstein condensates. They approach lasers in the regime of high gain. }
 \nb{Similarly to polaritons, gas of photons confined in a dye-filled optical microcavity  can form a macroscopic coherent state \cite{photoncondensate,photoncondensate2,photoncondensate3,photoncondensate4}. }%The rapid thermalization of rovibrational modes of the dye molecules by their collisions with the solvent and phonon dressing of the absorption and emission by the dye molecules leads to the thermal equilibrium distribution of photons and concomitant accumulation of low-energy photons. Such systems resemble microlasers \cite{microlasers}, but unlike microlasers, they exhibit a sharp threshold that occurs far below the inversion.

% Equations
\nb{Experiments have shown that polariton or photon condensate lattices can be constructed using various techniques. One such technique is optical engineering of polariton lattices by injecting polaritons into specific areas of the sample using an SLM \cite{wertz,manni,pendulum,baumbergGeometrical,BerloffNatMat2017}. Additionally, potential landscapes have been engineered to confine polaritons or photons  \cite{chneiderRev, amo,klaers17}.  The evolution of gain-dissipative condensates in a lattice is described by rate equations derived from the tight-binding approximation  \cite{NJP_Kalinin2018, OurLattices}, taking the form of Stuart-Landau equations: }

%Many techniques have been proposed and realised in experiments to construct the lattices of polariton or photon condensates. Polariton lattices can be optically engineered by injecting polaritons in specific areas of the sample using the SLM \cite{wertz,manni,pendulum,baumbergGeometrical,BerloffNatMat2017}. Various potential landscapes to confine polariton or photons have also been engineered  \cite{chneiderRev, amo,klaers17}. The rate equations describing the evolution of gain-dissipative  condensates in a lattice were derived using the tight-binding approximation of  the space and time-resolved mean-field equations \cite{NJP_Kalinin2018, OurLattices} and take the form of the Stuart-Landau equations
%
\begin{equation}
\dot{\Psi_i}=   (\gamma_i- |\Psi_i|^2) \Psi_i -i U |\Psi_i|^2\Psi_i+ \sum_{j\ne i} {\cal C}_{ij}\Psi_j,\label{ee0}
\end{equation}
\nb{where $\Psi_i$ is  the complex amplitude, $U$ is the polariton-polariton interaction strength, $\gamma_i$ is the effective injection rate (gain minus loss). The coupling strength is complex ${\cal C}_{ij}=J_{ij} + i G_{ij}$ where the dissipative part of coupling is $J_{ij}$ and the Josephson part is $G_{ij}$.  If $|J_{ij}|\gg |G_{ij}|$ and the injection feedback is introduced that modifies the injection rate to bring all amplitudes to the same value $A_{\rm th}$, for instance  by $\dot{\gamma_i}=\epsilon (A^2_{\rm th} - |\Psi_i|^2)$ then Eq. (\ref{ee0}) reaches the fixed point which is the minimum of the XY Hamiltonian \cite{NJP_Kalinin2018}. The total number of quasi-particles in the system becomes
\begin{equation}
{\rm Total}\equiv N A^2_{\rm th} = \sum_{i=1}^N \gamma_i + \sum_{i=1}^N\sum_{j<i}^N J_{ij} \cos(\theta_i-\theta_j),
\label{cond}
\end{equation}
where $\theta_i$ is the  phase of the $i-$th condensate. 
It follows from Eq.~(\ref{cond}) that if  the  total injection, $\sum \gamma_i$  to stimulate that  number of quasi-particles is minimized,   then  the $XY$ Hamiltonian is globally minimized.
 To minimise  the Ising Hamiltonian or $n$-states Clock Hamiltonians a resonant pump can be added to the system \cite{GD-resonant}. The resonant pump can be tuned with the condensate frequency, or $n$ times the condensate frequency for $n>1$ ($n:1$ resonance). In this case, the dynamics of the oscillators (e.g. photons or polaritons) obey
\begin{equation}
\dot{\Psi_i}= (\gamma_i- |\Psi_i|^2) \Psi_i -i U |\Psi_i|^2\Psi_i  + \sum_{j\ne i} {J}_{ij}\Psi_j + h(t) \Psi_i^{*(n-1)},\label{res}
\end{equation}
where $h(t)$ is the strength of the resonant excitation that grows in time until it reaches its stationary value $H$. At the fixed point, the total number of quasi-particles in the system reaches 
\begin{equation}
{\rm Total}\equiv N A^2_{\rm th} =\sum_{i=1}^N \gamma_i + \sum_{i=1}^N\sum_{j<i}^N J_{ij} \cos(\theta_i-\theta_j)-H A_{\rm th}^{n-2}\cos(n\theta_i).
\label{cond2}
\end{equation} 
}
\nb{At $n>1$, the last term of Eq.~(\ref{cond2}) forces the  phases to take discrete values $\theta_i=2\pi/n$ and, therefore, minimize the Ising Hamiltonian (for $n=2$) or   the $n$-state Clock Hamiltonian (for $n>2$).} %The minimization of HOBO may be achieved when the system operates much above the threshold, and higher-order terms must be addressed \cite {stroev2021discrete}.

\begin{figure}[!h]
	\begin{minipage}[c]{0.95\linewidth}
		\includegraphics[width=\linewidth]{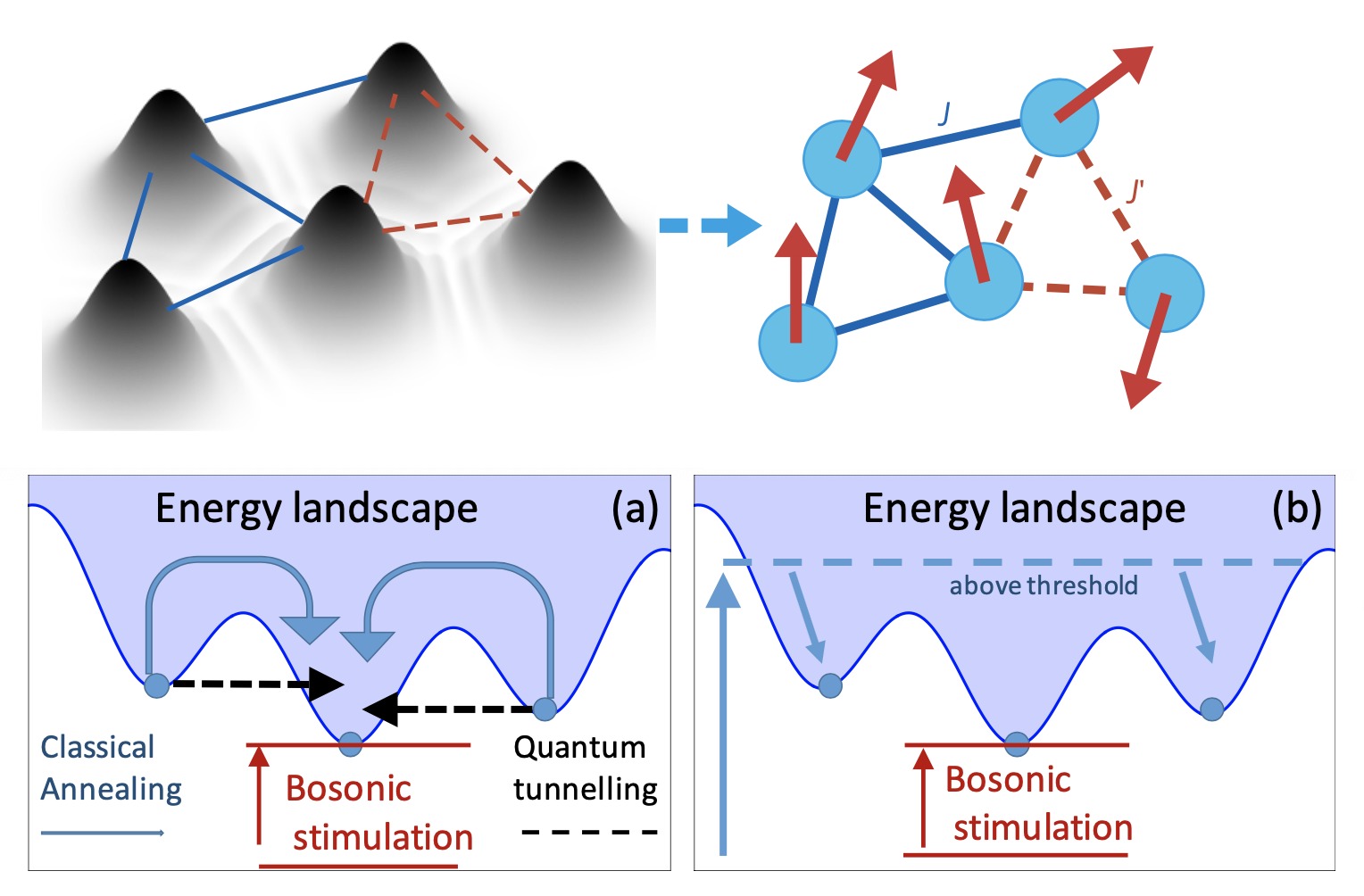}
		\caption{Top: Schematic of the condensate density map for a five-vertex polariton graph. The sign of the coupling depends on the separation distance between the sites and is either ferromagnetic (solid-blue lines) or anti-ferromagnetic (dashed-red lines). Each vertex of the graph polaritons represents a local phase mapped to a classical vector spin. Bottom: schematics of different types of annealing for finding the global minimum of the energy landscape of the simulated XY Hamiltonian \cite{BerloffNatMat2017}.}
		\label{9_polariton}
	\end{minipage}
	\hfill
\end{figure}

% Different dynamics
%If the time evolution of the reservoir of noncondensed particles is slow, the system of $N$ interacting coherent centres is better described by the following equations \cite{OurLattices}: 
%\begin{eqnarray}
% \dot{\Psi_i}&=& -i U |\Psi_i|^2\Psi_i  +(R_i- \gamma_c) \Psi_i + \sum_{j\ne i} { J}_{ij}\Psi_j,\label{ee1}\\
% \dot{R_i}  &=&\Gamma_i- \gamma_R R_i -R_i|\Psi_i|^2,\label{ee2}
 %\end{eqnarray}  
% where  $R_i$  is the occupation of the $i-$th  reservoir, $\Gamma_i, \gamma_R$ and $\gamma_c$ characterize the rate of particle injection into the reservoir and the linear losses of the reservoir and condensate, respectively.    If one replaces  $\Psi_i$  by the electric field and $R_i$ by the population inversion of the $i-$th laser, the result is a form of the  Lang-Kobayashi equations normally derived to describe the dynamical behaviour of coupled lasers from  Lamb's semiclassical laser theory  \cite{lang80,acebron}.  The total injection of the particles in the system of $N$ condensates at the fixed point is given by
%\begin{equation}
%\sum_{i=1}^N \Gamma_i=(\gamma_R+\rho_{\rm th})[N \gamma_c- \sum_{i=1}^N\sum_{j<i}^N J_{ij} \cos(\theta_i-\theta_j)].
%\end{equation}
%Similar to Eq.~(\ref{cond}), if the total injection into the system is minimal, the phases of coherent centres minimize the XY Hamiltonian.

\nb{
\subsection{Hybrid optical machines}
\label{Hybrid optical machines}
}
\nb{If physics provides the route to optimization we can ask if we can combine the best optimization algorithms with the optics functionality. The idea of using a hybrid approach to optimization, combining the best of both optical and electronic technologies, has the potential to greatly improve the efficiency and effectiveness of optimization processes. One possibility has been recently realized by Microsoft by building a hybrid opto-electronic platform that performs matrix-vector multiplication in the optical domain while iteratively applying the gradient descent to the minimum of the objective function using electronics \cite{kalinin2023analog}. This co-designing of unconventional hardware and algorithms paves the way for scalable architectures with compute-in-memory operation and spatial-division multiplexed representation of variables. }

\subsection{Mathematical description of optical optimisers}
\label{Mathematical description of optical optimisers}

\begin{figure}[!h]
	\begin{minipage}[c]{0.95\linewidth}
		\includegraphics[width=0.65\linewidth]{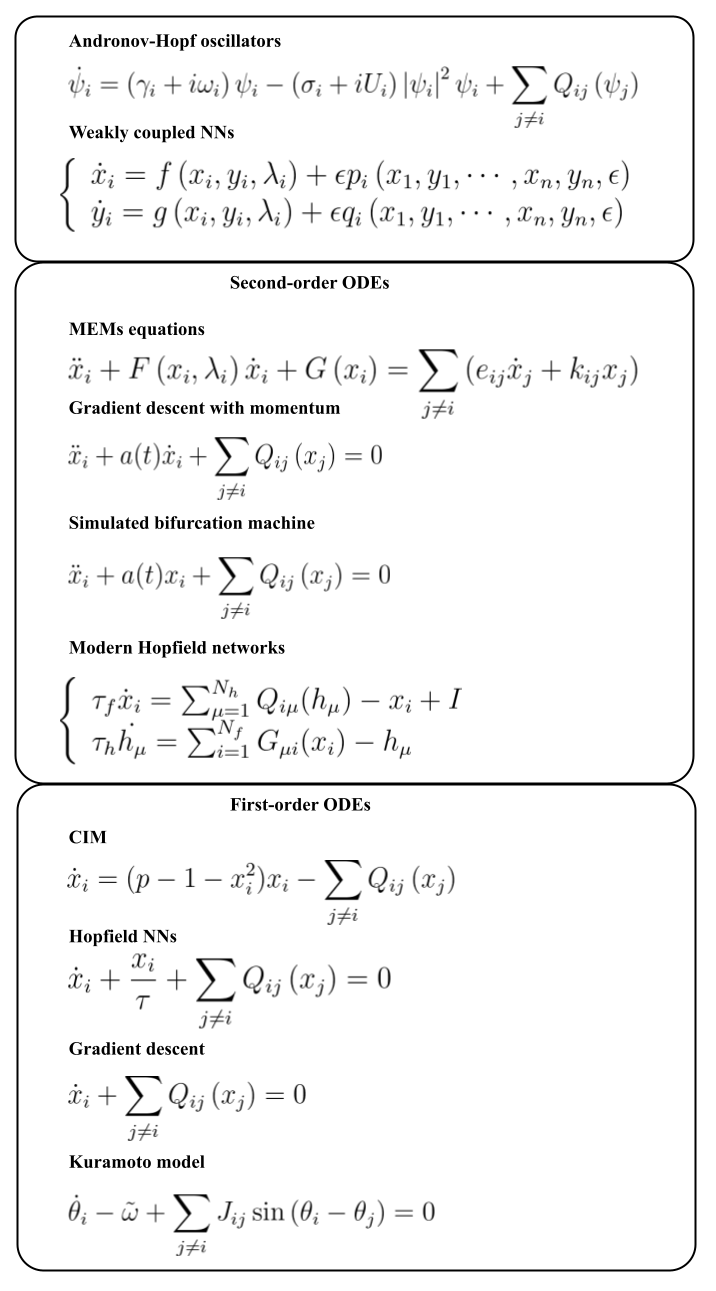}
		\caption{The ordered list of the main models from different branches of science used in the context of the optimization. The most general equations are closer to the top, starting with the canonical AHO model, which encompasses all equations below through certain transformations. In contrast, the simpler ones, like gradient descent, are located at the bottom. We non-rigorously group the models according to their use of the second-order derivative terms. The functions $\sum_{j \neq i} Q_{i j}\left(x_j\right)$ can have different forms such as $\eta \frac{\partial E}{\partial x_i}$ in gradient descent case, $Q_{i j}\left(x_j\right)=J_{ij} x_j$ or $Q_{i j}\left(x_j\right)=J_{ij} \varphi(x_j)$ in case of the Hopfield NNs.
		}
		\label{summary_mapping}
	\end{minipage}
	\hfill
\end{figure}

% Part 1
%Intro
Many existing optical machines can be described as the evolution of a set of $N$ classical degrees of freedom. The variety of optical platforms, such as atoms, polaritons, excitons, photons, etc., shares many similar features in their mathematical description. We present here the structured list of the main equations used in the context of nature-inspired physical systems and algorithms in Fig.~\ref{summary_mapping}. There are a few main reasons to highlight the unified picture of these equations. The first one is to show that all of the presented equations represent the same phenomena of the minimization principle and bifurcation dynamics, unifying many equations from math, physics, theory of neural networks, etc., see \cite{syed2022physics}. The second important feature is represented through the structure of the list. One can easily find the correct transformation between two chosen equations. This is the reason we non-rigorously placed them in the order so that the canonical Andronov-Hopf oscillators (AHO) model resides at the top of the list and the most straightforward gradient resides at the bottom. One can land at the required equations starting from the canonical model by using the proper transformation in the neighbourhood of the bifurcation leading to the solution or omitting some of the terms or derivatives.
Moreover, the difference between the presented equations appears to lie only in the chosen parametrization of the system. The optimization process can be done differently, even in the scope of classical dynamical systems depending on the chosen parameters. Such a unified framework allows one to merge many empirical results or to work in the same framework of a universal model for a better comparison of results. 

%Hamiltonian systems
The Principle of Least Action, the Principle of Minimum Power Dissipation (or Minimum Entropy Generation) or the Variational Principle are good demonstrations that ``optimization is built into physics'' \cite{vadlamani2020physics}. In Hamilton's formulation, the fundamental Principle of Least Action states that a true dynamical trajectory of a system between an initial and final configuration in a specified time is found by choosing the one among the set of possible imaginary trajectories that makes the action locally stationary (in other words have least action). Such a variational task is an excellent example of physics spawning complex problems. For even more complicated tasks, one can consider the formulations of the Principle of Least Action for classical and quantum field theories. We do not include the explicit Hamiltonian equations ($\dot{q}_i=\frac{\partial H}{\partial p_i}, \quad \dot{p}_i=-\frac{\partial H}{\partial q_i}$) in the second block in the Fig.~\ref{summary_mapping}. However, they also connected with the presented equations and served as a perfect entry point for considering the whole list from the physicist's point of view. Within the scope of this review, we restrict ourselves to classical systems with a discrete number of degrees of freedom and focus on PDEs that can be mapped into Newtonian-like equations of motion. Additionally, we do not pay much attention to the changes in the original equations of motion, such as Lagrange multipliers, holonomic constraints or relativistic factors.

% Entry equation - GD
Also depicted in  Fig.~\ref{summary_mapping} is the well-known classical gradient-descent dynamics with the target cost function defined by the gradients $-\sum_{j \neq i} Q_{i j}\left(x_j\right)$, which is the most straightforward equation among the presented dynamical systems. One can connect it with gradient descent with momentum (see the centre of Fig.~\ref{summary_mapping}) or the classical momentum (CM) method \cite{polyak1964some}, or it's improved version -- Nesterov accelerated gradient-descent \cite{nesterov1983method}).

% Kuramoto
Kuramoto model  is a well-known mathematical model used to describe synchronization phenomena occurring in a system of coupled oscillators \cite{kuramoto1975international,kuramoto1975formation,acebron2005kuramoto}. One can obtain this model from the AHO equations using the transformation that involved the eigenvalues and eigenvectors of the coupling matrix at the neighbourhood of the Hopf bifurcation, directly derive it from a nontrivial dissipative Hamiltonian or view this model as the gradient descent over the cost function corresponding to the classical XY Hamiltonian. 

%Kuramoto equations can be directly derived from a nontrivial dissipative Hamiltonian $\sum_{i=1}^N \frac{\omega_i}{2}\left(q_i^2+p_i^2\right)+\frac{K}{4 N} \sum_{i, j=1}^N\left(q_i p_j-q_j p_i\right)\left(q_j^2+p_j^2-q_i^2-p_i^2\right)$ after canonical transformation $I_i=\left(q_i^2+p_i^2\right) / 2$ and $\phi_i=\arctan \left(q_i / p_i\right)$, which gives the transformed Hamiltonian $\sum_{i=1}^N \omega_i I_i-\frac{K}{N} \sum_{i=1}^N \sum_{j=1}^N \sqrt{I_j I_i}\left(I_j-I_i\right) \sin \left(\phi_j-\phi_i\right)$ (with $I_j=1/2$) \cite{witthaut2014kuramoto}.

% Main equations
The bottom part of  Fig.~\ref{summary_mapping} consists of 
Hopfield NN and coherent Ising machine description. Hopfield NN is a recurrent artificial NN and can be viewed as the gradient descent variant with the effective projection term with characteristic time $\tau$ and the gradient terms $-\sum_{j \neq i} Q_{i j}\left(x_j\right)$, which are usually represented through  $-\sum_{j \neq i} J_{ij} \varphi(x_j)$, where $\varphi(x_j)$ is the projection function and $J_{ij}$ are the coupling strengths \cite{Hopfield1982}. CIM equations are very close to the Hopfield description. CIM is a network of OPOs, in which the ``strongest'' collective mode of oscillations corresponds to an optimum solution while going above the threshold of a particular Ising problem \cite{CIM1,CIM3}. The main difference between the classical description of CIM (which is debated to be essentially non-classical \cite{sankar2021benchmark,inui2021entanglement}) and Hopfield NN
is the additional pumping term $p$ and saturation mechanism $-x_i^{2}$. 
% \cite{CIM1,CIM3,CIM4,CIM5,CIM6,CIM_eqs2015,CIM_eqs2019} 

% SBM
The middle part of the Fig.~\ref{summary_mapping} contains simulated bifurcation machine (SBM) equations, which are inspired by the adiabatic evolution of classical nonlinear Hamiltonian systems exhibiting bifurcation phenomena \cite{goto2016bifurcation,goto2019combinatorial,goto2021high}. The higher derivative makes the connection with the physics more visible and improves the simulation algorithm's performance for specific parameters. 

% Lagrangian NNs
An alternative perspective on the connections between the physical Lagrangian/Hamiltonian systems and neural network evolution is given by the Modern Hopfield networks, or dense associative memories \cite{ramsauer2020hopfield,krotov2020large}.
Modern Hopfield networks operate with feature $x_i$ and memory (hidden) $h_{\mu }$ neurons that evolve as continuous variables in continuous time. The characteristic times for each group are $\tau_{f}$ and $\tau_{h}$. The symmetric coupling functions are chosen according to $Q_{i \mu}\left(h_\mu\right) = \xi_{i \mu} f_{\mu}$ and $G_{\mu i}\left(x_i\right) = \xi_{\mu i} g_{i}$ and connect only neurons from different groups, i.e. a feature neuron $i$ to the memory neuron $\mu$ and reverse. The outputs of the memory neurons and the feature neurons are denoted by non-linear functions $f(\{h_{\mu }\})$ and $g_{i}=g(\{x_{i}\})$ correspondingly. These functions can be represented as derivatives of the Lagrangian functions for the two groups of neurons $f_\mu=\frac{\partial L_h}{\partial h_\mu}$ and $g_i=\frac{\partial L_x}{\partial x_i}$. Choosing the specific Lagrangian will define the network's dynamics (or updates rule), which minimises the energy function. One can recover an effective theory of evolution by integrating out hidden neurons.

% Universal model and other equations
The upper part of Fig.~\ref{summary_mapping} contains the Andronov-Hopf oscillators model \cite{marsden2012hopf,andronov1971theory}, the canonical model describing the appearance of the bifurcations, which are among the essential phenomena observed in neuron dynamics, responsible for the periodic activity. The functions $Q_{ij}$ are accountable for the interaction between the $i$ and $j$ oscillators, while $\gamma_i$, $\omega_i$, $\sigma_i$, $U_i$ represent the effective gain, self-frequency, nonlinear dissipation and self-interactions respectively. Many lasers \cite{pal2020rapid}, photonic, polaritonic \cite{OurLattices}, and biological systems \cite{hoppensteadt1997weakly} exhibit the so-called Andronov-Hopf bifurcation at the threshold that can spawn the limit cycle behaviour. AHO can be an attractive choice for the unifying framework for many models presented here, which demonstrate a variety of collective phenomena \cite{syed2022physics}. Another important property of the AHO model is its canonicity, which means that in the vicinity of bifurcation, one can get every equation in Fig.~(\ref{summary_mapping}) below AHO by a certain transformation, which is not true in the reverse case.
One can also investigate the bifurcation phenomena and the time-dependent behaviour of the coefficients near the bifurcation point since it is the crucial mechanism for the system to find a solution to the optimization task. AHO shares its canonicity with another model - weakly interacting neural networks. The network consists of $N$ neural oscillators comprised of excitatory ($x_{i}(t)$) and inhibitory ($y_{i}(t)$), that evolve according to the presented dynamical equations \cite{hoppensteadt1997weakly}. In the local context, functions $f,g$ are responsible for the internal behaviour of the $i$th part of the system. At the same time, $p,q$ represents the external interactions, the strength of which is parametrized by the $\epsilon$ parameter. The explicit transformations between the equations can be found in \cite{hoppensteadt1996synaptic,hoppensteadt1996synaptic1,hoppensteadt2001synchronization}.

% Connections
We will omit the explicit description of the transformations that lead from the top equations to the bottom, while the detailed discussion and corresponding references can be found in \cite{syed2022physics}. Although the coupled microelectromechanical systems (MEMs) do not contain the optical elements, they are governed by similar optical second-order differential equations \cite{hoppensteadt2001synchronization}. The transition from the AHO to the CIM, Hopfield of SBM equations can also be found in \cite{syed2022physics}. It is important to remember that introducing sophisticated time dynamics of the parameters can improve the minimisation properties of each of the presented types of equations. For example, it is possible to introduce  specific time schedules (e.g. the chaotic amplitude method that anneals the coupling terms depending on the discrepancy between the oscillator amplitude and its saturation point \cite{leleu2021scaling}) or to introduce the high-order terms (e.g. $\dot{\psi_{i_k}} \sim \sum_{i_1, i_2, \ldots i_k-1}^N Q_{i_1, i_2, \ldots i_l, \ldots, i_k} \psi_{i_1} \psi_{i_2} \ldots \psi_{i_l} \ldots \psi_{i_k-1}^{*}$  \cite{stroev2021discrete}). 

An additional note should highlight the Principle of Minimum Power Dissipation and its role in analogue optimization machines. It was shown that many physical systems act through this principle and perform Lagrange function optimization \cite{vadlamani2020physics}. The Lagrange multipliers are given by the gain or loss coefficients or their time-varying parametrization; see, for example, the equations of the CIM. Depending on the characteristics of the machine, it can be helpful in many other applied domains. 

%~~~~~~~~~~~~~~~~~~~~~~~~~
% Part 2 - Quantum elements
The operation of optical machines consisting of $N$ elements can be described in a unified fashion as an evolution of a set of $N$ classical or quantum oscillators. The difference between classical and quantum comes from the system's initial state. It affects the speed and probability of finding the final state (usually a solution to a problem). If the occupation numbers of oscillators are large and somewhat uncertain and interactions are weak, then the system evolves as an ensemble of classical fields with corresponding classical-field action \cite{berloff2002scenario}. This analogy is valid for any bosonic oscillators, including optical: atoms, polaritons, excitons, photons, etc. For instance, the density matrix of a completely disordered, weakly interacting Bose gas with large and somewhat uncertain occupation numbers is almost diagonal in the coherent-state representation. The initial state can be viewed as a statistical ensemble of coherent states. To the leading order, each coherent state evolves along its classical trajectory. The evolution leads to an explosive increase of occupation numbers in the low-energy region of wave number space where the ordering process takes place \cite{berloff2002scenario}. Even if the occupation numbers are of order unity in the initial state, so that the classical matter field description is not yet applicable, the evolution, which can be described at this stage by the standard Boltzmann quantum kinetic equation, inevitably results in the appearance of large occupation numbers in the low-energy region of the particle distribution. Therefore, one can switch from the kinetic equation to the matter field description for the long-wavelength component of the field at a particular moment of the evolution when the occupation numbers become appropriately large. The optical system can be described using a classical matter field when this happens. However, the quantum dynamics before this moment plays a crucial role.
This fully quantum dynamics with entanglement and superposition of states allows for complete scanning of the high dimensional space of the system until the coherent state is found. After that, the system behaves classically. while this coherent state settles to a fixed point that is a solution to a problem. During the passage to the coherent state, the quantum effects should enhance the search for the optimal state and potentially lead to the quantum speed-up.

\subsection{Associative memory model}
\label{Associative memory model}

%Intro info
In this section, we present the associative memory model as one of the  NN models, which exploits the links with spin Hamiltonians.  This correspondence implies that many physical systems with nontrivial (nonzero) interaction potentials can be used as computational devices.

%Associative memory description
The standard model of associative memory \cite{Hopfield1982} uses a system of $N$ binary neurons, with values $\pm 1$. A configuration of all the neurons is denoted by a vector $\sigma_{i},i=1,.. , N.$ The model stores $K$ memories, denoted by $\xi_{i}^{\mu}, \mu=1,.. , K$, which are also binary. The model is defined by an energy function (or, further Lyapunov function), which is given by
\begin{equation}
E=-\frac{1}{2} \sum_{i, j=1}^{N} \sigma_{i} J_{i j} \sigma_{j}, \quad J_{i j}=\sum_{\mu=1}^{K} \xi_{i}^{\mu} \xi_{j}^{\mu},
\label{Associative memory}
\end{equation}
and a dynamical update rule that decreases the energy at every update. The fundamental problem is that when presented with a new pattern, the network should respond with a stored memory that most closely resembles the input. Many physical systems we considered in Section \ref{Description of physical optical platforms for optimization} can follow the gradient of this Lyapunov function, which automatically converts them into the ANN.

%Hebbian learning and model capacity
The theory of Hebbian learning addressed the associative memory \cite{hebb1940human,hebb1961distinctive} and describes how to prescribe the coupling coefficients between the neurons $J_{ij}$ (usually normalised by the number of patterns $K$). Usually, $J_{ij}$ is taken as the sum of the outer products of the stored patterns. One can find more ways to define coupling coefficients in the associative memory, e.g. pseudoinverse rule, Storkey learning rule or others. There has been a lot of work investigating this model's capacity, defined as the maximal number of memories that the network can store and reliably retrieve. It has been demonstrated that in the case of random memories, this maximal value is of the order of $K^{\max } \approx 0.14 N$ \cite{Hopfield1982,amit1985storing,mceliece1987capacity}. If one attempts to store more patterns, several neighbouring memories in the configuration space will merge, which produces a global minimum of the energy (\ref{Associative memory}), thus preventing recovery of the stored memories. It is possible to improve the capacity close to $K^{\max }=N$ by modifying the Hamiltonian (\ref{Associative memory}) in a way that removes second-order correlations between the stored memories \cite{kanter1987associative}.

%Functionality of the associative memory model
The simple associative memory model (\ref{Associative memory}) has many benefits. Firstly, it is quadratic in variables, which means that the energy gradient is linear to these variables. Therefore, one can easily calculate the corresponding updates of the neurons that lower the energy function (\ref{Associative memory}). The following consequence of this mathematical structure is that one can reproduce the energy function (\ref{Associative memory}) together with required dynamical behaviour using various physical hardware systems. To build an associative memory machine, one needs to connect the elements representing the analogue variables via nontrivial interaction potential proportional to the strength $J_{ij}$ and project the final stable state into the discrete domain to obtain the binary states of neurons. Furthermore, the model's simplicity allows one to easily modify and incorporate other extensions. Finally, the model's universality means it is possible to solve different tasks via associative memory by mapping between tasks; for example, the classification task can be reduced to pattern recognition/restoration.

%Background, connection with the Ising, NN and Optimization
Another well-known name for the associative memory model is the Hopfield NN, a form of recurrent ANN with binary threshold nodes. Moreover, Hopfield NN shares many other similarities with the physical spin-glass model and several combinatorial optimization tasks. For example, the Hopfield model is isomorphic to the Ising model of magnetism (for zero temperature) \cite{ising1925contribution}, which has been extensively analyzed in physical contexts. In combinatorial optimization, finding the ground state of the Ising model is NP-hard and can be related to the QUBO. Moreover, computing the statistical sum of the spin-glass has the same NP-hard complexity class, which was a significant obstacle in calculating its various thermodynamic quantities. Other examples of tasks are the Boolean satisfiability problem or SAT \cite{karp1972reducibility} and weighted MAX-2-SAT.

%Dynamical update rules
To fully define the associative memory model, one has to specify the dynamical update rule of the neurons. For instance, the update rule can describe the discrete state of neurons in discrete time steps:
\begin{equation}
\sigma_{i}(t+1)=\left\{\begin{array}{ll}
1, & \text { if } \Sigma_{j} J_{i j} \sigma_{j}(t)>0, \\
-1, & \text { otherwise,}
\end{array}\right.
\label{hopfield_discrete_1}
\end{equation}
with the same notation used in \ref{Associative memory}. The continuous version has the form:
\begin{equation}
\frac{d x_{i}}{d t}=-\frac{x_{i}}{\tau}+\sum_{j} J_{i j} g\left(x_{j}\right)+h_{i},
\label{hopfield_continuous_1}
\end{equation}
where $x_i$ denotes the mean state of the $i$-th neuron that can get continuous values in the initially defined range, $h_{i}$ is a direct input or bias coefficient in case the Lyapunov function (\ref{Associative memory}) has non-zero field, $g$ is a monotone function that bounds the continuous states and converts them into the discrete in the final state of convergence, e.g. makes the correspondence between the variables as $\sigma_i={\rm sign}(x_j)$, and $\tau$ is the characteristic time  of the convergence to an optimal or suboptimal solution.

%Computation/ Optimization vs NN
The analogue computation with the NN can be described as an evolution of the vector-state variables in the high-dimensional continuous space. One can precisely trace it using Eq.~(\ref{hopfield_continuous_1}). The vital aspect of such a differential equation structure is an existence of a Lyapunov function. This Lyapunov function $E$ behind the Hopfield NN can lead to the understanding of possible final states, which appear to be attractors of the system's dynamical behaviour. For both models, one can realise the dynamical state update using a particular hardware system described previously. However, one should differentiate between different regimes that can be realised on the hardware level: the task of finding the ground state (the global minimum) of the model and pattern restoration (descending on the surface of the Lyapunov function towards its nearest minimum).

%Lyapunov function
The explicit formula for the Lyapunov function in the discrete variant of the model with the non-zero field is:
\begin{equation}
E=-\frac{1}{2} \sum_{i, j=1}^{N} \sigma_{i} J_{i j} \sigma_{j}-\sum_{i=1}^{N} h_{i} \sigma_{i}.
\label{lyapunov_discrete_1}
\end{equation}
In the case of continuous variables Eq.~(\ref{hopfield_continuous_1}), the same function has a slightly different forms:
\begin{equation}
E=-\frac{1}{2} \sum_{i, j=1}^{N} \sigma_{i}
J_{i j} \sigma_{j}-\sum_{i=1}^{N} h_{i} \sigma_{i}+\frac{1}{\tau} \sum_{i=1}^{N} \int^{\sigma_{i}} g^{-1}(Z) d Z,
\label{lyapunov_continuous_1}
\end{equation}
where the last term appears due to the correspondence between the discrete and continuous state $\sigma_i = g(x_i)$. For $g(x)$, one usually picks the $g(x) = \operatorname{tanh}(x/\beta)$ function, where the time-dependent parameter $\beta$  tends to zero  during the evolution of the Hopfield NN forcing the last term of the Eq.~(\ref{lyapunov_continuous_1}) to disappear \cite{hopfield1985neural}. The essential property of the dynamical update rules is that the energy decreases through the system evolution, which leads to the final stable patterns in the phase space.

%Additional properties
The classical Hopfield NN has many modifications for the Lyapunov function, variables update rules and other features. One version is known as modern Hopfield NNs \cite{ramsauer2020hopfield}. Modern Hopfield networks with continuous states can be integrated into deep learning architectures because they are continuous and differentiable with respect to their parameters. Moreover, they retrieve patterns with just one update, conforming to deep learning layers. For these reasons, modern Hopfield networks can serve as specialised layers in deep networks to equip them with memories. Possible applications of Hopfield layers in deep network architectures find their way in multiple instance learning, defence against adversarial attacks \cite{krotov2018dense}, processing of and learning with point sets, sequence analysis and time series prediction, storing and retrieving reference data, e.g. the training data, outliers, high error data points, prototypes and many other purposes \cite{ramsauer2020hopfield}. Even more importantly, the functionality of the modern Hopfield networks can be compared with various methods from the ML domain, such as SVMs, random forest, boosting, decision trees, Bayesian methods and many others \cite{fernandez2014we,klambauer2017self}.

\begin{figure}[!h]
	\begin{minipage}[c]{0.95\linewidth}
		\includegraphics[width=\linewidth]{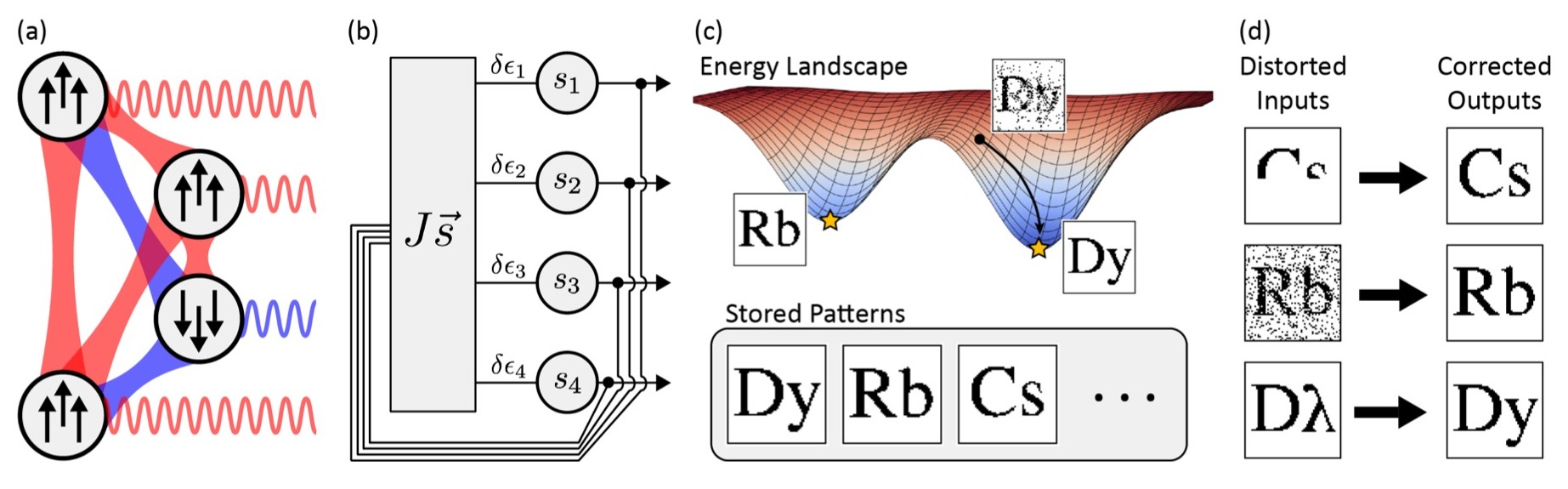}
		\caption{(a) Four nodes with the all-to-all coupling and sign-changing connectivity between spin ensembles. Blue and red show ferromagnetic versus antiferromagnetic $J_{ij}$ links. One can find the physical details in \cite{marsh2021enhancing}. (b) The realization of the Hopfield NN by the spin ensemble. Binary neurons $s_i$ of a single-layer network are recurrently fed back and subjected to a linear transform $J$ with the consequent element-wise threshold operation. (c) The Hopfield model exhibits an energy landscape with many metastable states. Energy-minimizing dynamics drive similar spin configurations to the stored local minimum, characterized by the basin of attraction. Too many memories make the basins of attraction vanish. (d) Schematic of the associative memory problem - recalling multiple stored patterns by completing distorted input images. Figure from \cite{marsh2021enhancing}.}
		\label{6_hopfield}
	\end{minipage}
	\hfill
\end{figure}

%Optical realization
As we mentioned above, many optical systems can perform optimization tasks. Since there are intrinsic similarities between this task and the associative memory model, one can exploit this relation to realize Hopfield NN using optical systems. Such realizations include previously discussed laser networks, Ising machines, photon \cite{leonetti2021optical} and polariton systems \cite{BerloffNatMat2017}, and confocal cavity QED NN \cite{marsh2020enhancing}, see Fig.~\ref{6_hopfield}. The connection between the optical networks and the Hopfield model is important since it allows one to incorporate such layers into more complex optical architectures without complicated adjustments.

\subsection{Higher-order systems}
\label{Higher order systems}

%Tensor extensions
One significant extension of the Hopfield model consists in  incorporating the tensor terms, which depend on the $\sigma_i$ variables polynomially in $n$ \cite{krotov2016dense}. The such extension allows one to increase the number of stored patterns to $K^{max}=\alpha_{n} N^{n-1}$, where $\alpha_{n}$ is a numerical constant. Moreover, it is possible to observe the so-called "feature to prototype transition" when increasing $n$ in the NN training. The prototype theory provides an alternative approach to learning in which objects are recognized as a whole. Although tensor terms are assumed not to be biologically plausible \cite{krotov2020large}, they can be reproduced on some artificial physical setups \cite{stroev2021discrete}. From this perspective, artificial tensor platforms can significantly benefit from such technological opportunities. The higher order Hopfield NNs \cite{joya2002hopfield} can be written as
\begin{equation}
\frac{dx_{l}}{dt}= -\frac{x_l}{\tau} +  \sum_{\bar{\Omega}} {\bf A}_{l,i_1,{\tiny{\cdot\cdot\cdot}}, i_k}^k s_{i_1}{\tiny{\cdot\cdot\cdot}} s_{i_k}; \hspace{3em}
s_{l}=  g\biggl(\frac{x_{l}(t)}{\beta}\biggr),\label{hopfield_tensor_1}
\end{equation}
where $x_{l}$ are real continuous variables, $g(x)$ is the threshold function and $\beta$ is the scaling parameter that can depend on time. Such systems can solve HOBO, see Eq.~(\ref{hobo}), because of the $k$-local coupling. 

%Experimental realizations
It was shown \cite{stroev2021discrete} that polariton systems above the threshold are described by
\begin{eqnarray}
\frac{d\Psi_{l}}{dt}&=& \Psi_l (\gamma_l(t)  - |\Psi_l|^2) +  \sum_{\bar{\Omega}} {\bf A}_{i_1,{\tiny{\cdot\cdot\cdot}}i_k}^k \Psi_{i_1} {\tiny{...}}\Psi_{i_k}^*,
\label{main1}\\
\frac{d\gamma_l}{dt}&= & \epsilon(\rho_{\rm th} - |\Psi_l|^2),
\label{main2}
\end{eqnarray}
where $\bar{\Omega}$ is the set of indices that  excluded index $l$.  Eq.~(\ref{main2}) describes the feedback mechanism that drives all $\rho_i=|\Psi_i|^2$ to a priori set values $\rho_{\rm th}$, $\epsilon$ characterizes how fast  $\gamma_i$ adjusts to changes in $\rho_i$. Next, we proceed with the different ways of connecting the practical computational tasks with the actual physical behaviour of the presented systems.

\section{Mathematical formulation of applications}
\label{Mathematical formulation of applications}

%Intro
This section considers a range of generic applications that follow from the network's ability to solve optimization problems or/and act as Hopfield networks. We start with the simple problems from classical computer science with the corresponding mapping to the QUBO \nb{or mixed-integer problems}. We then move to modern tasks that differ in information capacity and are considered to suffer from the so-called "curse of dimensionality", where it is more suitable to work with the probability distributions instead of the individual variables. However, in both cases, we do not pay attention to whether the presented mapping is efficient (like in the following subsection) or not (when one needs multiple sequential operations with a considerable amount of pre and post-processing in between). Some of the inefficient embeddings can still possess mathematical challenges and can be improved either in the general formulation or with task-specific information. At the end of this chapter, we discuss the NN architectures and their capabilities.

\subsection{Direct encoding/decoding}
\label{Direct encoding/decoding}

%Intro
This subsection describes the connections/correspondences between different computational tasks \cite{karp1972reducibility,cook1971complexity,garey1974some}.

%However, the big part of the classical computer science algorithms which are still important and relevant in a modern setting is missing. Nevertheless, the current focus is on the SAT formulation's universal approach. %SAT and its importance
The propositional satisfiability problem (SAT) lies at the heart of such correspondence. It is a fundamental problem determining whether a set of sentences in propositional logic is satisfactory. A clause is built as the disjunction, the logical OR (denoted by $\lor$) of some Boolean variables or their negations.
A set of several clauses, which must be satisfied simultaneously, is the conjunction, logical AND (denoted by $\wedge$) of the clauses. 
One can write a satisfiability problem in the general form:
\begin{equation}
(x_1 \lor x_2 \lor ... )\wedge (y_1 \lor y_2 \lor ...)\wedge... (... ),
\label{SAT1}
\end{equation}
where the $x_i,y_i$ are "literals", any of the original
variables or their negations. The form (\ref{SAT1}) is called a conjunctive normal form (CNF), and one can easily see that any logical statement between Boolean variables can be written as a CNF.
%The satisfiability problem has a universal structure and plays a central role in many fields.

%SAT variants
SAT is the first problem that was proven to be NP-complete \cite{cook1971complexity,karp1972reducibility}. Currently, no known algorithm efficiently solves each SAT instance. The question of its existence is equivalent to the famous P vs NP problem. Nevertheless, many heuristics SAT algorithms can solve problem instances involving a significant number of variables, sufficient for many applications. Additionally, many versions of the SAT problems exist, like 3-SAT and the generalization k-SAT, HORN-SAT, and XOR-SAT, which can better suit particular unconventional tasks.

%MAX-2-SAT
One specific SAT version - weighted MAX-2-SAT allows one to easily reformulate the task as QUBO. A simple 2-SAT has $m$ clauses of $2$ literals each. A MAX-2-SAT is the problem of assigning values that maximize the number of satisfied clauses. Weighted MAX-SAT gives each clause a positive weight so that the measure of violating the cost appears in the problem. To reformulate a weighted MAX-2-SAT problem as a QUBO, one has to use the fact that maximizing the weight of satisfied clauses is equivalent to minimizing the weight of unsatisfied clauses, and using the logic $\overline{x_i \vee x_j} = \overline{x_i} \wedge \overline{x_j}$. The final form looks then:
\begin{equation}
\max_{x_i} \sum_{i,j<i} w_{ij} x_{i} x_{j},
\end{equation}
which is the QUBO that has the same form as Eq.(\ref{ising}). Thus, the connection between the SAT (that can be easily converted into weighted MAX-2-SAT by use of the Boolean logic) and QUBO is revealed.

%Lucas
We know the Ising formulations for many NP problems \cite{lucas2014ising}. For example, one can find number partitioning, graph partitioning, clique existence, binary integer linear programming, exact cover, set packing (or maximal independent set), vertex cover, satisfiability (with the emphasis on 3SAT to MIS reduction), set cover, knapsack with integer weights, graph colouring, Hamiltonian cycles and paths, travelling salesman problem, Steiner trees, feedback vertex set, feedback edge set, graph isomorphisms among the covered problems, as well as some useful tricks for the near-term quantum adiabatic optimization devices. We mention some of them in a slightly different form below.

% minimal maximal matching
% clique cover
% job sequencing with integer lengths
% minimal spanning tree with a maximal degree constraint

\subsection{Logistics}
\label{Logistics}

%Definition
Logistic and planning problems are usually related to the well-known travelling salesman problem \nb{(TSP). TSP is a well-known optimization problem in which the goal is to find the shortest possible route that visits a given set of $N$ cities and returns to the starting city. In order to solve this problem, one approach is to use a spin matrix to represent the route.  Spin matrix has  elements $x_{v,i} \in \{0,1\}$  indexed by the vertex number  $v$  and its order in the path. 
The weighted edges $w_{uv} \ge 0$ from the edges set $E$  describe the cost of travelling between two cities. The Ising Hamiltonian can be written  as  }
\begin{eqnarray}
H_{\rm TSP} &=& A\sum_{i=1}^{N} \biggl(1 - \sum_{v=1}^N x_{v,i}\biggr)^2 +  A\sum_{v=1}^{N} \biggl(1 - \sum_{i=1}^N x_{v,i}\biggr)^2+ A\sum_{(uv)\notin E} \sum_{i=1}^N x_{u,i}x_{v,i+1} \nonumber \\
&+&B\sum_{(uv)\in E}w_{u,v} \sum_{i=1}^N x_{u,i}x_{v,i+1}.
\label{tsp1}
\end{eqnarray}
 \nb{The first two terms in the Hamiltonian ensure that each city is  in the route and appears only once. The third term in the Hamiltonian ensures that any adjacent cities in the route are connected, while the fourth term minimizes the sum of weights of all cities in the route. By choosing reasonable values for the constants $A$ and $B$ (e.g., $A$ should be large enough for $B>0$), it is possible to ensure that only valid routes are explored. }  %Fujitsu Digital Annealer has recently demonstrated that their performance is superior to the commercial solver, and comparable or superior to special-purpose solvers.

% This problem can be reformulated as the quadratic assignment problem. It consists of placing factories that minimizes the cost defined by the sum of the product of given flow and distance when there are N factories and N locations.

\nb{\subsection{Financial applications}}
\label{Portfolio optimization}

%Definition
Optimizing the portfolio selection means finding the most optimal combination of investments for an institution or individual. One of the modern portfolio optimization problem formulations has the following form \cite{markowitz1952portfolio}:
\begin{equation}
    \min_{0\le x_i\le 1}\lambda\bigg[\sum_{i=1}^N\sum_{j=1}^N J_{ij} x_i x_j\bigg]-(1-\lambda)\bigg[\sum_{i=1}^N\mu_i x_i\bigg], \quad \sum_{i=1}^N x_i=1,
    \label{optim}
\end{equation}
where $N$ is the number of different assets, and $x_i$ is the decision variable representing the proportion of capital invested in asset $i$. Here coupling coefficient $J_{ij}$ represents the covariance between returns of assets $i$ and $j$, $\mu_i$ is the mean return of asset $i$, and $\lambda \in [0,1]$ is the risk aversion parameter.  When $\lambda=0$, the model maximizes the portfolio's mean return, and the optimal solution will be formed only by the assets with the greatest mean return. When $\lambda=1$, only the total risk associated with the portfolio is minimized.

%Modifications
There are different modifications to the portfolio optimization problem. For instance, one can introduce bounding and cardinality constraints that specify that there should be $K$ different assets in the portfolio or/and the portion of some assets should be within certain bounds. This is achieved by \begin{equation}
    \sum_{i=1}^N z_i=K, \quad \epsilon_i z_i \le x_i \le \delta_i z_i, \quad z_i\in\{0,1\}.
\end{equation}
The cardinality-constrained mean-variance model is a mixed quadratic and integer programming problem in the NP-hard class of problems. \nb{It can be minimised if we follow the Hopfield dynamics \cite{fernandez2007portfolio,sadigh2012cardinality}.}
The discrete dynamics becomes
\begin{equation}
 x_i(t + 1) = G_i[-2 \lambda\sum_j  J_{ij} x_j(t) + (1-\lambda)\mu_i)],   
\end{equation}
where $G_i$ is a sigmoid with values in $[\epsilon_i, \delta_i].$ 
When solving any optimization problem, constraints usually appear in the energy function. However, in many cases of Hopfield networks, this is not necessary. Constraints on $x_i$ are satisfied using a sigmoid's activation function since its outputs already lie inside the desired interval. To fulfil the cardinality constraints, we begin with $3K/2$ neurons. After getting a minimum for the objective function, we remove the asset with the smallest output and repeat this process until the network has precisely $K$ assets. These remaining assets solve the original portfolio selection problem. To satisfy the constraint $\sum x_i=1$, one can use various adjustments, for instance, to evaluate the feasibility of every portfolio and change the proportions of capital $x_i$ to be invested in each selected asset \cite{fernandez2007portfolio}.

\nb{A variant of the Hopfield networks for solving mixed-integer programming was constructed to solve the financial transaction
settlement problem and implemented in an opto-electronic hardware \cite{kalinin2023analog}. The method is based on the discretised version of the MEMs equation shown in Fig.~\ref{summary_mapping} with annealing of the parameters $\gamma(t)$ and $\alpha(t)$ in
\begin{equation}
 \ddot{x_i} + \gamma(t) \dot{s_i} + \alpha(t) x_i = \sum_{j\ne i}  J_{ij} g(x_j), 
\end{equation}
where $g(x)$ once again caps the value of $x$ between $-1$ and $1$ (e.g. $g(x)=\tanh(x)$).
}
\nb{\subsection{Mixed-Integer and Box-Constrained Programming}
\label{Mixed-Integer Programming} }

\nb{The financial applications discussed in the previous section are not the only ones that utilize mixed-integer programming or box-constrained continuous optimization formulations. In fact, there is a strong correlation between discrete combinatorial problems and continuous non-convex programming, and it is often advantageous to represent discrete combinatorial problems in a continuous formulation that is conducive to many efficient methods. An early example is the Motzkin-Straus continuous formulation of the NP-hard MaxClique problem \cite{motzkin1965maxima}. The objective of the MaxClique problem is to identify the largest complete subgraph of a given graph such that all pairs of vertices within that subgraph are connected by an edge. If $J$ represents the graph adjacency matrix, then the MaxClique Ising Hamiltonian to be minimized is as follows:
\begin{equation}
 H= - \sum_i s_i + \lambda \sum_{i,j} (1-J_{ij})s_i s_j,
\end{equation}
where $s_i\in\{0,1\}$ and takes the value of $1$ if $i-$th vertex belongs to the maximum subgraph. The  first term represents the objective to have the largest subgraph possible and the second  penalizes the lack of the edge connecting the vertices in the subgraph. Parameters $\lambda$ is a Lagrange multiplier to enforce the constraint.  This problem, however, can be formulated as a continuous box-constrained quadratic optimization problem as \cite{motzkin1965maxima}
\begin{equation}
{\rm min_{x_i\ge 0}}\quad   - \sum_{i,j} J_{ij} x_i x_j+ \lambda (1-\sum_{i} x_i)^2. \label{ms}
\end{equation}
where $x_i$ is real. The optimal solution has nonzero entries with $x_i=1/d$, where $d$ is the size of the optimal subgraph. Moreover, the mixed-integer or box-constrained optimization is prevalent in numerous real-life applications. For instance, it can be used in scheduling and resource allocation problems, where decisions must be made regarding the allocation of limited resources to various tasks. It can also be applied in transportation and logistics, where it can help optimize routes and reduce costs. It can be utilized in manufacturing and production planning to optimize production schedules and minimize costs.
All of the optical platforms we considered above operate with continuous amplitudes, so it seems natural that they can natively encode and process these problems in analog fashion. Recently, there have been several proposals of solving these continuous problems. The mixed-integer programming can be addressed by a set of programmable bosonic quantum field modes since the eigenspectrum of the bosonic number operators consists of nonnegative integers so they can naturally represent integer variables \cite{khosravi2021mixed}. Box-constrained quadratic programming  problems can be solved by using modifications to the dynamics of CIMS based on OPOs \cite{ronagh2022non} or opto-electronic iterative machine \cite{kalinin2023analog}.
}

\subsection{Phase retrieval}
\label{Phase retrieval}

%Definition
\nb{The minimization of the XY model, which is used to solve the Quadratic Constrained Optimization (QCO) problem, is closely related to the challenging phase retrieval problem. The objective of the phase retrieval problem is to recover a signal or image from the magnitude of its Fourier transform \cite{PhaseRetrieval1,PhaseRetrieval2,PhaseRetrieval3}. This problem arises because signal detectors are typically only able to record the modulus of the diffraction pattern, resulting in a loss of information about the phase of the optical wave. The task is to determine  a complex vector ${\bf x}$ from the measurement real vector ${\bf b} = | {\bf A} {\bf x} |$, where the matrix ${\bf A}$ is complex-valued so that    \cite{PhaseRetrievalToXY} 
\begin{equation}
	\min_{x_j, u_i} \sum_i \bigg( \sum_j A_{ij} x_j - b_i u_i \bigg)^2. \label{pr1}
\end{equation}
Here  ${\bf u} $ is a complex vector of components with amplitude $1$ such that ${\bf A} {\bf x} = diag( {\bf b}) {\bf u}$. Introducing ${\bf J} = diag({\bf b}) ({\bf I} - {\bf A} {\bf A}^\dagger) diag({\bf b})$ (where the the Moore-Penrose inverse of a matrix ${\bf A}$ is denoted as ${\bf A}^\dagger$)  we rewrite Eq.~(\ref{pr1}) as minimization of the XY Hamiltonian 
\begin{equation}
\min_{\theta_i} \sum_{ij} J_{ij} {\bf s}_i  {\bf s}_j,
\end{equation}
where ${\bf s_i}=(\cos(\theta_i), \sin(\theta_i)).$}

\subsection{Machine learning}
\label{Machine learning}

%Intro
\nb{The exponential growth of data has surpassed our capacity to process it using human and computational resources. This has led to the development of data-driven methods and a shift from classical computer science paradigms to modern machine learning (ML) approaches. In ML, the focus is on predicting outcomes from given data, with the richness of data significantly influencing performance. Three crucial components in ML are data, features, and algorithms. Data can be collected in various ways and can be of great value depending on the context. Features represent the properties of considered objects and are essential for the success of ML approaches. However, determining and selecting the right features can be time-consuming. The choice of algorithm depends on the context and influences accuracy, speed, and computational complexity. These components are presented in order of their significance in the ML pipeline.}

The components were presented according to their significance in the ML pipeline. Simply saying, one can only extract useful information from a noisy but  meaningful dataset. The following subsection starts the discussion with the simple classical algorithms, which are the basis of many existing applications. Then, we outline the central ideas behind the main ML methods that will be the centre of attention for transferring into the special-purpose hardware. At the end of this chapter, we cover the wide range of capabilities of the NNs.

\subsubsection{Regression}
\label{Regression}

%Linear regression model 
\nb{Regression analysis is a statistical method that has been used for many years to estimate the relationships between a dependent variable and one or more independent variables. This technique allows for the modeling of the relationship between these variables, and can be used to make predictions about the dependent variable based on the values of the independent variables. One of the most commonly used forms of regression analysis is linear regression, which assumes that the relationship between the dependent and independent variables is linear. In linear regression, a line is fit to the data in such a way as to minimize the sum of squared errors between the observed values of the dependent variable and the predicted values based on the line.}
%Regression analysis is one of the earliest methods in statistical modelling that allows estimating the relationships between a dependent variable and independent variables. The most common form of regression analysis is linear regression. 
This model assumes that the dependent variables denoted by $y_i$ have a linear relationship depending on the m-vector of points ${\{x_{i1},\ldots ,x_{im}\}_{i=1}^{n}}$ with an addition of the disturbance terms $\epsilon_i$ in each case. This relationship can be written in the following form:
\begin{equation}
y_{i}=\beta_{0}+\beta_{1} x_{i 1}+\cdots+\beta_{m} x_{i m}+\epsilon_{i} = \sum_{j=0}^{m} \beta_{j} x_{i j} + \epsilon_{i}.
\label{linear_regression}
\end{equation}
To shorten notation we use the matrix form $\mathbf{y}=\boldsymbol{X} \boldsymbol{\beta}+\boldsymbol{\epsilon}$ where: $\mathbf{y} = \{y_i\}, \boldsymbol{X} = \{x_{ij}\},\boldsymbol{\beta} = \{\beta_{j}\} ,\boldsymbol{\epsilon} = \{\epsilon_i\}, (i=1,\ldots ,n), (j=0,\ldots ,m)$, with $x_{i0}=1$. The linear regression task is the estimation of the values of the regression coefficients $\beta_j$ given the data points $x_{ij}$ and observables $y_i$, so that the error term $\boldsymbol{\epsilon} = \mathbf{y}-\boldsymbol{X} \boldsymbol{\beta}$ is minimized. One can use different metrics for that purpose, such as the sum of squared errors of $\epsilon_i$ or others.

%Least-squares estimation
The most common parameter estimation technique is called the least-squares estimation. Here, the optimum parameter is defined through the minimization of the sum of the mean squared loss
\begin{equation}
\underset{\beta_j}{\min } \sum_{i=1}^{n}\left( \sum_{j=0}^{m} \beta_j x_{ij}-y_{i}\right)^{2},
\label{least_squares}
\end{equation}
which can be connected with the conventional QP. The optimal solution can be obtained by differentiating Eq.~(\ref{least_squares}) and equating it to zero with respect to parameters $\beta_j$. In matrix notation, the solution can be written as
\begin{equation}
\boldsymbol{\beta} = (\boldsymbol{X}^{T} \boldsymbol{X})^{-1} \boldsymbol{X}^{T} \mathbf{y}.
\label{least_squares_optimal_parameters}
\end{equation}

%Other estimates and regularization
There exist different modifications of the proposed procedure: generalized least squares, where one introduces a certain degree of correlation between the residuals $\epsilon_i$ (\ref{least_squares}), or the weighted least squares, where the knowledge of the variance of observations is incorporated as the coefficients $w_{k}$ before each of the residual. Moreover, intrinsically different techniques can be based on maximum likelihood estimation, Bayesian methods, or regularization. 

%Regression extensions (nonlinear)
A natural extension of linear regression is in replacing linear dependence with a polynomial. In the case of one argument, it is possible to rewrite Eq.~(\ref{linear_regression}) as
\begin{equation}
y_{i}=\beta_{0}+\beta_{1} x_{i}+\beta_{2} x_{i}^2+\cdots+\beta_{m} x_{i}^{m}+\epsilon_{i} = \sum_{j=0}^{m} \beta_{j} x_{i}^{j} + \epsilon_{i}.
\label{polinomial_regression}
\end{equation}
Given the data points $x_{i}^{j}$, the task is the same as Eq.~(\ref{least_squares}) except for the change in variables $x_{ij} \rightarrow x_{i}^{j}$. Similarly, it is possible to replace the polynomial basis with a set of some nonlinear functions $f(x_i)_j$, so that $x_{i}^{2} \rightarrow f(x_i)_j$.

%Regression extensions (multivariate)
Multiple linear regression is a generalization of linear regression with more than one independent variable. The basic model for multiple linear regression can be written in a similar form:
\begin{equation}
\mathbf{y}_{i}=\beta_{0}+\beta_{1} \boldsymbol{X}_{i1}+\beta_{2} \boldsymbol{X}_{i2}+\cdots+\beta_{m} \boldsymbol{X}_{im}+\boldsymbol{e}_{i} = \sum_{j=0}^{m} \beta_{j} \boldsymbol{X}_{ij} + \boldsymbol{e}_{i},
\label{multilinear_regression}
\end{equation}
where instead of variables $x_{ij}$ one has a set of matrix elements $\boldsymbol{X}_{ij}$ of size $k \times k$. Depending on the chosen norm for the matrix, it is possible to formulate the task of finding the regression coefficients. Taking the square Frobenius norm of the matrix, the search for optimal coefficients $\beta_i$ is equivalent to solving Eq.(\ref{least_squares}), except for the additional sum over the $k^2$ matrix elements:
\begin{equation}
\underset{\beta_j}{\min} \sum_{l=1}^{k^2} \sum_{i=1}^{n}\left( \sum_{j=0}^{m} \beta_j x_{ij}^{l}-y_{i}^{l}\right)^{2}.
\label{least_squares_multinomial}
\end{equation}
This can be extended further for multivariate linear regression or combined with the nonlinear basis with minor consequences concerning the parameters search and hardware operations, except for the much more complicated procedure for preprocessing the coefficients for any modification. Regression can be considered the simplest form of supervised learning.

\subsubsection{Classification}
\label{Classification}

%Intro
Classification is one of the popular tasks for ML. The purpose of classification is to sort the objects among the initially defined classes. The earliest algorithms include naive Bayes and decision trees. Here, we only consider Markov random field (MRF) encoding, which is the general case for such models.

%k nearest neighbours
The $k$-nearest neighbours algorithm is a non-parametric classification method used in statistics \cite{fix1985discriminatory,altman1992introduction}. It aims to classify the objects by considering their $k$ nearest neighbours with the defined class. The consequent attaching objects to a particular group is repeated until the convergence. We omit the explicit corresponding formulas because of their similarity with the $k$-means, the unsupervised clusterization algorithm, presented below. Both methods are usually based on Euclidean distances and can easily be transferred to special-purpose optimization hardware.

%Support vector machines
Another classification method is called a support vector machine (SVM). SVM is a supervised learning model that analyses data for classification purposes. It aims to construct a hyperplane between the classes of training data points in a high-dimensional space, emphasising a good separation achieved by maximising its margin. SVM was introduced in \cite{vapnik1963recognition} and standardised in \cite{cortes1995support}.

%Hard margin scenario
Linear SVM deals with the $n$ points $\bf{x}$ in the $m$-dimensional space, where each point has been assigned a binary class $y_i = \pm 1$. The task is to construct a hyperplane that divides these two groups with the maximum distance between them. The so-called "hard margin" scenario assumes that the initial data is linearly separable. One can start by constructing two parallel hyperplanes, separating groups of different classes with the largest distance between these two surfaces. The target surface between these hyperplanes is called the maximum margin hyperplane. To mathematically describe these surfaces, one can write:
\begin{equation}
\mathbf{w}^{T} \mathbf{x}^{i} -b = \sum_{j} w_j x_j^{i} - b = \pm 1,
\label{SVM}
\end{equation}
where $w_j$ are the components of the normal vector for both of the hyperplanes, $x_j^{i}$ are $m$-dimensional coordinates of the vector with the serial number $i$, $b$ defines the surface shift concerning the zero coordinates and $\pm 1$ defines the class. Everything above $y=1$ belongs to one class, and everything below $y=-1$ belongs to another. The offset of the hyperplane is determined by $b/\left \| \bf{w} \right \|$, while the marginal distance equals $2/\left \| \bf{w} \right \|$. To maximize the marginal distance, one has to minimize the norm of $\left \| \bf{w} \right \|$ and hence its square $\left \| \bf{w} \right \|^2$. This task can be reformulated as the optimization problem, adding the constraints that prevent data points from being positioned into the margin
\begin{equation}
\begin{array}{c}
\text {min} \left \| \bf{w} \right \| \\
\text { s.t. } y_{i}\left(\mathbf{w}^{T} \mathbf{x}^{i}-b\right) \geq 1, \text{ for } i=1,...,n
\label{SVM_opt}
\end{array}
\end{equation}

%Soft margin scenario
The natural extension of SVM is in considering a so-called "soft margin" case. It is assumed that the given data points are not linearly separable. In this case, one has to introduce a new kind of variable $\xi_i = \operatorname{max}(0,1-y_{i}\left(\mathbf{w}^{T} \mathbf{x}^{i}-b\right))$ for each point $i$, which is usually referred to as the hinge loss function, playing a regularizer role. Thus, it is possible to rewrite Eq.~(\ref{SVM_opt}) as
\begin{equation}
\begin{aligned}
&\operatorname{min} \frac{1}{n} \sum_{i=1}^{n} \xi_{i}+C\|\mathbf{w}\|^{2}\\
&\text { s.t.} y_{i}\left(\mathbf{w}^{T} \mathbf{x}^{i}-b\right) \geq 1-\xi_{i} \text { and } \xi_{i} \geq 0, \text { for all } i,
\label{SVM_opt_soft}
\end{aligned}
\end{equation}
where the constant $C$ regulates the interplay between the pure hard margin classifier and the soft margin one. We can reformulate the problem using the Lagrangian duality:
\begin{equation}
\begin{aligned}
&\displaystyle \max_{a_i} \sum_{i=1}^{n} a_{i}-\frac{1}{2} \sum_{i=1}^{n} \sum_{j=1}^{n} y_{i} a_{i}\left(\mathbf{x}_{i}^{T} \mathbf{x}_{j}\right) y_{j} a_{j}\\
&\text { s.t. } \sum_{i=1}^{n} a_{i} y_{i}=0, \text { and } 0 \leq a_{i} \leq \frac{1}{2 n C} \text { for all } i ,
\label{SVM_opt_Lagrange}
\end{aligned}
\end{equation}
where the norm vector $\bf{w}$ is expressed through the new variables $a_i$, so that $\mathbf{w} = \sum_{i=1}^{n} a_i y_i \mathbf{x}^{i}$, and the initial task of determining the offset of the surface is expressed via $\mathbf{b} = \mathbf{w}^{T} \mathbf{x}^{i} - y_i$. Thus, it is possible to obtain the problem, which has an exact QP formulation. This problem can be solved with the standard quadratic algorithms, thus, can be solved using special-purpose hardware.

%Extensions
It is helpful to mention the nonlinear extension of the SVM, which solves nonlinear classification task and can exploit the different functional forms of kernels. One can modify the scalar dot product in the quadratic form Eq.~(\ref{SVM_opt_Lagrange}) by a different kernel function $k(\mathbf{x}_i,\mathbf{x}_j)$ depending on the properties of the analogue hardware.

\subsubsection{Finding the principal eigenvector} 
\label{Finding the principal eigenvector}

% PageRank
Finding the principal (dominant) eigenvector of a given matrix ${\bf J}$ belongs to the $\mathbb{P}$-class of problems. However, finding such a dominant eigenvector on an ever-growing large matrix becomes a computationally intensive task incompatible with Moore's law. At the same time, a range of real-life problems would benefit from fast calculation of the principal eigenvector. \nb{The PageRank algorithm \cite{brin1998anatomy,page1999pagerank} is a well-known method for evaluating the relative importance of web pages based on the structure of the links between them. The web network is represented as a directed graph, where each page is a node and each hyperlink is an edge connecting one page to another. The PageRank algorithm computes a single score vector, known as the PageRank, for the entire database of web pages. The key assumption underlying this algorithm is that pages transfer importance to other pages via links, and thus the components of the PageRank vector determine the importance of each page. Mathematically, finding the PageRank vector is equivalent to calculating the principal eigenvector of the link-structure matrix, also known as the Google matrix. Calculating the principal eigenvector is also required in other fields such as social network analysis, bibliometrics, recommendation systems, DNA sequencing, bioinformatics,  and distributed computing systems \cite{ermann2015google,gleich2015pagerank,kalinin2021large}.}

%For instance, the PageRank algorithm \cite{brin1998anatomy,page1999pagerank} evaluates the relative importance of pages by exploiting the web link structure. The web network is represented as a directed graph, where each page is a node of the graph, and each hyperlink is an edge connecting one page to another. For the entire database of web pages, the PageRank algorithm computes a single score vector, the PageRank. The algorithm's key underlying assumption is that pages transfer the importance to other pages via links; hence, PageRank components determine the importance of pages. Mathematically, finding the PageRank vector is equivalent to calculating the principal eigenvector of the link-structure matrix, Google matrix. Besides, calculating the principle eigenvector is required in social network analysis, recommendation systems, bibliometrics, bioinformatics, DNA sequencing, and distributed computing systems \cite{ermann2015google,gleich2015pagerank,kalinin2021large}.

% Applications
There are numerous applications of PageRank to chemistry and engineering sciences networks to investigate and analyse complex systems. \nb{As systems grow in size and complexity, the interactions between their networks and subnetworks can become increasingly complicated and difficult to track. In order to organize and study these complexities, network analysis methods such as PageRank can be used. These methods provide a way to analyze the structure of the network and identify important nodes or relationships within it \cite{gleich2015pagerank}. }
For instance, MonitorRank diagnoses root causes of issues in a modern distributed system: error logs and tracing debugging information \cite{kim2013root}. %MonitorRank returns a ranked list of systems based on the likelihood that they contributed to, or participated in, an abnormal situation. Given an anomaly detected in a system, MonitorRank solves a personalised PageRank problem on a weighted, augmented version of the call graph. The weights and augmentation depend on the anomaly detected. 
PageRank has been used for road and urban space networks, which help predict traffic flow and human movement. It was shown that PageRank is the best network measure in predicting traffic on individual roads \cite{jiang2008self}. 
%Finding the principal eigenvector is a vital subtask and routine in more complicated assignments, which we will consider below as other methods.

% Optical advantages
\nb{Recent research has shown that optical systems can provide significant advantages for calculating the principal eigenvector \cite{kalinin2021large}. By choosing appropriate control parameters for these optical systems, the steady state of optical networks can be used to solve an eigenvalue maximization problem \cite{Aiyer1990}. This results in the identification of the energy state dictated by the signs of the eigenvector corresponding to the largest eigenvalue of the interaction matrix, i.e., the principal eigenvector. In particular, estimates presented in \cite{kalinin2021large} suggest that special-purpose optical machines for PageRank calculations may offer dramatic improvements in power consumption compared to classical computing architectures.}
%The advantage of using optical systems for calculating the principal eigenvector has been recently shown  \cite{kalinin2021large}. For a certain choice of control parameters of these optical systems, the steady state of optical networks can solve an eigenvalue maximization problem \cite{Aiyer1990}, which results in finding the energy state dictated by signs of the eigenvector corresponding to the largest eigenvalue of the interaction matrix, i.e. principal eigenvector. In particular, the estimates presented \cite{kalinin2021large} show that special-purpose optical machines for PageRank calculations may provide dramatic improvements in power consumption over classical computing architectures.

\subsubsection{Dimensionality reduction}
\label{Dimensionality reduction}

%Intro
Dimensionality reduction involves the transformation of data from 
the space with many dimensions into a low-dimensional space, usually preserving meaningful and valuable properties from the original data. It isn't easy to handle high-dimensional data in practice due to the growth of the space volume. Dimensionality reduction is standard in data-intensive fields. It can be used in signal processing, neuroinformatics, and bioinformatics \cite{van2009dimensionality,sorzano2014survey}. One can find its applications in recommender systems \cite{sarwar2000application}, semantic search \cite{jensen2004semantics} or as a primary tool in many domains involving numerical analysis.

%PCA model
One of the well-known methods for dimensionality reduction is the 
principal component analysis (PCA) \cite{pearson1901liii}. The idea behind PCA is to approximate particular data with linear manifolds of lower dimensions. PCA can be alternatively interpreted as finding subspaces of lower dimension in the orthogonal projection on which the data variation is maximum.

%Linear manifolds, argmin
The initial task behind the PCA is to find the best approximation of the data points using lines and surfaces. Given the set of vectors ${\bf{x}}_{1},{\bf{x}}_{2},\dots ,{\bf{x}}_{m}\in \mathbb{R}^{n}$, the aim is at finding the sequence of $k$ $k$-dimensional affine spaces $L_k \subset \mathbb {R} ^{n}$ that find
\begin{equation}
\displaystyle \min_{L_{k}} \sum_{i=1}^{m} \operatorname{d}^{2} \left({\bf{x}}_{i}, L_{k}\right) =\displaystyle \min_{a_{jl}} \sum_{i=1}^{m} \sum_{l=1}^{n}\left(x_{i l}-a_{0 l}-\sum_{j=1}^{k} a_{j l} \sum_{q=1}^{n} a_{j q}\left(x_{i q}-a_{0 q}\right)\right)^{2},
\label{PCA}
\end{equation}
for each $k$, where $\operatorname{d} \left({\bf{x}}_{i}, L_{k}\right)$ is the Euclidean distance from the point ${\bf{x}}_{i}$ to the $L_{k}$. Affine spaces $L_k$ are defined as the sets of linear combinations $L_k = \{{\bf{a}}_{0}+\alpha_{1} {\bf{a}}_{1}+\dots +\alpha_{k} {\bf{a}}_{k}\} $ with coefficients $\alpha_i \in \mathbb {R}$, while the vectors $\{{\bf{a}}_{1},{\bf{a}}_{2},\dots,{\bf{a}}_{k}\} \subset \mathbb {R} ^{n}$ form orthonormal basis in $\mathbb {R} ^{n}$.

% Steps
Eq.~(\ref{PCA}) is an optimization problem. The initial vector ${\bf{a}}_{0}$ is simply defined as the solution to
\begin{equation}
\underset{{\bf{a}}_{0}}{\operatorname{min}} \sum_{i=1}^{m} \operatorname{d}^{2}\left({\bf{x}}_{i}, L_{0}\right) = \frac{1}{m} \sum_{i=1}^{m} {\bf{x}}_{i}.
\label{starting_vector_PCA}
\end{equation}
The next component is found iteratively by subtracting the projection $\mathbf{x}_{i}={\mathbf{x}}_{i}-{\mathbf{a}}_{0}\left({\mathbf{a}}_{0}^{T} {\mathbf{x}}_{i}\right)$ (with the scalar product ${\mathbf{a}}_{0}^{T} {\mathbf{x}}_{i}$) for the vectors corresponding to $L_j$:
\begin{equation}
{\bf{a}}_{j}=\underset{\left\|{\bf{a}}_{j}\right\|=1}{\operatorname{argmin}}\left(\sum_{i=1}^{m}\left({\bf{x}}_{i}-{\bf{a}}_{j}\left({\bf{a}}_{j}^{T} {\bf{x}}_{i}\right)\right)^{2}\right).
\end{equation}
The iterations continue until the number of the affine space $k$ reaches the $n-1$ of the initial problem space dimension. Using the identity $||{\bf{x}}_{i}-{\bf{a}}_{j}\left({\bf{a}}_{j}^{T} {\bf{x}}_{i}\right)||^{2} = ||{\bf{x}}_{i}||^{2}-\left({\bf{a}}_{j}^{T} {\bf{x}}_{i}\right)^{2}$, one can easily map this task into the QP in ${\mathbf{a}}_{i}$ variables with the normalization constraints and the coupling matrix $J_{ij} = - x_{i}x_{j}$. To shorten the presented notation, the iterative procedure can be written similarly to maximization tasks 
\begin{equation}
\hat{\mathbf{X}}_{k}=\mathbf{X}-\sum_{s=1}^{k-1} \mathbf{X} \mathbf{w}_{(s)} \mathbf{w}_{(s)}^{\mathrm{T}},
\end{equation}
\begin{equation}
\mathbf{w}_{(k)}=\underset{\|\mathbf{w}\|=1}{\arg \max }\left\{\left\|\hat{\mathbf{X}}_{k} \mathbf{w}\right\|^{2}\right\},
\end{equation}
where $k$ is the number of principal component, $\mathbf{X}$ is the data matrix of size $n \times m$, $\mathbf{w}_{s} = \left(w_{1}, \ldots, w_{m}\right)_{(s)}$ are the weight coefficients. If the sequential operation is limited on the specific hardware system, one can still use the first iteration of the PCA method to obtain the largest eigenvalues of a matrix. One can find many alternative formulations of the PCA task, such as cancelling correlations between coordinates, i.e. covariance matrix diagonalization or singular value decomposition. 

%The maximization form is actually a Rayleigh quotient, because
%\begin{equation}
%\mathbf{w}_{(k)}=\underset{\|\mathbf{w}\|=1}{\arg \max }\left\{\left\|\hat{\mathbf{X}}_{k} \mathbf{w}\right\|^{2}\right\}=\arg \max \left\{\frac{\mathbf{w}^{T} \hat{\mathbf{X}}_{k}^{T} \hat{\mathbf{X}}_{k} \mathbf{w}}{\mathbf{w}^{T} \mathbf{w}}\right\},
%\end{equation}
%and the quotient's maximum possible value is the largest eigenvalue of the matrix $\hat{\mathbf{X}}_{k}^{T} \hat{\mathbf{X}}_{k}$. 

%Connection with the SVD
Singular value decomposition (SVD) is a special form of a rectangular matrix decomposition in the form
\begin{equation}
\mathbf{X}=\mathbf{U} \mathbf{\Sigma} \mathbf{V}^{\top},
\label{SVD}
\end{equation}
where $\mathbf{U}$ is the unitary matrix (representing the rotation as the linear transformation of the space in the geometrical interpretation), $\mathbf{\Sigma}$ is the rectangular diagonal matrix with non-negative real numbers on the diagonal (which are called the singular values, the action of the matrix has the interpretation of the corresponding scaling by diagonal elements) and $\mathbf{V}^{\top}$ is another unitary matrix (with the same additional rotation interpretation).

%Relation to the LSA
SVD is essentially vital in the standard techniques of the latent semantic analysis (LSA) \cite{hofmann2000learning,landauer2007lsa}, which purpose is to process documents and detect the relationship between libraries and terms. %Correctly performed SVD allows one to come up with a compact representation of this correspondence. 

%Correspondence
There is a direct correspondence between PCA and SVD decomposition. To perform the PCA, one has to find the eigenvectors of the covariance matrix $\mathbf{X} \mathbf{X}^{\top}$ (without the appropriate scaling factor $\frac{1}{n-1}$). The covariance matrix is diagonalizable, and with the normalized eigenvectors, one can write 
\begin{equation}
\mathbf{X} \mathbf{X}^{\top}=\mathbf{W D W}^{\top}.
\end{equation}
Applying SVD to the same data matrix $\mathbf{X}$ gives
\begin{equation}
\mathbf{X} \mathbf{X}^{\top}=\left(\mathbf{U} \mathbf{\Sigma} \mathbf{V}^{\top}\right)\left(\mathbf{U} \mathbf{\Sigma} \mathbf{V}^{\top}\right)^{\top}=\left(\mathbf{U} \mathbf{\Sigma} \mathbf{V}^{\top}\right)\left(\mathbf{V} \mathbf{\Sigma} \mathbf{U}^{\top}\right),
\end{equation}
which gives
\begin{equation}
\mathbf{W D W}^{\top}=\mathbf{U} \boldsymbol{\Sigma}^{2}. \mathbf{U}^{\top},
\end{equation}
Using this correspondence, one can perform the SVD decomposition as PCA on the special-purpose hardware.

\subsubsection{Clusterization}
\label{Clusterization}

%Intro
The most detailed description of clusterization is the separation of the objects on a specific basis. The goal can be defined as a classification without any prior information about the classes. The machine can set the number of clusters in advance or define them automatically. The algorithm determines objects' similarity by their marked features and puts the objects with many similar features in the same class. There are successful applications of clusterization in market analysis (consumer analytics), image compressing, data analytics, and anomaly detection.

%k-means clustering
$K$-means clustering is a clustering method that aims to partition $n$ observations into $k$ clusters. Each of these observations is located in the cluster with the nearest mean, also called a centroid \cite{lloyd1982least,steinhaus1956division,macqueen1967some}. There are heuristic algorithms that deal with such an assignment; however, the problem is NP-hard.

%Definition
Given a set of observations $\{\textbf{x}_1, ..., \textbf{x}_n\}$ in a $d$-dimensional space k-means algorithm aims to partition these observations into $k$ sets $\{S_1, S_2, ..., S_k\}$ to minimise the within-cluster sum of squares (or variance):
\begin{equation}
\underset{S_i}{\arg \min } \sum_{i=1}^{k} \sum_{\mathbf{x} \in S_{i}}\left\|\mathbf{x}-\boldsymbol{\mu}_{S_{i}}\right\|^{2},
\label{k-means}
\end{equation}
where $\boldsymbol{\mu}_{S_{i}}$ is the mean of points in the set $S_i$. One usually uses an iterative technique consisting of two steps to perform such an optimisation task. Given an initial set of k means $\boldsymbol{m}_{1}^1,...,\boldsymbol{m}_{k}^1$, the first step is to assign each observation to the cluster with the nearest mean, according to the Euclidean distance.
The next step is to recalculate the centroids:
$\boldsymbol{m}_{i}^{t+1}= \sum_{x_{j} \in S_{i,(t)}} \boldsymbol{x}_{j}$. Finally, the loop is run until the convergence. The algorithm uses the assigning of objects to the nearest cluster by Euclidean distance, and it is a suitable method for transferring its sequential operations to the specific hardware.

%Mean Shift
Mean shift is a high-dimensional-space analysis method for locating the maximum density function given a discrete number of data sampled from this arbitrary density function. It is helpful in complex hierarchical algorithms and is used in different computer vision or image processing domains.

%MS definition
Given data points $\boldsymbol{x}_i$ in $n$-dimensional space, one can use the kernel function $k(r)$, acting on the norm value $r$, to determine the mean shift's value. The kernel function has to be non-negative, non-increasing and continuous. One can use the flat kernel so that $k(r)=1$ if $r<r_0$ and $0$ outside. Each iteration consists of calculating the function
\begin{equation}
F(x) = \sum_{i} k\left(\frac{(\boldsymbol{x}-\boldsymbol{x}_i)^{2}}{\alpha^2}\right),
\label{mean_shift}
\end{equation}
where there are $\alpha$ states. The maximum of $F(x)$ is computed using the square norm. 

\subsection{Neural networks}
\label{Neural networks}

%Intro
ANNs are often associated with ML. We considered the associative memory model, a simple recurrent shallow NN in the subsection \ref{Associative memory model}. This model can be extended to higher-order systems, simultaneously gaining many useful properties. However, optical systems are not tied only to this type of architecture \cite{genty2021machine}.

%Definition
Any NN can be defined as a set of neurons and connections between them. An artificial neuron's task is to take input numbers, process them in a certain way (executing a special function), and output the results. The standard mathematical transformation of one NN layer can be written as $\varphi(\sum_{i=0}^{N} w_i x_i + b)$, where $w_i$ denote the weights for the input data points $x_i$ (or independent variables), and the constant $b$ is the shift called bias. Here, the $\varphi$ is a nonlinear activation function. A single-layer NN that performs a similar transformation and produces a single output number is called a perceptron. The perceptrons, assembled into multilayered structures, are called multilayer perceptrons. The introduction to the NNs is presented in \cite{gurney1997introduction,anderson1995introduction,krose1993introduction} with more modern work \cite{yegnanarayana2009artificial} and the latest results after the deep learning breakthrough \cite{goodfellow2016deep}.

%Parameters of the NN
The activation function $\varphi$ plays an essential role in the NN design because the output signal would be simply a linear function in its absence. There are many functional activation functions, such as binary step function, sigmoid (or logistic function), hyperbolic tangent, etc. They allow a NN to map an input to the output appropriately. Thus, NN is considered a universal function approximator \cite{cybenko1989approximation}. To choose the NN weights, one usually uses the backpropagation procedure \cite{hecht1992theory,chauvin1995backpropagation}, although there are many alternatives. Backpropagation consists of tuning the NN weights according to the difference between the actual output value of the network and the predicted one, with the final goal of minimizing this discrepancy or the cost function. The tuning procedure involves computing the total discrepancy gradients on each layer, starting from the final one and updating the corresponding weight values. Through the extensive number of such iterations, there is a chance that the weights will be tuned in the desired way.

% Deep and shallow correspondence
Many deep NN (NN with many layers) can be mapped into a shallow one with a significant overhead on the number of neurons in the standard layer. That means that any deep NN functionality can, in principle, be performed on a physical device suitable to a shallow architecture. With an appropriate mapping, both networks will have the same approximation qualities \cite{ba2013deep,poggio2017and,mhaskar2016deep,khrulkov2017expressive}.

%Mimicing the algorithms
A well-trained NN can approximate many complicated algorithms, some of which are presented in this review. However, one has to provide enough input conditions and good output answers, especially when the problem is of high computational complexity. In addition, however, the resulting correlations need to be better understood. The valuable properties of the NNs go far beyond the optimization domain. We will consider some of them below.

\subsubsection{Neural networks and dynamical systems}
\label{Neural networks and dynamical systems}

%Physics-informed machine learning
Using ML models in the domain of physical sciences, i.e. incorporating physical laws and domain knowledge into neural architectures, is called physics-informed machine learning (PIML). It provides a powerful approach to modelling different physical phenomena. This rapidly growing field can pursue many other goals. Among them are constructing better predictive models with high accuracy and reliable generalization abilities, increasing data processing rate, accelerating the dynamical processes through optimized architecture, and solving inverse problems with interpretable models. One should expect that emulating complex nonlinear dynamics should benefit from the PIML. This can be seen in the applications in weather forecasting \cite{kashinath2021physics}, modelling of turbulence \cite{wu2018physics,mohan2020spatio}, nonlinear dynamics \cite{chen2021physics,karniadakis2021physics}, applications of the ML to the Koopman operator theory \cite{brunton2021modern}. Optical hardware can be potentially used to speed up these applications.

%Correspondence with the dynamical systems
The correspondence between NN architectures and dynamical systems is straightforward. Some of the NN can be viewed as discretizations of dynamical systems, which is true in reverse order - one can design NNs to have specific properties, such as invertibility \cite{celledoni2021structure}. This correspondence can broaden the applicability of the potential optical hardware. Their connection with dynamical systems and deep learning can be found in \cite{weinan2017proposal}. The generalization of the optimization algorithms inspired by different optical systems has canonical universality property \cite{bravetti2019optimization}.

\subsection{Probabilistic graphical models}
\label{Probabilistic graphical models}

%Intro
Graphical models provide a natural tool for dealing with uncertainty and complexity \nb{in a wide range of tasks and  establish a powerful framework for representing and analyzing complex data structures. The graph theoretic approach used in these models offers an  interface for modeling data and designing efficient general-purpose algorithms. Many models used in disparate fields such as statistics, systems engineering, information theory, and pattern recognition can be considered special cases of the general graphical model formalism. Graphical models are particularly useful for representing joint probability distributions and performing inference based on observed data  \cite{murphy2001introduction,lauritzen1996graphical,jordan2004graphical}.}

%Definition
Probabilistic graphical models (PGMs) are graphs with the nodes represented by random variables, while edges connecting them represent conditional independence assumptions. \nb{Probabilistic graphical models (PGMs) provide a compact representation of joint probability distributions. There are two main types of graphical models: undirected models, also known as Markov random fields (MRFs), which are widely used in the physics and vision communities, and directed models, also known as Bayesian networks (BNs), belief networks, or causal models, which are more popular with the artificial intelligence and machine learning communities }\cite{murphy2001introduction}.

%Definition and Ising Similarity
The spin Hamiltonians are particularly useful for PGMs. We recall the Ising spin model of Eq.~(\ref{ising}) . Each spin variable $s_{i}$ can be treated as a random binary variable so that their coupling strengths serve as the connections between random variables. Certain configurations of spin variables $X = \left(x_{1}, \ldots, x_{N}\right) \in\{-1,+1\}^{N}$ is called an assignment. The probability of an assignment in the PGM is given by
\begin{equation}
\text{P}(s_i=x_i)=\frac{1}{Z} \exp \left(-\sum_{i=1}^{N} \sum_{j=1, j<i}^{N} J_{i j} x_{i} x_{j} \right),
\label{probability_assignment}
\end{equation}
where
\begin{equation}
Z=\sum_{X \in\{-1,+1\}^{N}} \exp \left(-\sum_{i=1}^{N} \sum_{j=1, j<i}^{N} J_{i j} x_{i} x_{j} \right)
\label{partition_function}
\end{equation}
is the so-called partition function.

%%% Some works on the computational complexity of the Ising-related algorithms include \cite{schraudolph2008efficient}, where authors give polynomial-time algorithms for the exact computation of lowest-energy states, worst margin violators, partition functions, and marginals in PGM in case of the planar Ising model. Another recent work investigates tractable Ising models from the sampling and statistical inference complexity \cite{likhosherstov2019inference} with the possible extensions of algorithms for $K_{33}$ (minor)-free topologies of zero-field Ising models, which is the generalization of the previous results on planar graphs.

%Quantities of interest
There are several quantities of interest in the PGMs. First, it is the inference task - the computation of the quantity $Z$ given by Eq.~(\ref{partition_function}). The exact inference is the computation of $Z$ with all possible assignments, which is a hard problem for an arbitrary graph. The running time of the exact algorithms of finding $Z$ is exponential in the size of the largest cluster of corresponding graph nodes. There are rare cases of the Ising model graphs when it is possible to compute its partition function in polynomial time, but the problem of computing $Z$ is generally hard. Hence, the approximate inference is widely used. Other quantities of interest can include finding the low-energy states (low energy sampling), worst margin violators, constituents of partition functions - assignment likelihood and marginal probabilities and certain moments concerning the partition function and the target value.

%Spin machines connection and programmability
Some popular approximate inference methods include sampling (Monte Carlo), variational methods and message-passing algorithms \cite{murphy2001introduction}. Since many optical spin machines are not flexible in terms of programmability compared to conventional computers, one can hardly exploit sophisticated methods in hardware operations. That is why the sampling procedure is the most promising application from the hardware perspective, especially for inference tasks. Expanding the spin machines' functionality is a promising direction, given the speed and energy efficiency of the optical efficiency domain.

%Universality
The physical system often realises the symmetric coupling coefficients, making the model undirected. Using the system-specific devices that redirect light, it is possible to introduce the asymmetry in the variable connections, which opens the path to the directional PGM. In addition to the universality concept, one can see many practical tasks encoded into the Ising model (such as portfolio optimisation) as special cases of the PGMs. Moreover, the hardware's ability to realise the high-order interaction terms allows one to encode complicated conditional dependencies with little or no overhead on the number of variables. 
%(for example, SLM in the optical context)

%Applications
%The famous example of sprinkler and rain system, both depending on the weather and influencing the state of the grass, can be encoded into directed graphical models. 
%Applications
However, the application scope of optical machines aimed at simulating PGMs is far beyond the scope of this problem. One can encode complicated large graphs with many factors, representing large-scale practical problems and efficiently use them as supporting decision-making networks. There are also applications in control theory and game theory. For example, PGM can compactly model joint probability distributions using sparse graphs to reflect conditional independence relationships in complex systems. It is possible to decompose similarly multi-attribute cost functions (or utility functions). For instance, let the general cost function be a sum of local cost functions. Each local term has parental nodes (random variables or factors), which it depends upon.
Moreover, some of the utility nodes will also have action (control) nodes similar to parent nodes because they depend on the state of the environment and the performed actions. The resulting graph is called an influence diagram. Using such a diagram, one can perform sampling, similar to the inference task, and compute the optimal (sequence of) action(s) to maximise or minimise the cost function \cite{murphy2001introduction,cowell2006probabilistic}. The application of the same strategy was used in multi-person game theory \cite{kearns2001graphical}. In such a way, one can exploit optical spin machines to investigate dynamical systems and decision policy on factor graphs. There are many more applications of such correspondence between spin system functionality, control theory, and decision-making. The advantages of optical systems will benefit large complex graphs with complex connections between units \cite{murphy2001introduction}. Exploring the functionality of optical machines with respect to different paradigms is a promising research direction.

\subsection{Image processing}
\label{Image processing}

%Similarity
Several problems in computer vision can be formulated as binary quadratic programs, a particular case of  QUBO. One can also see the similarity with PGMs. The conventional approach to such problems is to use the semidefinite relaxation technique, which appeared to be quite efficient \cite{wang2016large}. The problems discussed include image co-segmentation, image segmentation with different constraints, graph matching, image deconvolution, graph bisection, and others. The computational complexity of these problems is high, which makes it necessary to propose an improved version of the semidefinite programming approach, which is more efficient and scalable. Some of these formulations are listed with little corresponding details, and we refer the reader to the original work \cite{joulin2010discriminative}.

\begin{equation}
\begin{array}{rl}
\min_{{\bf x}\in\{-1,+1\}^N} & \mathbf{x}^{\top} \mathbf{A} \mathbf{x} \\
\text { s.t. } & \left(\mathbf{x}^{\top} \mathbf{t}_{i}\right)^{2} \leq \kappa^{2} n_{i}^{2}, i=1, \ldots, s,
\end{array}
\label{image_co-segmentation}
\end{equation}
is the image co-segmentation task with the matrix $\mathbf{A}$ \cite{joulin2010discriminative}, $s$ is the number of images, $n_{i}$ is the number of pixels for $i$-th image, and $n=\sum_{i=1}^{s} n_{i} . \mathbf{t}_{i} \in\{0,1\}^{n}$ is the indicator vector for the $i$-th image, $\kappa \in(0,1]$ is a parameter.

\begin{equation}
\begin{array}{rl}
\min _{\mathbf{x} \in\{0,1\}^{KL}} & \mathbf{h}^{\top} \mathbf{x}+\mathbf{x}^{\top} \mathbf{H} \mathbf{x} \\
\text { s.t. } & \sum_{j=1}^{L} \mathbf{x}_{(i-1) L+j}=1, i=1, \ldots, K \\
& \sum_{i=1}^{K} \mathbf{x}_{(i-1) L+j} \leq 1, j=1, \ldots, L
\end{array}
\label{graph_matching}
\end{equation}
is the graph matching task and $x_{(i-1) L+j}=1$ if the $i$-th source point is matched to the $j$-th target point; otherwise it equals to $0$. $h_{(i-1) L+j}$ records the local feature similarity between source point $i$ and target point $j$; $H_{(i-1) L+j,(k-1) L+l}=\exp \left(-\left(\mathrm{d}_{i j}-\mathrm{d}_{k l}\right)^{2} / \sigma^{2}\right)$ encodes the structural consistency of source point $i, j$ and target point $k, l$. The corresponding details can be found in \cite{schellewald2005probabilistic}.

\begin{equation}
\min _{\mathbf{x} \in\{0,1\}^{n}}\|\mathbf{q}-\mathbf{K} \mathbf{x}\|_{2}^{2}+\mathrm{S}(\mathbf{x})
\label{image_deconvolution}
\end{equation}
is the image deconvolution task, where $\mathbf{K}$ is the convolution matrix corresponding to the blurring kernel $\mathbf{k}, \mathrm{S}$ denotes the smoothness cost, $\mathbf{x}$ and $\mathbf{q}$ represent the input image and the blurred image respectively \cite{wang2016large}.

\begin{equation}
\begin{aligned}
\min _{\mathbf{x} \in\{-1,1\}^{n}} &-\mathbf{x}^{\top} \mathbf{W} \mathbf{x}, \\
\text { s.t. } & \mathbf{x}^{\top} \mathbf{1}=0
\end{aligned}
\label{graph_bisection}
\end{equation}
is the graph bisection task with $W_{i j}=\exp \left(-\mathrm{d}_{i j}^{2} / \sigma^{2}\right), \text { if }(i, j) \in \mathcal{E}$; and $0$ otherwise, where $\mathrm{d}_{i j}$ denotes the Euclidean distance between $i$ and $j$. These tasks can potentially be mapped into the special-purpose hardware dealing with quadratic assignments or low-level programmable tasks.

\subsection{Several examples of hardware embeddings}
\label{Several examples of hardware embeddings}

%Characteristic of the embeddings
Here we consider the hardware representation of several tasks we considered previously. We characterise each assignment stating its possible embedding on the particular hardware type - spin machines and characterise several parameters of such embedding. We consider discrete and continuous variables (the latter can require additional overhead on the number of discrete operational units), direct mapping of the problem coefficients or partial concerning other factors, whether the hardware requires the consequent manner of operations or not, incorporating additional constraints into the coefficients of the problem and other details. Overall, these factors determine whether the possible embeddings are efficient or not (significant overhead, consequent operations, etc.). We present the examples in Table \ref{embeddings}.

\begin{table}
\caption{Example of possible encoding for the several tasks}
\centering
    \begin{tabular}{ | l | p{8cm} | p{6.8cm} |}
    \hline
    Assignment & Formulation & Details \\ \hline
    MIN-2-SAT & \begin{equation}
\min _{x_{i}\in\{0,1\}^{n}} \sum_{i, j<i} w_{i j} x_{i} x_{j}
\end{equation} & Discrete variables, direct mapping,
    straightforward operation with respect to dynamical updates, efficient. \\ \hline
    Phase retrieval & \begin{equation}
\begin{aligned}
\min \sum_{i j} M_{i j} u_{i} u_{j} 
\\
\text { s.t. } \left|u_{i}\right|=1, i=\overline{1, n}
\end{aligned}
\end{equation} & Continuous variables, direct mapping on the QCO (Eq.(\ref{xy})), straightforward operation, efficient concerning the QCO hardware. \\ \hline
    Regression & \begin{equation}
\min _{\beta_{j}} \sum_{i=1}^{n}\left(\sum_{j=0}^{m} \beta_{j} x_{i j}-y_{i}\right)^{2}
\end{equation} & Continuous variables, partial mapping, straightforward operation, inefficient concerning the variables mapping. \\
    \hline
    SVM & \begin{equation}
\begin{array}{c}
\text {min} \left \| \bf{w} \right \| \Rightarrow \text {min} (\sum_{j} w_{j}^{2}) \\
\text { s.t. } y_{i}\left(\mathbf{w}^{T} \mathbf{x}_{i}-b\right) \geq 1, \text{ for } i=1,...,n
\end{array}
\end{equation} & Continuous variables, partial mapping, straightforward operation, inefficient concerning the variabes mapping. \\
    \hline
    k-means & \begin{equation}
\underset{S_{i}}{\min } \sum_{i=1}^{k} \sum_{\mathbf{x} \in S_{i}}\left\|\mathbf{x}-\boldsymbol{\mu}_{S_{i}}\right\|^{2}
\end{equation} & Continuous variables, partial mapping, consequent operation, inefficient variabes mapping and operation setup. \\
    \hline
    Graph bisection & \begin{equation}
\begin{aligned}
\min _{x_i \in\{-1,1\}} &-\sum_{i, j} w_{i j} x_{i} x_{j} \\
\text { s.t. } & \sum_{i} x_{i}=0
\end{aligned}
\end{equation} & Discrete variables, partial mapping due to the constraints, straightforward operation, inefficient representation of the constraints. \\
    \hline
    Image \\ co-segmentation & \begin{equation}
\begin{aligned}
\min _{x_i \in\{-1,+1\}} & \sum_{i, j} a_{i j} x_{i} x_{j} \\
\text { s.t. } &(\sum_{j} t_{i,j} x_{j})^{2}=0 \leq \kappa^{2} n_{i}^{2}, i=1, \ldots, s
\end{aligned}
\end{equation} & Discrete variables, partial mapping due to the constraints, straightforward operation, overhead on auxiliary variables, inefficient concerning the constraints and overhead. \\
    \hline
    \end{tabular}
\label{embeddings}
\end{table}

\begin{figure}[!h]
	\begin{minipage}[c]{0.95\linewidth}
		\includegraphics[width=\linewidth]{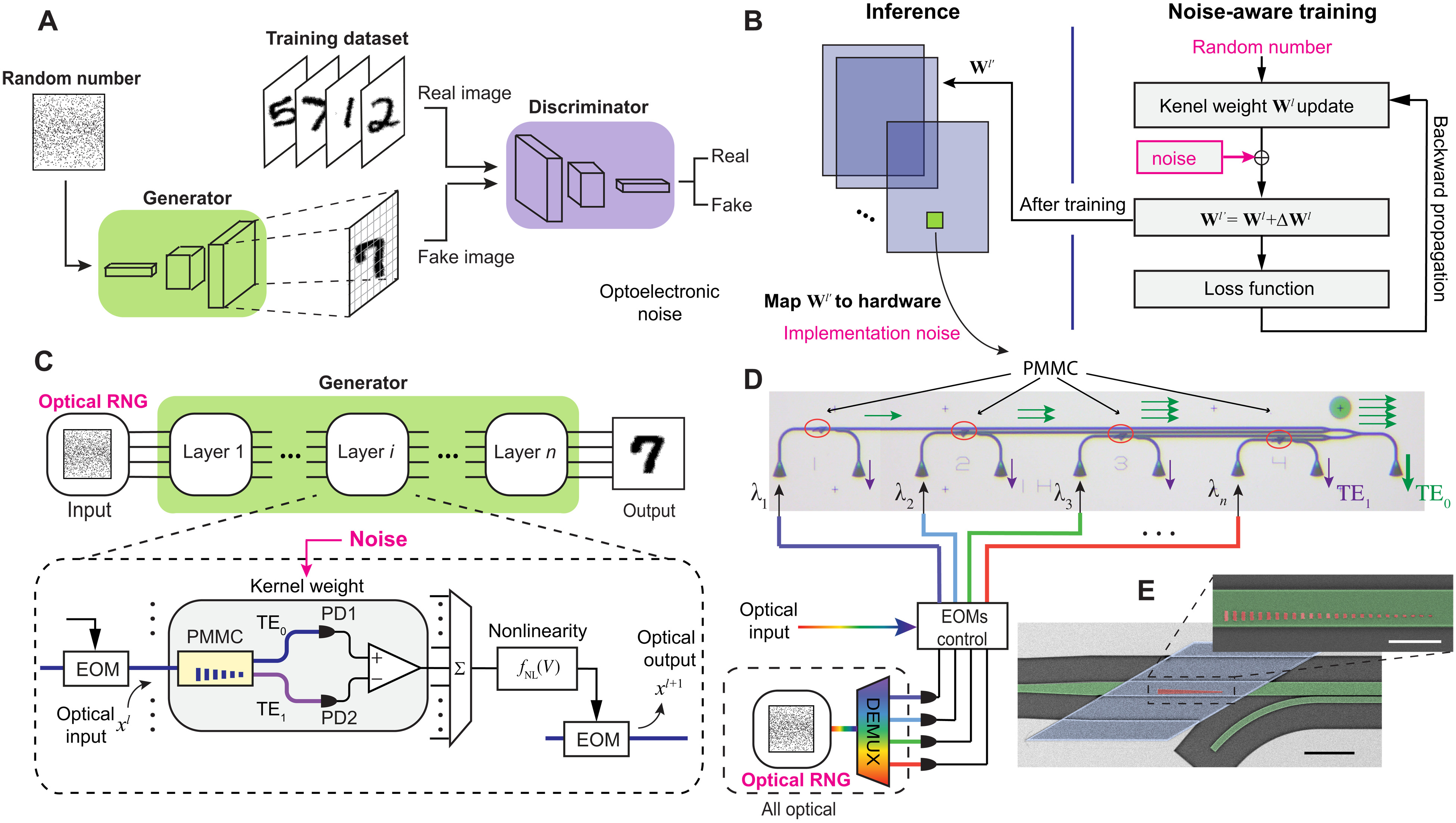}
		\caption{(A) A GAN architecture scheme consisting of the discriminator and generator.
(B) The offline noise-aware training and inference processes flow of the generator. 
(C) Decomposition of the generator into individual layers. In each layer, the input signals pass through the photonic tensor core and are converted to the electrical domain by photodetectors (PDs). After postprocessing, the data are converted back into the optical domain and transferred to the next layer. EOM, electro-optic modulator.
 (D) Optical microscopic image of the photonic tensor core consisting of four input channels. The optical RNG is input to the photonic tensor core through O/E and E/O conversion in our experiment. DEMUX, demultiplexers. 
(E) The false-colored scanning electron microscopy (SEM) image of the photonic tensor core. The Si3N4 waveguide, the GST metasurface, and the Al2O3 protection layer are colored green, red, and blue, respectively. Scale bar, 10 $\mu$m. Inset: The zoomed-in SEM image of the phase-gradient metasurface on the waveguide. Scale bar, 2 $\mu$m. Picture and its description are taken from \cite{wu2022harnessing}.
}
		\label{GAN}
	\end{minipage}
	\hfill
\end{figure}

\nb{A significant part of the current review is devoted to the description of the reductionism approach to the optimization problem, where one has to transform the target assignment into the known formulation (for example Ising model) or combinations of conditions (see graphical models) and then to map it into the particular hardware to get a proper solution. Although many presented approaches are similar, we are not restricted to covering only them. To support this statement, we present a few high-complexity realizations of specific computational problems.}

\begin{figure}[!h]
	\begin{minipage}[c]{0.95\linewidth}
		\includegraphics[width=\linewidth]{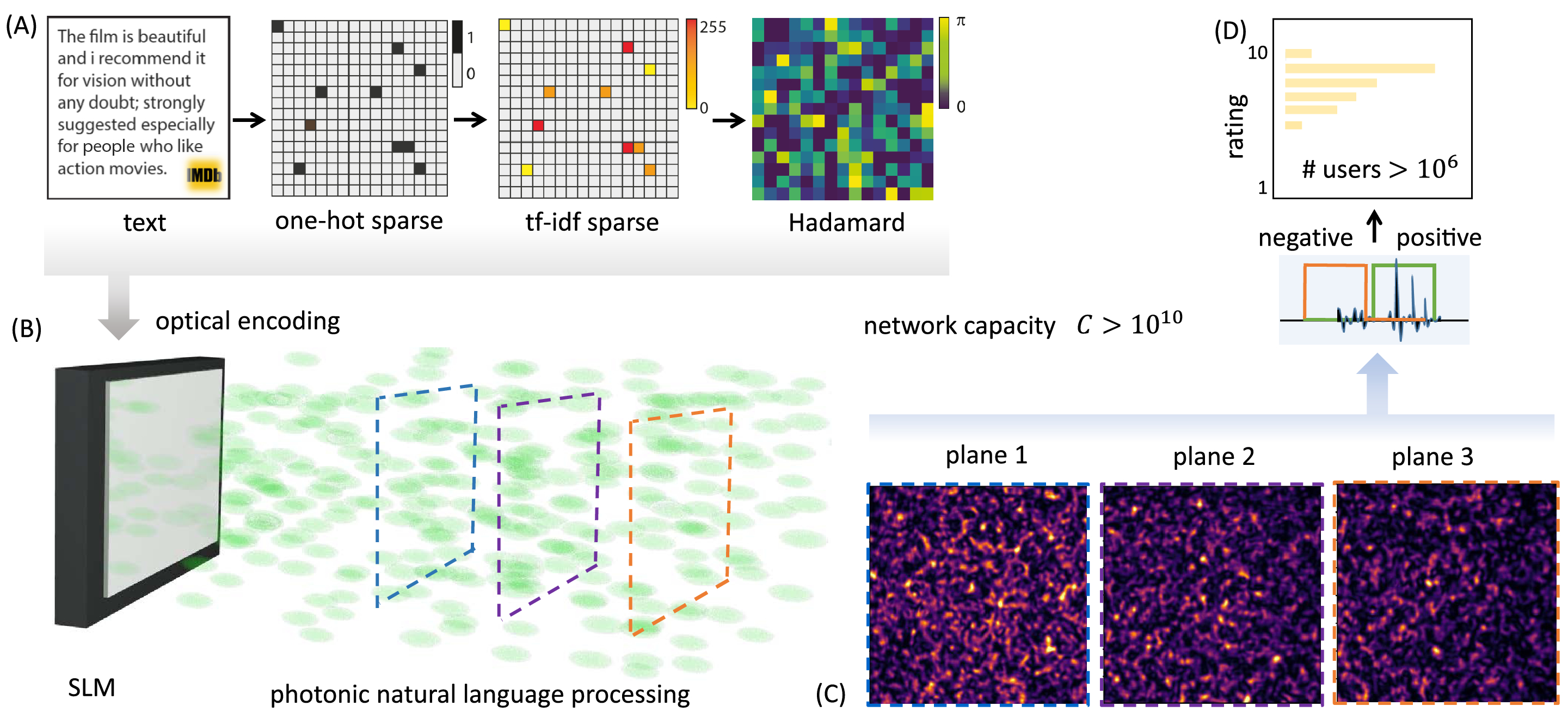}
		\caption{Three-dimensional PELM for language processing. (A) The text database entry is a paragraph of variable length. Text pre-processing: a sparse representation of the input paragraph is mapped into a Hadamard matrix with phase values in $0,\pi$.
(B) The mask is encoded into the optical wavefront by a phase-only SLM. Free-space propagation of the optical field maps the input data into a 3D intensity distribution (speckle-like volume).
 (C) Sampling the propagating laser beam in multiple far-field planes enables upscaling the feature space. 
 (D) The example shows a binary text classification problem for large-scale rating. Pictures and their description are taken from \cite{valensise2022large}.
}
		\label{NLP}
	\end{minipage}
	\hfill
\end{figure}

\nb{
Among them, we can highlight the recent demonstration of natural language processing on a photonic processor \cite{valensise2022large}, see Fig.~\ref{NLP}.  The significant advance is achieving capacity exceeding $1.5 \times 10^{10}$ optical nodes, which enables large-scale applications. In another work \cite{garg2022dynamic}, authors realized an optical neural network to simulate inference at an optical energy consumption of 2.7 aJ/MAC for computer vision model Resnet50 (Residual Network) and 1.6 aJ/MAC for natural language processing model BERT (Bidirectional Encoder Representations from Transformers) with little accuracy degradation. The third example \cite{wu2022harnessing} demonstrated a realization of a complicated generative network on the basis of a photonic computing core consisting of an array of programmable phase-change metasurface mode converters. The corresponding scheme can be found in Fig.~\ref{GAN}.
Overall, one can see more and more sophisticated architectures being reproduced using optical platforms.}

\nb{\subsection{Spin glass simulators}}

\nb{ The inherent relationship between spin glass models and optimization tasks has been a longstanding subject of research. Considerable attention has been devoted to examining the complexity of both paradigms. In this regard, simulating a realistic glass model can be approached as a computational problem. Hence, it is necessary to mention several works on these simulations using optical platforms. A concise overview of the measurements of various parameters of artificial disordered systems through optical setups, focusing on overlap distributions and replica symmetry-breaking realizations, can be found in \cite{conti2022replica}. Additionally, the chapter includes the construction of a statistical model of light modes dynamics in a random laser with a new equivalent for the overlap distribution. One of the first experimental measurements of overlap distributions in a random laser system was reported in \cite{ghofraniha2015experimental}. The subsequent work by Basak et al. in 2016 \cite{basak2016large} describes the observation of strong intensity fluctuations in standard ordered cavities, with subsequent analyses highlighting the replica symmetry breaking phenomena. Another example is the work by Moura et al. in \cite{moura2017replica}, where the intensity fluctuation overlap distribution is measured in the spontaneous mode-locking regime of a multimode Q-switched Nd:YAG laser. Research on spin glass simulators goes hand in hand with the optimization domain, motivating the search for unconventional hardware that is the primary focus of this review. This results in the realization of optical computing platforms that can address spin-glass problems on a large scale, such as the platform based on spatial light modulation and multiple light scattering \cite{pierangeli2021scalable}.
}

\section{Main directions of technological development in optical computing}
\label{Main directions of technological development in optical computing}

% Section description
The demand for computational resources is gaining momentum due to their use in many practical applications. This trend is supported by a growing industrial interest from prominent IT companies (Microsoft, Google, IBM, Amazon, etc.)  and fast-growing start-ups. To get a better global picture, one must understand the current paradigm of conventional heavy calculations and what advantages the optical machines can offer. First, we present several key metrics of the standard approaches and then show the benefits of optical devices. After that, we describe the strategies pursued in photonic neuromorphic computing.

\subsection{Performance of information processing systems}
\label{Modern electronic devices}

% Moore's law restatement
In general, Moore’s law concerns several metrics. All of them are reaching saturation but at a different paces. To maintain the same effectiveness of the hardware, new technology is required. However, the more significant demand for superior hardware is caused by the explosive growth of AI applications, which puts much more pressure on research and development performance. For example, the need for computational capabilities has increased by more than five orders of magnitude from $2012$ to $2018$ because of the AI developments shown in the OpenAI report \cite{li2021challenges}.

%Metrics, MAC and FLOP
Several key metrics characterise the performance of information processing systems. We will use MAC (multiply-accumulate operation containing one multiplication and one addition) and FLOP (floating-point operation). The relationship between them is that $1$ MAC counts as $2$ FLOP. One usually uses MACs and FLOPs to measure the speed performance of the device, which depends on the frequency or the characteristic intrinsic operation time on the hardware. Alternatively, one can use the operations per second (OPs), be it conventional mathematical operation or hardware state switching, but this notation is rarely used. Another important metric is energy consumption or efficiency, which can be measured in FLOPs/W (FLOPs per watt). One can consider alternative metrics, such as the total training energy in joules in the case of the training NN or J per spike in the operations performed on the spiking NN (SNN) architecture. Many combined metrics and their variations exist, such as speed per area (Op/s/mm), that are used to describe some other energy characteristics. Other important parameters of the hardware setup may include the analogue level of noise, scalability properties, specific architecture parameters, etc.

\begin{figure}[!h]
	\begin{minipage}[c]{0.95\linewidth}
		\includegraphics[width=\linewidth]{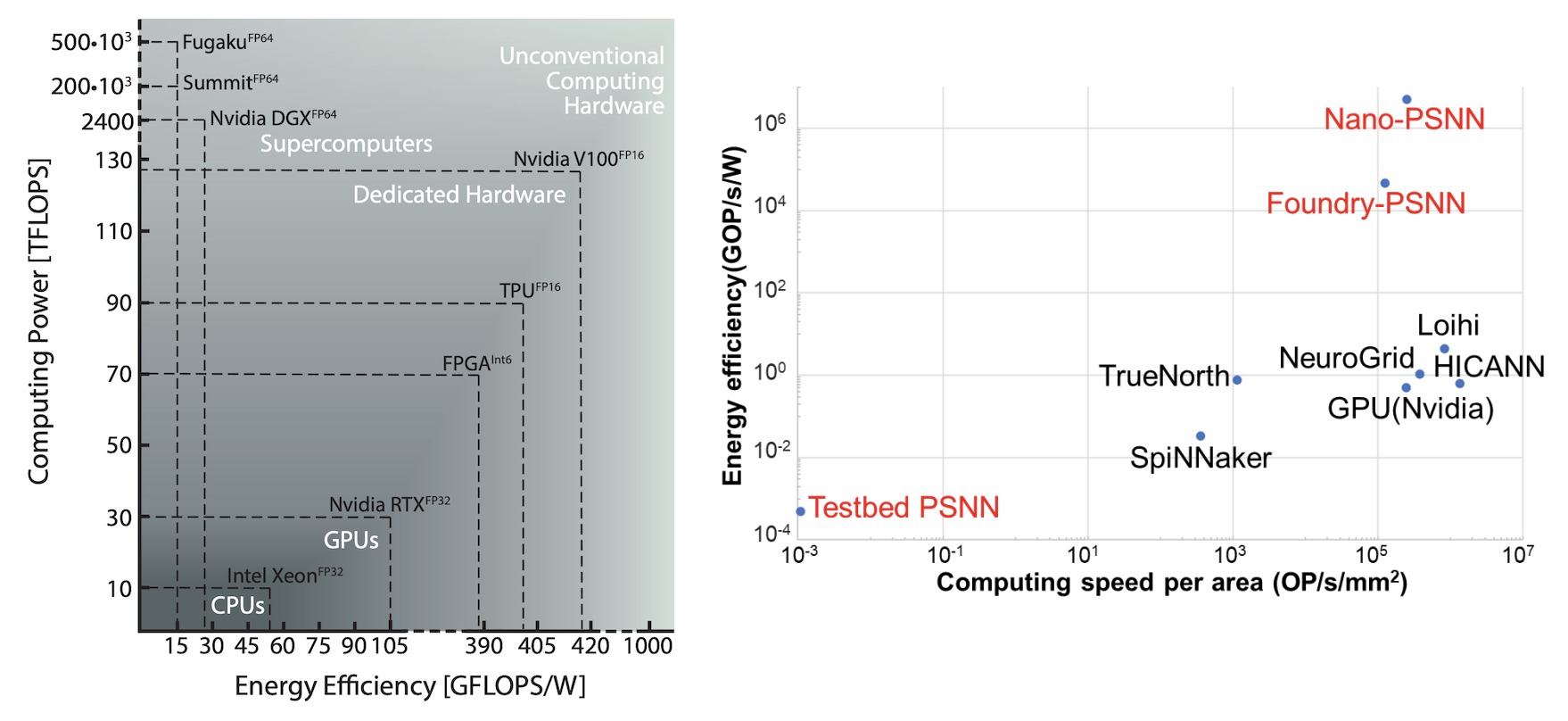}
		\caption{Left panel: computing power and energy efficiency of different types of computing hardware. The schematic distribution of the processing power versus energy efficiency is shown for several CPUs, GPUs, FPGAs, supercomputers and potential unconventional computing devices based on optical systems, reproduced with permission from \cite{kalinin2021large}. Right panel: energy efficiency values versus computing speed per area for spike-event hardware compared with results described in the literature. Reproduced with permission from \cite{lee2021izhikevich}.}
		\label{11_performance}
	\end{minipage}
	\hfill
\end{figure}

% Description of the conventional systems and reviews
Data centres that use thousands of CPUs and hundreds of GPUs consume megawatts of power. Despite the versatility of conventional computers, their characteristics are not enough to achieve high performance in the key metrics. Thus,  application-specific hardware that differs in architecture and logic reduces this gap between the desired efficiency and computer capabilities. One can find several discussions of these devices in \cite{lee2021izhikevich,kalinin2021large} with the corresponding comparison of the key metrics; see also Fig.~\ref{11_performance}. In addition, we mention some of these electronic devices as reference points for comparison with optical devices.

% Key metric
Classical computing architectures can differ in the details within one type of device. However, it is common to characterise them using two key metrics -- the processing power or the computing speed using FLOPS and the energy efficiency. One can use the ratio of the FLOPS to the power consumption in watts (W) to get the energy efficiency metrics \cite{kalinin2021large}.

% CPU, GPU
The standard estimate of the modern CPU efficiency is $2$ TFLOPs, while the power efficiency is about $10$ GFLOPs/W. Therefore, we can use the Intel Xeon processor as one of the top devices in terms of efficiency for working with the double-precision format, which has $4.8$ TFLOPs and $29$ GFLOPs/W \cite{sun2019summarizing}. Graphics processing units (GPU) are the advanced specialised electronic architectures and workhorses of the current ML tasks in real applications because of the parallel computing options. Most of the GPUs operate at near $0.3$ kW power consumption with the range of $0.5$ to $7$ TFLOPs and corresponding $1.6$ to $23$ GFLOPs/W energy efficiency for the work with the double-precision format.

% Supercomputers
Another type of classical hardware is powerful non-distributed computer systems that are not so energy efficient but have enormous computing power. The top $10$ list starts with NVIDIA DGX SuperPOD with $2356$ TFLOPs and nearly $26.2$ GFLOPs/W. The most powerful supercomputer in processing power is Fujitsu's Fugaku, with $442000$ TFLOPs and $14.8$ GFLOP/J. One can link several devices into the powerful distributed system to achieve much higher processing power with additional energy costs. 

% Dedicated hardware
Another class of electronic devices can be named "dedicated hardware". Although the GPU is not usually attributed to this class, it performs a similar role. A good example is the field-programmable gate array (FPGA), an integrated circuit that can be configured by a customer using a hardware description language. On average, FPGA can achieve
$10$ TFLOPs with near $50$ GFLOPs/W energy consumption rate and several times more ($\operatorname{x} 5,\operatorname{x} 7$) by working with lower precision numbers \cite{nurvitadhi2017can}. Another example of dedicated hardware is Google's Tensor processing unit (TPU). TPUs are custom-developed application-specific integrated circuits (ASICs) to accelerate ML workloads. The efficiency can be estimated as $90$ TFLOPs, and $400$ GFLOPs/W \cite{jouppi2017datacenter}.

% Neuromorphic electronic hardware
Further improvements in electronic special-purpose devices are expected to come from analogue architectures based on memristors \cite{jeong2018memristor}, non-volatile memories, compact low-voltage field-effect transistors and engineering of heterostructures of two-dimensional materials taking into account the quantum effects. Another option is to explore the different architecture of the dedicated hardware. For example, IBM claims to achieve $176 000$ times better energy efficiency with their bio-inspired neuromorphic chip TrueNorth chip than the conventional general-purpose Intel i7 system for specific applications \cite{merolla2014million}. Nevertheless, TrueNorth has a relatively slow frequency rate of $1$ kHz and an approximate energy efficiency of $2.3$ pJ/bit. Moreover, it requires additional connections for the incoming neural spikes. One can further explore Intel’s Loihi \cite{davies2018loihi} or NeuroGrid \cite{benjamin2014neurogrid} devices, which are close to the modern GPU \cite{izhikevich2007dynamical}.

% Reference point and future requirements
Despite impressive and innovative developments, more than the presented classical architectures are needed to satisfy the need. For example, some estimates on demand from future autonomous vehicles require the information processing at $100$ TOps rate with the energy consumption of less than $100$ Watt with the additional low latency \cite{poulton2019long}.

\subsection{Optical energy consumption}
\label{Optical energy consumption}

% Intro
Optical devices can process information instantaneously. Additional advantages include negligible energy consumption and heat generation. State-of-the-art for CPUs and GPUs metrics can be converted into $20$ pJ/MAC \cite{keckler2011gpus}. The dedicated hardware and application-specific circuits can achieve $1$ pJ/MAC with reduced precision of the calculations \cite{sze2017efficient}. The same so-called ``ideal'' benchmark is supported by the work \cite{shen2017deep}, where authors used a programmable nanophotonic processor with a cascaded array of $56$ programmable MZIs in a silicon photonic integrated circuit to perform the vowel recognition task. Modern AI chips can reach the $100$ mW/GOps operation power per second, but the future competitive requirements should be $\sim 10-1$ mW/GOps \cite{li2021challenges}.

%The hybrid electro-optical DSP coprocessor \cite{tamir2009high} is estimated to perform $16,384$ Giga 8-bit integer instructions per second,
%the number of instructions per power unit, measured in
%Giga Int8 (Gint8) operations per watt is 16,384/10
%=1638.4/watt. 
%1 pj/mac
%10-1 mw/Gops
%1mac = 2 flops

%Examples_1
We can consider several examples of photonic hardware and highlight their technical characteristics, such as speed and energy consumption. The photonic accelerator architecture based on coherent detection \cite{hamerly2019large} enables a new class of ultra-low-energy processors operating at very low (sub-aJ) energies for MAC operation. These structures can be reprogrammed and trained on the fly and have good scalability of up to one million elements. Additionally, \cite{hamerly2019large} discusses the “standard quantum limit” for optical NNs that can be bounded with $50$ zJ/MAC values for irreversible digital computation.

%Examples_2
Optical NNs can achieve accurate results with extremely low optical energies \cite{wang2022optical}. It was shown experimentally that optical NN with dot product calculated optically achieved high accuracy on the MNIST digits classification using few photons (of the order $10^{-19}$ J of optical energy) per weight multiplication. The essential idea was to reduce the noise from accumulating scalar multiplications in dot-product sums.

%Examples_3
Some optical machines can use pre-optimized mathematical structures for architectural benefits. A good example is energy-efficient, high-throughput, and compact tensorised optical NN exploiting the tensor-train decomposition \cite{xiao2021large}. Such a NN can improve the energy efficiency by a factor of $1.4 \times 104$ compared with digital electronics ANN hardware and by a factor of $2.9 \times 102$ compared with silicon photonic technologies. Moreover, it was possible to achieve better energy efficiency with fewer elements for footprint-energy efficiency calculation \cite{xiao2021large}. In general, neuromorphic photonic systems potentially offer petaMAC per second per mm${}^2$ processing speeds \cite{nahmias2019photonic} and attojoule per MAC energy efficiencies \cite{nozaki2019femtofarad}.

% Architecture role and SNNs
Energy consumption is closely related to the physical properties of the neural architecture. For example, event-driven spiking neural networks (SNNs) outperform ANNs in energy efficiency. The dynamic of many models can be described using the universal Izhikevich model \cite{izhikevich2007dynamical}. Event-driving neuromorphic computing overcomes traditional von-Neumann architectures' limitations but has several specific problems with the throughput, scalability, training methods, etc.

% Photonic SNN device
The successful implementation of an optoelectronic spiking neuron inspired by the Izhikevich model was reported in \cite{lee2021izhikevich}. A nanoscale optoelectronic neuron with $200$ aJ/spike input can trigger the output from on-chip nanolasers with $10$ fJ/spike. This neuron can support a fanout of $\sim 80$ or overcome $19$ dB excess optical loss while running at $10$ GSpikes/second in the NN. Such a scheme corresponds to $100$ throughput and $1000$ times energy-efficiency improvement compared to state-of-art electrical neuromorphic hardware such as Loihi and NeuroGrid \cite{lee2021izhikevich}. The hybrid systems of quasiparticles can be another potential platform for spiking architectures. Exciton-polaritons can achieve $1$ pJ/spike with $100$ ps timescale \cite{tyszka2021leaky,opala2019neuromorphic}.

\subsection{Evaluation of speed}
\label{Evaluation of speed}

% Vector-matrix multiplication
A universal optical computer was not a viable option to compete with classical computers. Instead, a specified optical computer or optical block as a part of a hybrid classical/nonclassical architecture has become a focus of recent research. One of the first realisations of simple mathematical operations, such as a free-space fan-in/out vector-matrix multiplication, was introduced by Goodman in 1978 \cite{goodman1978fully}. It is the essential linear algebra operation, where the input vector is loaded into an array of light sources, and the multiplication matrix is encoded into the SLM. The light propagation is analogous to broadcasting the initial vector into SLM, which performs element-wise multiplication, after which the lens gathers all the beams in the horizontal direction and summates the intensities. One can evaluate the performance of this device as $N^2$ MAC for one multiplication of the vector with $N$ elements and a square matrix $N^2$. However, the effective performance is limited by the system's frequency $f$, mainly of the SLM, resulting in  $f N^2$ MACs; see also \cite{tamir2009high}. Nevertheless, using $256$-length input vector and $125$ MHz frequency rate, the device's performance can reach impressive $\sim 8$ TMACs. 

% Vector-matrix multiplication 2, more examples
Other schemes based on different forms of the free-space matrix-vector multiplication can reach similar values. In 2020, Lightmatter presented an optoelectrical hybrid chip 'Mars' with $0.4-4$ TMACs depending on the frequency of weights \cite{ramey2020silicon}. A massively parallel convolution of $ 16$x$16$' tensor core' scheme based on crossbar architecture has been built on a chip with $13$ GHz modulation speed for the inputs, and approximate 2 TMACs \cite{feldmann2021parallel}. Another scheme based on the electro-optical Mach–Zehnder modulators represents a universal optical vector convolutional accelerator and achieves more than ten TOPS speed, with a consequent successful use as an optical convolutional neural network in facial and handwritten digit images recognition  \cite{xu202111}. Most of the photonic hardware with the feed-forward architecture can operate at high (GHz) speeds and usually have good scalability characteristics \cite{hamerly2019large,wang2022optical,xiao2021large}.

% Reservoir computing
Another critical factor affecting the optical device's speed performance is the hardware's architecture. From this perspective, RC might improve many aspects of optical computing devices. One could expect several orders of magnitude speed-up compared to the typical ANN structure. RC optoelectronic/optical implementations are usually divided into spatially distributed and time-delayed \cite{van2017advances}.
The RC scheme on a silicon photonic chip with optical waveguides, splitters and optical combiners can achieve the data processing rate of $0.12$ and up to $12.5$ Gbit/s \cite{vandoorne2014experimental}. Moreover, more exotic physical systems, such as exciton-polaritons, can reach similar performance so that the SNN architectures can achieve the characteristic operation time of the order of $100$ ps with the energy efficiency of $1$ pJ/spike \cite{opala2019neuromorphic,tyszka2021leaky}

%Spin machines
Optic-based spin machines also enjoy competitive speed characteristics. CIM evolved from having just $4$ spins and $12$ connections in 2014 (Stanford) to $16$K spins and $256$M connections in 2021. The $2000$-node version achieves semidefinite relaxation minimum of a cost function in $0.1$ ms and further improves the solution \cite{cim2000nodes}. The new generation of CIMs based on Thin-Film LiNbO3 (TFLM) photonic circuits will be released in 2022. It will feature an OPO network with $\sim \mu W$  pump power, $\sim$ fs pulse duration, $100$ GHz – $1$ THz clock frequency and the synchronized operation of multiple CIMs on a chip. 

% EP spin machines
Exciton-polaritons possess even better ultrafast timescales. For example, the polariton graph simulator \cite{BerloffNatMat2017} is easily scalable to $10$K elements and shows $\sim 100$ ps operational times respectively, while the degenerate lasers \cite{nir2018} system have $\sim \mu$s characteristic timescale. However, all-to-all controllable couplings have yet to be experimentally implemented.

\subsection{Other important properties}
\label{Other important properties}

% Intro, accuracy and NN noise
Other essential factors are undoubtedly affecting optical devices' attractiveness and performance, such as intrinsic noise and analogue accuracy of the hardware. For example, the recognition results using MNIST handwritten digits can show different accuracy on different devices, which can be a good measure of how well a particular NN is adjusted to a specific task \ cite{lee2021izhikevich}. The comprehensive analysis of the error sources and their classification for the electro-optical device can be found in \cite{li2021challenges}.

% Architectures
The essential part of the hardware is its structure/architecture. It affects many other properties of the optical devices, be it the accuracy, scalability, the potential for future optimization, etc. The interplay between the hardware's electronic and photonic components depends on the architecture. It directly affects optical/electronic conversion, storing and reading the data, and logic operations cost in the case of a hybrid architecture.

% Derivative metrics
Scalability is one of the key metrics and is the consequence of the architecture choice. It measures the ability of a system to keep its algorithmic performance with a growing number of variables.

% Additional degrees of freedom
The optical setups enjoy additional degrees of freedom compared to the conventional electronic hardware. For example, two independent variables in the complex plane can parameterise short optical impulses. In addition, one can explore optics-specific degrees of freedom such as polarisation and orbit angular moments of light.

% Post optimisation
Lastly, current optical hardware is used to employ classical algorithms and NN architectures that are conventional for standard electronic architecture. These algorithms are designed using Boolean logic, which is suitable for a digital computing system. However, they are not always optimal for optics implementation. Therefore, developing specialised algorithms optimised for optical computer platforms is necessary, further reducing the operational complexity and execution time.

\nb{\subsection{Noise in analog optical computing}
\label{Noise in analog optical computing}
In many analogue devices, noise plays a crucial role in their operation. The investigation of noise, its sources, and its properties in optical systems is a fundamental subject that has been addressed extensively. The classical theory of laser noise \cite{arnaud1995classical} encompasses fundamental concepts of fluctuations caused by atomic transitions between lower and upper levels and the independence of classically-prescribed optical fields, considering both moderate and high-power laser cases. These ideas were expanded to include sources of quantum noise (such as momentum fluctuations of electrons at optical frequencies and uncertainty-related fluctuations of the electromagnetic field), shot noise, the transition to classical noise in high-power lasers, the distinction between lasing and non-lasing modes, and intensity fluctuations at different frequencies and their corresponding distributions \cite{henry1996quantum}.
The transition between classical and quantum noise has been extensively studied \cite{schenzle1989classical}, with discussions on the origin of quantum noise emerging from the reversible or irreversible part of dynamics and comparisons with purely classical fluctuations and corresponding physical examples \cite{haus1995classical}. While noise in analog photonics is generally considered a harmful effect, it is possible to mitigate its impact or make the system robust towards specific types of perturbation using system-specific techniques. Analog deep learning platforms experience both deterministic and non-deterministic noise sources, with the amount of noise increasing with operational speed. To address this issue and efficiently deploy Multiply-Accumulate (MAC) operations, the authors of \cite{mourgias2022noise} introduced and experimentally demonstrated a noise-resilient deep learning scheme with a record-high 10GMAC/sec/axon compute rate. This approach uses a coherent silicon integrated circuit that combines a noise-tolerant linear neuron architectural scheme with noise-aware training methods.}

\nb{
There are many other alternative ways of dealing with “noisy” analog computations. For example, in \cite{giamougiannis2023analog}, authors discuss the advantages of the pre-trained analog optical processors with bit precision operations. To surpass the performance of digital processors because of the confined photonic hardware size and the limited bit precision of high-speed electro-optical components, engineers usually used post-training techniques such as inference averaging, dynamic precision inference etc., to compensate for the “noisy” analog computations. Hence, the authors proposed and experimentally demonstrated a speed-optimized dynamic precision neural network (NN) inference via tiled matrix multiplication (TMM) on a silicon photonic processor with the aim of targeting high-accuracy and speed-optimized classification tasks.
The advantages of optical computing over digital computing for accelerating deep learning lie in operations executed at low precision. The key metric here is the effective number of bits of precision of analog processors, which is limited by noise. Dynamic precision analog computing for neural networks was proposed in \cite{garg2022dynamic}. It lies in repeating operations and averaging the result, decreasing the impact of noise. This method reduces energy consumption by up to 89\% for a particular computer vision model and 24\% for a natural language processing model, overcoming the weight noise, thermal and shot noises. }

\nb{The computational errors from different sources tend to accumulate and severely impair the large-scale photonic neural networks' performance. Usually, one can not expect something valuable from the noise realizations. Counterintuitively to the current belief, it was  demonstrated that a photonic generative network can act as a part of a generative adversarial network for generating handwritten numbers \cite{wu2022harnessing}. The implementation was realised  with a photonic core consisting of an array of programable phase-change memory cells, applied noise-aware training by injecting additional noise, and led to the demonstration of the network’s resilience to hardware nonidealities. Several offline noise-aware training schemes for discriminative models were proposed, such as injecting noises to layer inputs, synaptic weights, and preactivation \cite{wu2022harnessing}.
In summary, noise is considered a significant obstacle to efficient optical computation; however, it can be leveraged or exploited smartly.
}

\subsection{Optical minimizers of spin Hamiltonians}
\label{Optical minimisers of spin Hamiltonians}

% Characteristics of OMSH
\nb{Optical systems designed to minimize spin Hamiltonians have the potential to find the global minimum of hard optimization problems. These systems offer several advantages, including the ability to find better solutions to a wide variety of nonlinear optimization problems within a fixed time, to find solutions of a given precision more quickly, or to solve more complex problems at a fixed and limited cost. However, these machines also have their limitations and vary in terms of scalability, the ability to engineer the required couplings, the flexibility of tuning the interactions, the precision of read-out, and other factors that facilitate the approach to the global minimum rather than local minima. Despite these limitations, all of these machines have some aspects of their operation that promise increased performance over classical computations. To solve an optimization problem using optical minimizers of spin Hamiltonians, it is necessary to find an optimal mapping of the real-life problem onto a spin Hamiltonian. Some optimal mappings are already known, while for others finding an optimal mapping is a crucial step towards successfully solving the problem.}

%The optical systems described in this review as optical minimisers of spin Hamiltonians aim to find the global minimum of hard optimisation problems. They offer the potential for finding a better solution to a wide variety of nonlinear optimisation problems for a fixed time, finding a solution of a given precision faster, or solving more complex problems at fixed and limited cost. All these machines have advantages and limitations. They vary in scalability, ability to engineer the required couplings, the flexibility of turning the interactions, the precision of read-out, and factors facilitating the approach to global rather than the local minimum. However, they all have some parts of their operation that promise increased performance over the classical computations. To solve an optimisation problem on optical minimisers of spin Hamiltonians, one needs to think of an optimal mapping of the real-life problem onto a spin Hamiltonian, some of which are known \cite{lucas2014ising}, for others finding an optimal mapping will mean half of the success in solving. 

% Sampling of the landscape
\nb{Combinatorial optimization is a field of study that focuses on finding the absolute minimum configuration of a given problem. However, in many applications it is desirable to not only find one absolute minimum, but also to obtain multiple or all degenerate absolute minima and, in some cases, to sample many low-energy excited states \cite{zhu2019fair}. This sampling capability can be useful for applications that require distributional information about optimal solutions, such as the implementation of Boltzmann machines as generative models for machine learning \cite{hinton2002training}. In industrial settings, having access to a pool of candidate solutions to an optimization problem can make processes more efficient and flexible. For example, in drug discovery \cite{sakaguchi2016boltzmann}, structure-based lead optimization could generate many candidate molecules for simultaneous testing.}
\nb{One approach to solving large optimization problems is to decompose them into smaller subproblems that can be solved separately, for example to accommodate hardware limitations. By using multiple low-energy samples rather than just the optimum for each subproblem, it is possible to construct better solutions to the original problem \cite{bian2016mapping}. However, a spin minimizer designed for combinatorial optimization may not be well-suited for sampling all ground and low-energy states. The nonlinear stochastic dynamics of such machines in the presence of quantum noise can be exploited to sample degenerate ground and low-energy spin configurations of spin models. When these optical machines operate in a quantum-noise-dominated regime with short photon lifetimes (i.e., low cavity finesse), homodyne monitoring of the system can efficiently produce samples of low-energy spin configurations that are better than their classical counterparts \cite{ng2021efficient}.}

%An additional advantage of discovered principles of operation of optical minimisers of spin Hamiltonians leads to the opportunity of formulating new optimisation algorithms to be realised on specialised but classical computing architectures: FPGAs, GPUs, etc. For example, the principle of operation of the CIM was implemented as the network of nonlinear oscillators described by simplified equations \cite{CIM_eqs2017}. A similar approach has been recently realised using FPGAs using a network of Duffing oscillators \cite{toshiba}. 

% Emulator
To properly access the properties of such systems, one can use computer simulations in several scenarios. Such emulations allow one to avoid extensive labour experiments to predict properties of such systems properly, tune and optimise the parameters for optimal performance and even inspire new classes of algorithms for conventional computers. The emulation algorithms can be found in \cite{GainD_sciRep_2018,tiunov2019annealing}. Such techniques can apply to a broad type of NNs.

\subsection{Efficiency of Artificial neural networks}
\label{Artificial neural networks}

%Intro
Artificial neural networks are powerful tools for processing large data sets and analyzing vast amounts of information quickly and without explicit instructions. As a result, a wide variety of neural network architectures have been developed and implemented in various applications. The development of different neural networks is important because each architecture can represent different systems while maintaining a certain level of universality in approximating and representing complex systems. This expands the already significant scope of applicability of neural networks.

% Progress
\nb{Passive optics can perform many linear transformations without power consumption and minimal latency at rates over 50 Gb/s. Optical logic gates have been demonstrated to be feasible \cite{bontempi2012multifunctional,ballarini2013all,chen2013all,zasedatelev2019room,baranikov2020all}. However, attempts to replicate classical boolean electronic logic circuits in photonics have not been successful. Analog photonic computing devices are suitable for NNs due to their fast and energy-efficient computations. Optical nonlinearities can be used to implement various nonlinear functions \cite{nozaki2010sub}. Recent developments suggest that optical implementations of NNs can surpass electronic solutions in terms of computational speed and energy efficiency. However, the challenge of developing truly deep NNs with photonics remains. Photonic multilayer perceptrons and photonic spiking neural networks have potential for realizing all-optical artificial neural networks. Photonic accelerators for convolutional NNs are the most promising photonic solutions for enhancing inference speed and reducing power consumption in the near term.}

%Many linear transformations can be performed with passive optics without power consumption and minimal latency at rates over 50 Gb/s. The feasibility of optical logic gates has also been demonstrated \cite{bontempi2012multifunctional,ballarini2013all,chen2013all,zasedatelev2019room,baranikov2020all}. However, the attempts to replicate the classical boolean electronic logic circuits in photonics did not prove to be successful. Analog photonic computing devices are especially suitable for NNs, which require fast and energy-efficient (although approximate) computations. Furthermore, in principle, many optical nonlinearities can be used to implement various nonlinear functions  \cite{nozaki2010sub}. Recent developments suggest that optical implementations of NNs can overcome electronic solutions in terms of computational speed and energy efficiency. However, as discussed in our review,  the challenge of developing truly deep NNs with photonics still needs to be solved. Photonic multilayer perceptrons and photonic SNNs have a lot of potential to realize all-optical ANNs. In the near term, photonic accelerators for convolution NNs (multiple layers, weight sharing, sparse topology) are the most promising photonic solutions to enhance inference speed and reduce power consumption.

% Further research
However, there are still many opportunities to explore and improve the implementation of photonic NNs. For example, more research is needed to assess whether specific types of deep NNs can be implemented optically in an efficient manner that provides advantages over fully electronic implementations. Furthermore, some deep NNs, such as long-short-term memory NNs, generative adversarial nets, geometric deep NNs, and deep belief networks, have not yet been implemented in photonics.
%However, there are still many other opportunities to explore and improve the implementation of photonic NN. For example, more research is needed to assess whether a specific type of deep NN can be implemented optically efficiently, i.e., in a way that provides advantages concerning fully electronic implementations. Furthermore, photonics has not yet implemented some deep NNs (long-short-term memory NNs, generative adversarial nets, geometric deep NNs, deep belief networks, etc.).

% Scalability
\nb{The ultimate goal is to realise large photonic NNs with thousands of nodes and interconnections across many hidden layers. To achieve this, it is essential to work on the cascadability and robustness of photonic NNs to fabrication imperfections and parameter drifts over time  \cite{de2019machine}. Resonant structures like microring resonators are susceptible to manufacturing deviations \cite{tait2017microring}. Linear optical processors based on Mach-Zehnder interferometers  appear more robust due to their reconfigurability. Some studies discuss achieving reliable photonic computations with imperfect components \cite{miller2015perfect}.}

%The ultimate goal is to demonstrate large networks with thousands of nodes and interconnections across many hidden layers, i.e., truly deep architectures. Therefore, it is essential to work on the photonic NN cascadability (enabled by low propagation losses, crosstalk, and noise) and robustness to fabrication imperfections and parameter drifts over time \cite{de2019machine}. For instance, resonant structures like microring resonators are susceptible to manufacturing deviations \cite{tait2017microring}. At the same time, because of their reconfigurability, linear optical processors based on MZI appear more robust to process inaccuracies. Some studies discuss how to achieve reliable photonic computations even with imperfect components \cite{miller2015perfect}.

% Nonlinearities
Further investigation is needed to implement nonlinear activation functions in an all-optical manner in photonic NNs. While software can emulate nonlinearities, integrating nonlinear elements into hardware remains a challenge. Several approaches have been reported to address this issue, including the use of MZIs \cite{mourgias2019all}, graphene and quantum well electro-optic absorption modulators, and photonic crystals \cite{sharifi2016new}. Technological breakthroughs would greatly benefit photonic NNs, particularly the implementation of an integrated, non-volatile, and energy-efficient photonic memory element. Phase-change materials are a promising approach for achieving such photonic memories due to their potential for multi-level storage \cite{wuttig2017phase}. These materials’ cells have been exploited in photonic NNs, particularly for spiking neural networks (SNNs) \cite{feldmann2019all}.

%The all-optical implementation of the nonlinear activation function requires further investigation. Nonlinearities can be emulated in software, but integrating nonlinear elements into hardware is still challenging. Several approaches to address this issue have been reported using MZIs \cite{mourgias2019all},  graphene and quantum well electro-optic absorption modulators,  and photonic crystals \cite{sharifi2016new}. Some technological breakthroughs would benefit photonic NN, particularly implementing an integrated, non-volatile and energy-efficient photonic memory element. In this scenario, using phase-change materials seems the most promising approach to achieve such photonic memories since they have also shown the potential for multi-level storage \cite{wuttig2017phase}. Moreover, the cells of such materials have been recently exploited in photonic NN, mainly for SNNs \cite{feldmann2019all}.

\subsubsection{All-optical backpropagation}
\label{All-optical backpropagation}

% Definition
When training NNs, one usually considers the backpropagation algorithm by default. The essential idea behind the backpropagation is to compute the gradient of the loss function with respect to each weight by the chain rule and doing it consequently, one layer at a time, iterating backwards from the last layer to avoid redundant calculations of intermediate terms in this sequence of steps \cite{hecht1992theory,goodfellow2016deep}. Such a procedure allows one to fit the weights of a NN for a given task. Still, the complexity of the backpropagation is enormous. It grows linearly with the number of training examples or butches, the number of iterations, which is not known in advance and the basic complexity of feedforward input propagations, which can be estimated as a consequent series of matrix-vector multiplications. These evaluations hold for many cases, assuming batch gradient descent algorithm and simple matrix multiplication for the input propagation. However, one can reduce the number of steps with some approximate schemes. At the moment, there are many different ways to train NNs, including variants of backpropagation or alternatives, such as learning without backpropagation \cite{wilamowski2010neural}.

% BP and optics
Thus, the backpropagation algorithm remains one of the most expensive components to compute. The significant power and time consumption happens due to the sequential computation of gradients in the backpropagation procedure of NN training. Backpropagation through nonlinear neurons is another challenge to the field of optical NNs and a significant conceptual barrier to all-optical training schemes. Although there exist several practical, simple solutions, such as using approximation provided in a pump-probe scheme that requires only passive optical elements \cite{guo2021backpropagation} or by measuring the forward and backwards propagated optical fields based on light reciprocity and phase conjunction principles \cite{zhou2020situ}, the schemes still involve digital electronics or programming a high-speed SLM respectively. Therefore, having incomplete solutions, the work on the end-to-end optical training of NNs is in progress. Achieving the efficient all-optical backpropagation training method (besides the realization of depth and nonlinearities) will be a major achievement in the field. The question of such realisation is just a matter of time since there are no fundamental restrictions on such a development  \cite{lopez2022design,marquardt2022time}.

\begin{figure}[!h]
	\begin{minipage}[c]{0.95\linewidth}
		\includegraphics[width=\linewidth]{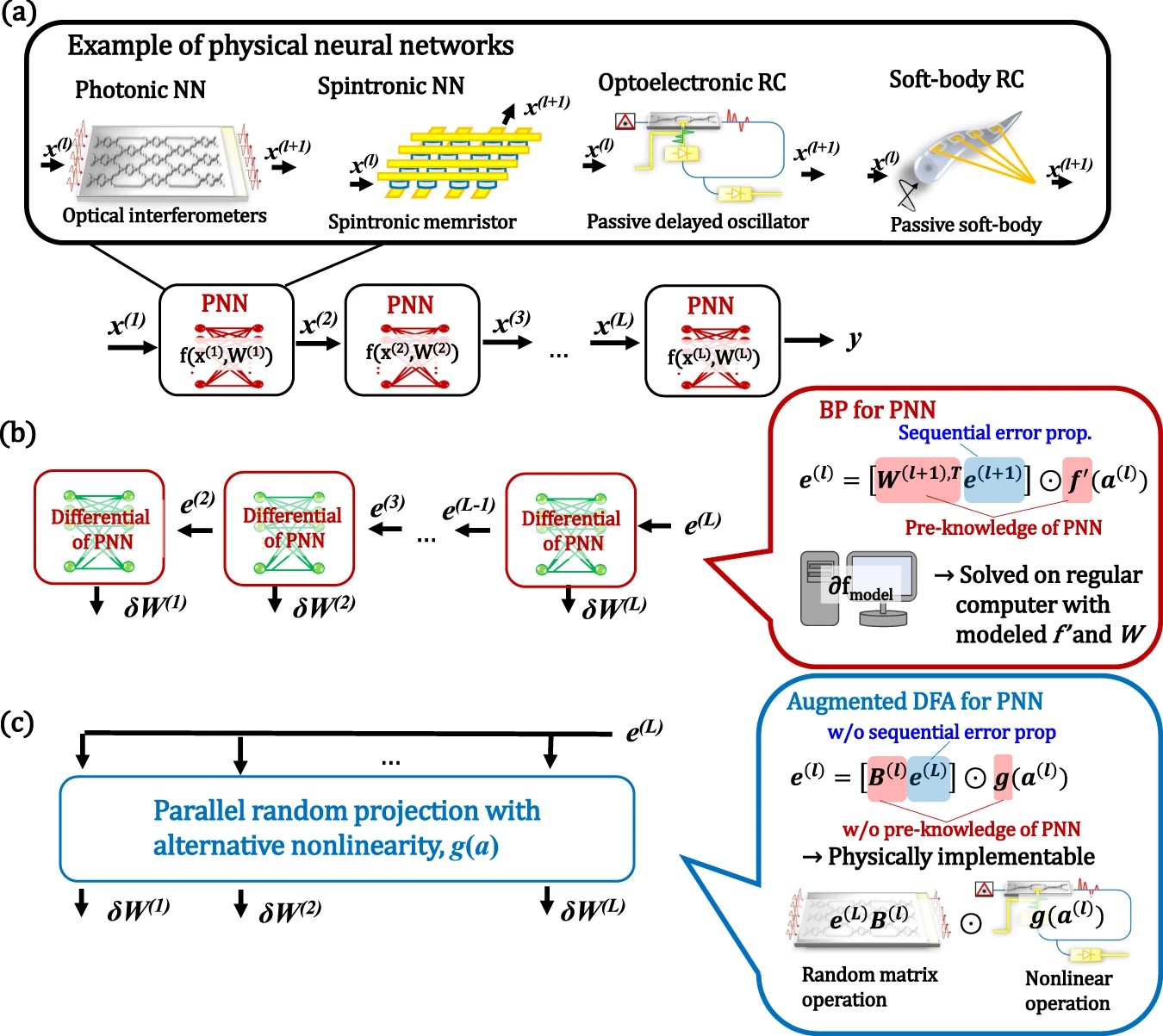}
		\caption{Schematics of physical neural networks (PNNs). Training sequence of PNN with (b) back-propagation, and (c ) augmented biologically plausible training called direct feedback alignment (DFA). Augmented DFA enables parallel, scalable, and physically accelerable training of deep physical networks based on random projection with alternative nonlinearity g(a). Picture and its description are taken from \cite{nakajima2022physical}.}
		\label{alternative_learning}
	\end{minipage}
	\hfill
\end{figure}

\nb{\subsection{Alternative learning methods}
The most computationally intensive part of the NN operations is the learning process, with backpropagation being the standard procedure behind many cases of training the weights. Exploiting the physical mechanism to reduce the energy requirements for such a training procedure (e.g. in the case of optical schemes) should be considered one of the significant achievements of optical NNs.
A hybrid in situ–in silico universal algorithm called physics-aware training was introduced in \cite{wright2022deep} with a few examples (including an optical one) as demonstrations. 
Since the analogue computation of the backpropagation terms according to the direct chain rule is complicated, scientists and engineers have devised many alternatives for the learning procedures. For example, the training algorithm called direct feedback alignment was introduced in \cite{nakajima2022physical}, see Fig.~\ref{alternative_learning}. This is a universal method based on random projection with alternative nonlinear activation and requires no information about the nature of the physical system. Another approach lies in a simple mechanism that can transmit teaching signals across neuronal layers by multiplying them by random synaptic weights and performs similarly as backpropagation on many tasks \cite{lillicrap2016random}.
}

\subsection{Statistical sampling}
\label{Statistical sampling}

% Universality
Statistical sampling is another essential domain where using optical machines can be beneficial. PGMs can effectively represent the probability distributions of different factors in complex systems. Moreover, due to its universal structure, one can model complicated large graphs with many factors for various practical problems.

% PGMs utility
The correspondence between the Ising model and the probability measure of the pairwise PGM allows one to solve many tasks, such as inference based on the given observations or sampling. For example, the latter can obtain the most and the least probable states by exploiting the sign before the energy function. Furthermore, the additional specific mechanism presented in several types of hardware can enhance the sampling procedure to efficiently use them as a source of additional information for particular problems.

% Optical
Unfortunately, the simulation of PGMs using optical machines needs to be better investigated. The obvious directions will be to increase the programmability of the optical spin models to access more options for manipulating the Ising/XY/Potts etc., states or decompose large and rich PGMs into their discrete approximations accessible by spin Hamiltonian simulators. Another option is to investigate additional hardware improvements in the context of PGMs. Finally, there are many more applications of such correspondence between spin system functionality, control theory, and decision-making.

\subsubsection{Neural architectures and transfer learning}
\label{Neural architectures and transfer learning}

%Different architectures
Many NN architectures can differ in forms (deep and shallow, feed-forward and recurrent), training methods, network topologies, and operational principles. Some photonic architectures were mentioned before; see Section \ref{Optical neural networks}. Moreover, some of these structures are best suited for one purpose than another. For example, recurrent NNs are good at tackling temporal dependencies, while convolutional NN is a standard architecture in image processing tasks. However, one can not easily realise all of the architectures on particular hardware due to its physical limitations or the ineffectiveness of the design.

%Transfer learning
To deal with the transfer of functionality between different architectures, one can pay attention to the domain of transfer learning. Originally, transfer learning was a research direction in ML that aimed at gaining knowledge from solving one type of problem and using it in a different but related domain; see recent reviews \cite{weiss2016survey,zhuang2020comprehensive}. However, transfer learning is a way to transfer features of one architecture to another and make the problem more hardware-friendly.

%Future perspective
Transfer of functionality will dramatically influence the ML domain and benefit the hardware computing field. It is of essential importance for optical devices, which have certain engineering limitations on the realisations of some architectures. Many more related research directions, like neural architecture search, can be adjusted to optimise the hardware systems.

\section{Optical quantum computing}
\label{Optical quantum computing}

% The classical devices
ANN in photonic integrated circuits and optical minimisers of spin Hamiltonians are the main paradigms for optical platforms that have already established an engineering base and clear development directions. Compared with emerging quantum technologies, a high-risk endeavour, classical optical devices offer advantages in speed, parallelism, energy consumption, or operational policy in short to medium term. Therefore, we can say that optical technologies are repeating their electronic special purpose hardware analogues development, with the technological progress making "another loop in its spiral development". 
% Preface 1
\nb{Quantum computers have emerged from exciting developments in physics and the theory of computation. There are several hardware platforms for developing quantum computing, and it is still being determined which technology or combination of technologies will be most successful.}
%The story of quantum computers is related to exciting developments in physics and the theory of computation. There has been a recent surge of investment by large public and startup companies. Such ramping up of industrial activity requires careful examination of the commercial potential of quantum computing technology. There are many hardware platforms on which quantum computing can be developed, and it is still being determined which technology, or a combination of technologies, will prove most successful.

% Preface 2
This section assesses the current status and future potential of quantum computing based on photons. The current view of the academic community is to exert caution when discussing future practical applications of quantum computing technology because it is so different from the information technology we use now. Many believe that quantum technology will substantially impact society in the decades ahead \cite{fedorov2022quantum}. Still, not many are confident about the commercial potential of quantum technology in the near term (five to ten years) \cite{preskill2018quantum}. Others are sceptical that quantum computers will ever become useful \cite{kalai2020argument}. At the core of critics' argument against the feasibility of quantum computers lies the notion of complexity. So far, a very low-level complexity class of probability distributions has been identified and described by noisy intermediate-scale quantum computers. Such computers would allow neither good-quality quantum error correction nor a demonstration of "quantum supremacy" -- the ability of quantum computers to make computations that are impossible or extremely hard for classical computers \cite{kalai2020argument}.

\subsection{Quantum optical devices}
\label{Quantum optical devices}

% Intro
The operation of quantum computers relies on three principles:  quantum entanglement, quantum complexity and quantum error correction. Therefore, quantum computers exploit the characteristic correlations among the parts of a quantum system that make them robust and scalable to large devices solving hard problems.

% Non-optical approaches to quantum technologies
By $2022$, many advances in quantum computing were announced (but some were also refuted). The leading technology is based on superconducting qubits (Google, IBM, Rigetty) and trapped ions (IonQ, Honeywell). Google team has announced quantum supremacy using $53$ qubits in $2019$; IBM entangled $65$ qubits while revealing a road map to more than $1000$ by $2023$. The advantages of superconducting qubit systems are that they are based on well-developed semiconductor technology; however, they have to be kept cold ($10$mK) and have a short decoherence time ($<10\mu$s). In contrast, trapped ions are very stable with much longer decoherence times (minutes), longer range interactions (beyond nearest neighbours) and report the best quantum volume among any quantum computer systems. However, many lasers are needed to be controlled simultaneously, the operation could be faster, and it would be hard to put many ions on a chip. So far, IonQ has achieved $32$ trapped ions in a chain, promised to achieve quantum supremacy by $2025$ and solve interesting real-life problems by $2028$. There are other proposals and small-scale realisations using silicon quantum dots \cite{fuechsle2010spectroscopy}, diamond vacancies \cite{bradley2019ten}, neutral atoms \cite{bernien2017probing,henriet2020quantum}, etc. One of the biggest disappointments was experienced by Microsoft in $2021$ invested into topological qubits. \nb{A topological qubit created from a pair of Majorana zero modes could theoretically benefit from topological protection. This protection could lead to stability and a lack of decoherence, potentially allowing topological quantum computers to scale up in power more easily than other approaches. The theoretical existence of Majorana zero modes was claimed to have been realized experimentally in 2018 \cite{zhang2018quantized}, but the paper was later retracted due to the discovery of erroneous data.}
%In theory, a topological qubit created from a pair of Majorana zero modes could benefit from topological protection. The topological protection leads to stability and a lack of decoherence that could help topological quantum computers scale up in power more easily than other approaches. The theoretical existence of Majorana zero modes was realised experimentally in $2018$ \cite{zhang2018quantized}, but the paper was retracted following the discovery of erroneous data presented.  

% Early-stage photon computers
Quantum computers based on photons had been considered impractical in the early ages of quantum computer developments because of difficulties in generating and controlling the required quantum states. However, such computers are being developed by photonic companies such as Xanadu (Toronto) and PsiQuantum (Palo Alto, CA) in addition to intensive academic research. The advantages of photon-based quantum computers are room temperature operation, much longer decoherence times (from ms to hours), and the systems being cheaper and easier to build. However, they become large quickly (although PsiQuantum claims that one million qubits would still be possible).

% Boson sampling
For a photon-based quantum computer, boson sampling was proposed as a counterpart to a random quantum circuit of superconducting qubit systems. A sampling task is one where the computer generates samples from a specific probability distribution. Quantum algorithms allow sampling from probability distributions well beyond the capabilities of classical computers. The most famous example is Shor’s factorisation algorithm which exploits the ability to sample a probability distribution efficiently based on the Fourier coefficients of a function on a quantum computer.

% Uncertainty principle
\nb{Squeezed states of light have an unequal distribution of quantum uncertainty between their amplitude and phase. The more a state is squeezed, the more photons it contains. Multi-photon squeezed light is found in many quantum-optics experiments and has been studied for over two decades in quantum computing models \cite{menicucci2006universal}. It has been proposed that a relatively simple optical circuit consisting of beam splitters and photon counters that exploits the properties of squeezed light could carry out a sampling algorithm faster than classical computers \cite{aaronson2011computational, hamilton2017gaussian}. Such an algorithm has many practical applications, including finding matching configurations between molecules \cite{banchi2020molecular} or different states of a molecule \cite{huh2015boson}.}
%There is a quantum uncertainty associated with the amplitude and phase of any state of light. In squeezed states of light, this quantum uncertainty is unequally distributed between the amplitude and phase and the more the state is squeezed, the more photons it contains. Multi-photon squeezed light is found in many quantum-optics experiments, and quantum computing models based on these states have been studied for over two decades \cite{menicucci2006universal}. 

%In particular, it was proposed that even a relatively simple optical circuit that exploits the properties of squeezed light and consists of beam splitters and photon counters could carry out a sampling algorithm at speed beyond the reach of classical computers \cite{aaronson2011computational, hamilton2017gaussian}. It was also proposed that such an algorithm has many practical applications \cite{bromley2020applications} such as finding matching configurations between molecules \cite{banchi2020molecular} or the different states of a molecule \cite{huh2015boson}.

% Device
More rigorously, the boson sampler is a quantum optical device in which a linear optical network mixes many non-classical photon sources. As a result, the photons are indistinguishable and, when originating from different sources, lead to complex photon counting statistics of the output detectors. When the number of the input/output channels of the boson sampler is large, the emulation of such a device with a classical computer is believed to be $\sharp$ $\mathbb{P}$-hard \cite{aaronson2011computational,wu2020speedup}. In the original formulation, the boson sampler was introduced as a device consisting of single-photon sources,  a linear interferometer and photon-counting detectors at the output channels. Several experiments implemented variations of this set-up: $5$ input photons in $21$-mode optical circuit \cite{carolan2014experimental}, and  $20$ input photons in $60$-mode interferometer \cite{wang2019boson}.

% Report 
Using single-photon sources creates various technological complications that reduce the scalability necessary to overcome the classical computer calculations (that roughly scale as $2^k$ in the number of operations where $k$ is the number of input photons). The lack of scalability in single-photon-based experiments on integrated platforms is due to non-deterministic state preparation and gate implementation. \nb{Using deterministically prepared squeezed states and linear optics with non-Gaussian operations provided by photon-counting detectors allows for significant scaling up in the number of input/output channels.} Therefore, the Gaussian boson sampling was proposed, where the single-photon sources are replaced by the single-mode squeezed light generated by parametric down-conversion sources \cite{aaronson2011computational}.
The achievement of  "quantum computational advantage" while implementing Gaussian boson sampling using $50$ input channels and a $100$-mode interferometer was recently reported \cite{zhong2020quantum}. The authors state  that their device provides $200$ seconds samples requiring classical computers billions of years. Specifically, the paper reports a Gaussian boson sampling experiment representing a quantum state in $10^{30}$-dimensional Hilbert space and a sampling rate that is $10^{14}$ faster than that of using digital supercomputers. This paper was described as the first independent verification of Google's quantum advantage claims and claimed to surpass Google's supremacy by several orders of magnitude.

% Accuracy of the supremacy claim
This huge computational advantage reported \cite{zhong2020quantum} is based on specific statistical tests measuring the proximity of the measured samples to the outcomes of noiseless simulations of the quantum experiment that were performed on a classical digital supercomputer. 
%As a result, the supremacy claims are now largely disputed. 
It was previously shown that a classically sampled distribution might pass the same statistical tests by only reproducing small-scale correlations of the actual theoretical distribution \cite{kalai2014gaussian}. Moreover, a polynomial-time algorithm based on taking a truncated Fourier–Hermite expansion on the boson sampling distribution \cite{kalai2014gaussian} may achieve similar or better sampling quality for the statistical methods of \cite{zhong2020quantum}. Another method for attaining similar sampling quality based on an algorithm of Clifford and Clifford \cite{clifford2018classical} was also proposed \cite{renema2020marginal}. Finally, very recently, a series of approximations were introduced to generate the probability distributions of any specific measurement outcome in a polynomial complexity \cite{popova2021cracking}. The accuracy of the experiment was achieved at the fourth-order approximation using a laptop computer. The algorithm was tuned towards the actual experiment and applies only to the  Gaussian boson sampling (not Fock-state boson sampling) \cite{aaronson2011computational}, only for threshold detectors (not photon-counting detectors), and only for a small number of modes (not quadratic in the number of photons as in the original proposal \cite{aaronson2011computational}). \nb{Subsequent experiments reported nontrivial genuine high-order correlations in GBS samples, providing evidence of robustness against possible classical simulation schemes} \cite{zhong2021phase}. 
%Therefore, it transpires that the "quantum supremacy" experiment of 2020 was not hard to emulate with a classical computer and the quantum advantage claims were issued prematurely. Additionally, these experiments were not scalable because of the set-up's bulkiness or photon losses. Moreover, the circuitry of these experiments was not reconfigurable, and therefore only a single, random algorithm could be executed.

So far,  experimental implementations of GBS lack programmability (reconfigurability of the circuitry) or have prohibitive loss rates that limit the scalability. There is a need for rigorous theoretical evidence of the classical hardness of GBS, althought some progress was recently made \cite{deshpande2022quantum}. 

% Other experiments
In 2021, Xanadu and NIST attempted to remedy this by implementing a programmable and potentially highly scalable circuit \cite{arrazola2021quantum}. The system uses eight modes of strongly squeezed vacuum initialized as two-mode squeezed states in single temporal modes. These pass through a fully programmable four-mode interferometer and are read out using photon number-resolving detectors on all outputs. This was achieved using strong squeezing and high sampling rates. The interferometer implemented a user-programmable gate sequence based on a network of beam splitters and phase shifters. The resulting eight-mode Gaussian state was measured on the Fock basis using eight independent photon-number-resolving detectors. The total device was composed of a 10 mm × 4 mm photonic chip coupled with a high-level application programming interface running on a classical computer.
%A more recent (2021) experiment by Xanadu and NIST attempted to remedy this \cite{arrazola2021quantum}. The circuitry they implemented is programmable and potentially highly scalable. The system uses eight modes of strongly squeezed vacuum initialized as two-mode squeezed states in single temporal modes that pass through a fully programmable four-mode interferometer and photon number-resolving readout on all outputs. This technological advance was achieved by using strong squeezing and high sampling rates. The interferometer implemented a user-programmable gate sequence based on a network of beam splitters and phase shifters. The resulting eight-mode Gaussian state is measured on the Fock basis using eight independent photon-number-resolving detectors. The total device was composed of a 10 mm × 4 mm photonic chip. The chip is coupled with a high-level application programming interface running on a classical computer.

\begin{figure}[!h]
	\begin{minipage}[c]{0.95\linewidth}
		\includegraphics[width=\linewidth]{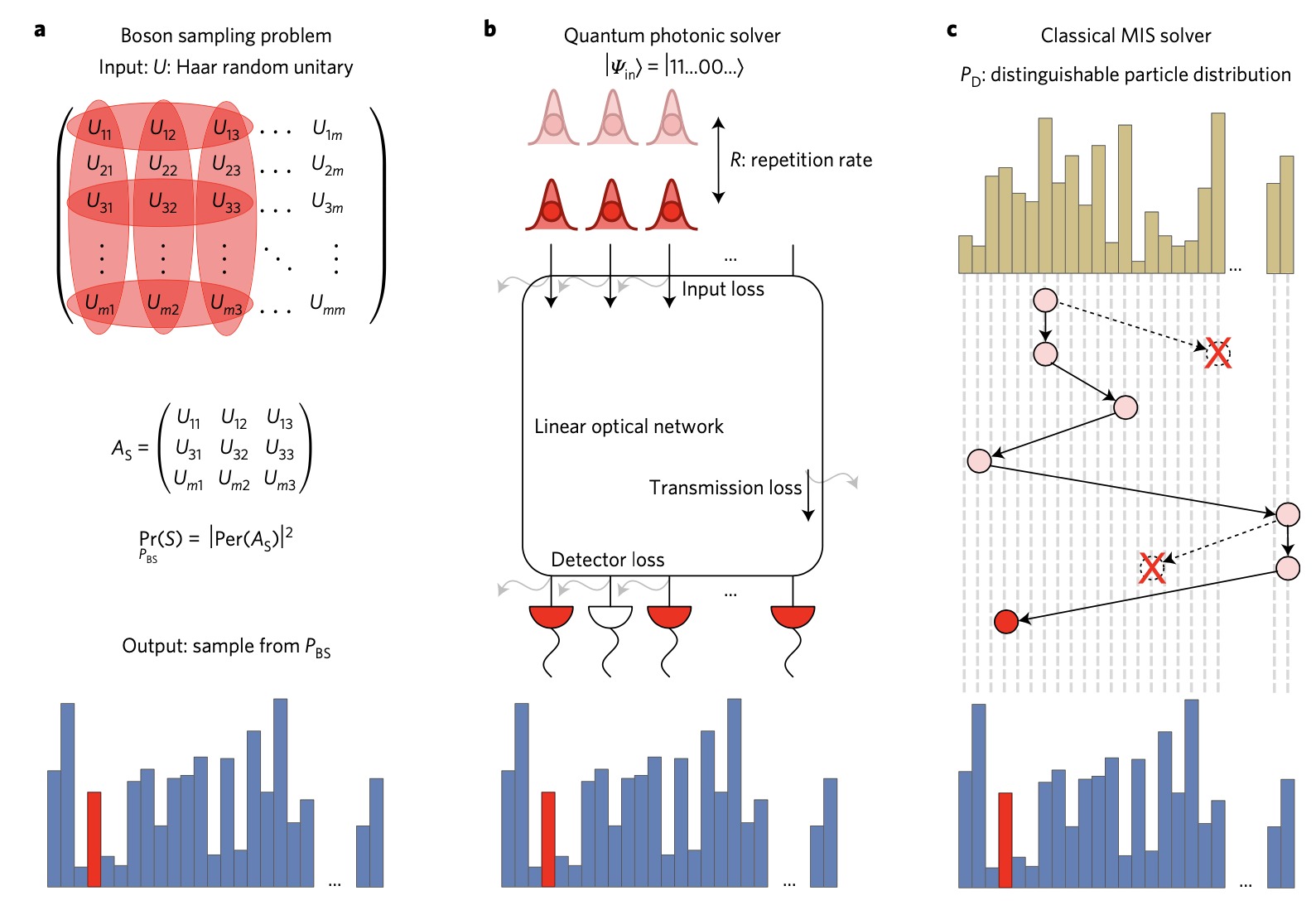}
		\caption{ a) Definition of the problem. Calculate the sample from the specific distribution defined by the modulus squared permanents of submatrices of a Haar random unitary matrix $U$. b) The scheme of the photonic experiments, i.e. single photons propagate through a linear optical network followed by its detection. c) A classical boson sampling algorithm based on Metropolised independence sampling using the distinguishable particle transition probabilities as the proposal distribution. Reproduced from \cite{neville2017classical} with permission.}
		\label{14_boson_sampling}
	\end{minipage}
	\hfill
\end{figure}

% Perspectives
There are many problems to overcome before the quantum sampling implemented on a quantum computer becomes useful for real-world applications. Photon losses need to be controlled and significantly decreased to enable photon travel through the circuitry to improve scalability. The improvement in the sampling fidelity and the quality of the squeezed states must be increased. The most exciting application would require individual control of the degree of squeezing and the amount of optical power in each squeezed state. However, the number of possible real-life applications that can be addressed using the current architecture is limited. The Xanadu group implemented two potentially practical algorithms by encoding problems into the beam splitter network. They used the generated samples to determine energy spectra for transitions between molecular states and to find the similarity between graph representations of different molecules. Specifically, a graph can be encoded in a photonic circuit by mapping its adjacency matrix into the structure of a linear optical interferometer with squeezed light \cite{bradler2018gaussian}. The photon-counting statistics can be used to specify so-called "feature vectors", which represent the graphs in Euclidean space \cite{schuld2020measuring} so that the distance between them can be used to quantify the similarity of the corresponding graphs. Such similarity measure between the graphs derived from a Gaussian boson sampling device is important, for instance, for classification in ML. Recently, other optical computing platforms based on squeezed states have been theoretically proposed on the route to a useful optical quantum computer \cite{bourassa2021blueprint,larsen2021fault}; see Fig.~\ref{14_boson_sampling}.

\subsection{Boson sampling and graph isomorphism}
\label{Boson sampling and graph isomorphism}

% The initial idea
Using the light interference network for quantum analogue calculation has many practical advantages. We mentioned that operating with the boson sampling setup allows one to calculate the permanent of a specific matrix, which is extremely hard from the computational perspective. However, it is more complex to make this helpful computation and was an open question for some time with a few remaining debates. Recently the connection between a Gaussian boson sampler and the graph isomorphism problem was established \cite{bradler2021graph}. The graphs are encoded into quantum states of light, and then their properties are probed with photon-number-resolving detectors. Using a complete set of graph invariants, the authors prove that the probabilities in the setup can be combined, and the isomorphism between the two graphs can be established only in the case of equal detection probabilities on the output.

% Short note and review
It is still an open question whether graph isomorphism has a specific complexity type. It is believed to belong to the class of $\mathbb{NP}$-intermediate computational problems. The existence of a polynomial-time algorithm that can determine whether two graphs are isomorphic is under question; however, there are quasi-polynomial types of algorithms. One can find the recent advances in photonic boson sampling with the description of both the technological improvements and future challenges \cite{brod2019photonic}. The proposed connection between the graph isomorphism and boson sampling can be further extended to other practical tasks, such as constructing graph kernels for the ML applications operating with the graph-structured data \cite{schuld2020measuring}.

\subsection{Quantum ML}
\label{Quantum ML}

% The concept
\nb{Programmable waveguide meshes possess the capability to execute arbitrary linear transformations between sets of input and output waveguides, a fundamental operation in photonic quantum computing. In this paradigm, quantum information is encoded in the quantum states of light propagating through photonic integrated circuits \cite{wang2020integrated}. A prevalent scheme encodes a qubit as a single photon in a superposition of two rail waveguides \cite{carolan2015universal}. Noisy intermediate-scale quantum (NISQ) devices have demonstrated potential in the field of quantum ML, offering the prospect of processing vast data sets at a significantly faster rate than classical computers \cite{biamonte2017quantum}. Quantum ML draws parallels with classical photonic deep NN accelerators, consisting of stages of linear waveguide meshes interconnected by activation layers exhibiting strong reversible nonlinearities \cite{steinbrecher2019quantum}. In a quantum optical neural network (QONN), programming a NISQ computer entails training the phases in the waveguide mesh via supervised learning on input and output quantum states. QONNs can be trained to perform an array of quantum information processing tasks, including quantum optical state compression and reinforcement learning. Recently, a QONN successfully programmed a one-way quantum repeater \cite{steinbrecher2019quantum,miatto2018hamiltonians}. Nonetheless, these concepts and numerous other ideas in neural quantum architectures remain far from practical implementation with current experimental capabilities.}
%Programmable waveguide meshes can be configured to execute any linear transformation between sets of input and output waveguides. These operations are also at the core of photonic quantum computing. Here, the quantum information is represented by quantum states of light propagating through the photonic integrated circuits \cite{wang2020integrated}. A typical scheme encodes a qubit as a single photon in a superposition of two rail waveguides \cite{carolan2015universal}. Noisy intermediate-scale quantum (NISQ) devices have now shown potential in quantum ML that promises to process large data sets vastly faster than classical computers \cite{biamonte2017quantum}. In these proposals, quantum ML parallels classical photonic deep NN accelerators: stages of linear waveguide meshes connected by activation layers. Still, these activation layers have strong reversible nonlinearities \cite{steinbrecher2019quantum}. In such a ‘quantum optical neural network’ (QONN), programming a NISQ computer reduces to training the phases in the waveguide mesh through supervised learning on input and output quantum states. The QONN can be taught to perform a range of quantum information processing tasks, including quantum optical state compression and reinforcement learning. Recently, a QONN managed to program a one-way quantum repeater \cite{steinbrecher2019quantum,miatto2018hamiltonians}. However, these concepts and many other ideas in neural quantum architectures developed earlier are far from useful in practice with the current experiments state.

\subsection{Comparison with other quantum approaches to optimization}
\label{Comparison with other quantum approaches to optimization}

% Main approaches
As previously discussed, CIM has shown several orders of magnitude time-to-solution advantages compared to D-Wave2000Q quantum annealer on similar dense matrix instances. \nb{Recent comparisons of the quantum approximate optimization algorithm (QAOA) with competing methods such as quantum annealing and simulated annealing \cite{streif2019comparison}, the D-Wave2000Q quantum annealer with the IBM Q Experience system that implements QAOA \cite{willsch2020benchmarking}, and benchmarking of QAOA on Google’s “Sycamore” \cite{harrigan2021quantum} enable us to compare the performance of optical spin machines with QAOA to some extent.}

% QAOA intro
In the QAOA, the variational wavefunction resembles a trotterised version of the quantum annealing procedure:
\begin{equation}
    |\Psi({\bf \beta},{\bf \gamma})>=\prod_{i=1}^N e^{-i\beta_i H_0} e^{-i\gamma_iH_{\rm objective}},
\end{equation}
where the starting state is $|+>$ is the product state of eigenstates of $\sigma_x$ with eigenvalue $1$, $|+> =\prod_i  (|0>_i + |1>_i )/\sqrt{2}$ which is simultaneously the superposition of all computational basis states. In contrast to a trotterised version of quantum annealing, the parameters $\beta_i$ and $\gamma_i$ are adjusted in a classical learning loop to minimize the objective function. Such adjustments are considered as $\mathbb{NP}$-hard problems themselves. As $P \rightarrow \infty$, QAOA approaches smooth $QA$.
\nb{According to the results of \cite{streif2019comparison}, the quantum approximate optimization algorithm (QAOA) can deterministically find the solution to specially constructed optimization problems where both quantum annealing and simulated annealing fail due to wide and tall energy barriers of the function being minimized.}However, there exists an efficient classical algorithm for these instances.

\begin{figure}[!h]
	\begin{minipage}[c]{0.95\linewidth}
		\includegraphics[width=\linewidth]{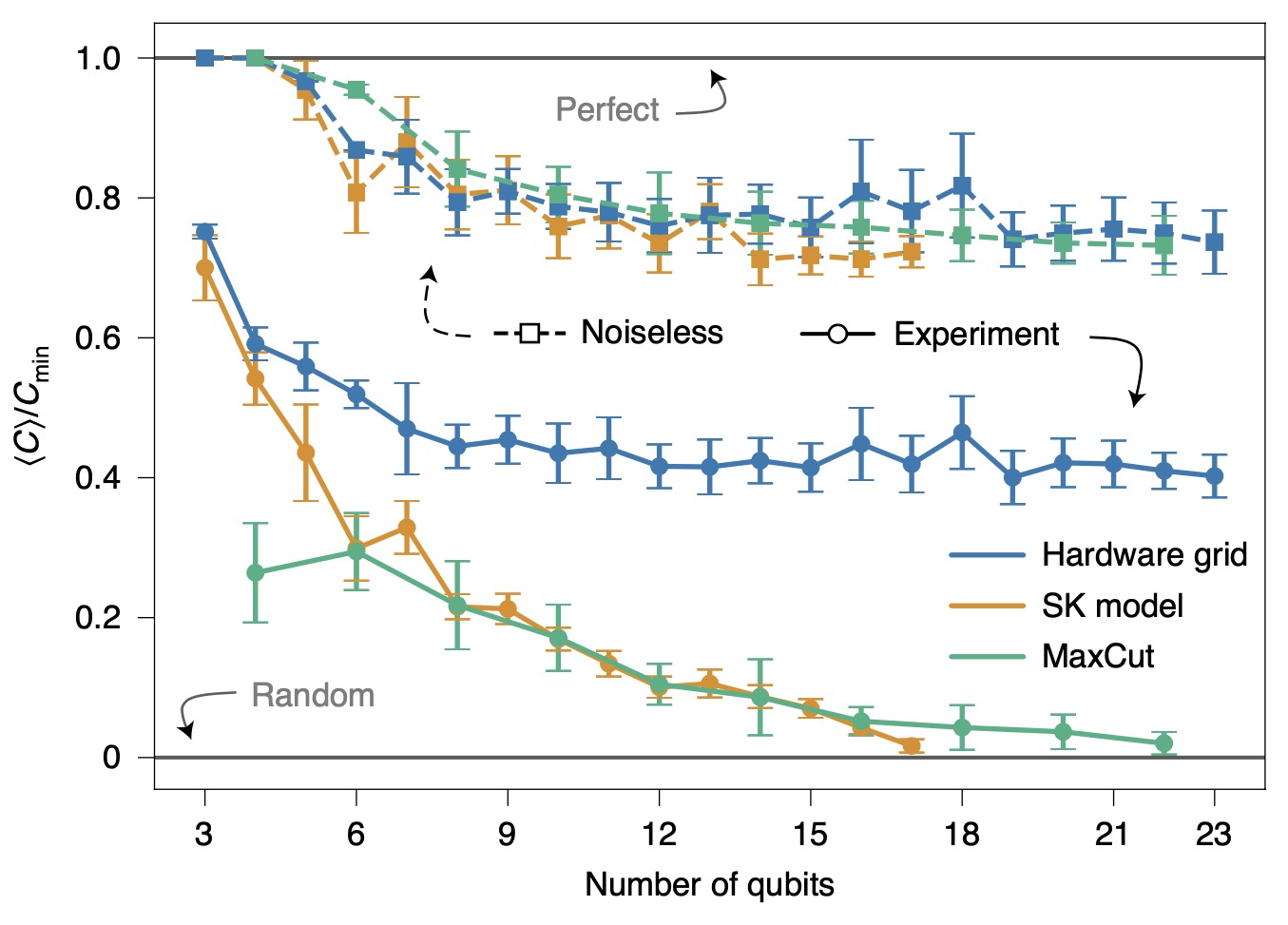}
		\caption{The ratio of the found solution to the best solution when using QAOA for various problem sizes $N$. Each solution is averaged across ten random instances (standard deviation is given as error bars). The experimental solutions of the SK and Max-Cut models approach random as $N$ increases. The figure is taken from \cite{harrigan2021quantum}.}
		\label{13_QAOA}
	\end{minipage}
	\hfill
\end{figure}

% Comparison
In \cite{willsch2020benchmarking}, small size (up to $N=18$) weighted Max-Cut problems and $2$-SAT problems were tested using D-Wave2000Q quantum annealer with IBM Q Experience. The actual machine IBM Q on 16 qubits gave such poor solution quality that the real physical experiment on D-Wave200Q was compared to the simulation of QAOA. Even in this case, physical QA has shown much better success probabilities than QAOA ($99.92$ vs $8.84 (p=1)$ and $42.39 (p=3)$, respectively, on as small matrices as $N=8$!) The conclusion drawn was that, for the set of problem instances considered and using success probability as a measure, "the QAOA cannot compete with quantum annealing". The corresponding plots can be found in Fig.~\ref{13_QAOA}.

% Results
In \cite{harrigan2021quantum}, the authors used the Google Sycamore superconducting qubit quantum processor to run the quantum approximate optimization algorithm (QAOA) for combinatorial optimization problems on a planar graph matching the hardware connectivity. They also applied QAOA to the Sherrington-Kirkpatrick model and Max-Cut, both high-dimensional graph problems that require significant compilation for QAOA. The problems were solved up to $N=23$ numerically (without noise) and experimentally. For QAOA the theoretically optimal ${\bf \beta, \gamma}$ and $p\in{1,2,3,4,5}$ were used in experiments. The success probabilities on average of the problem on graph matching hardware reached a plateau for $N>8$ at about $80$ \% (numerically) and about $45$ \% experimentally. For SK and Max-Cut problems, performance deteriorated quickly (for any $p$) to the probability of finding a solution by random guessing (for $N>15$). \nb{The authors deduced that although current quantum processors are unable to surpass classical optimization heuristics, utilizing prevalent techniques such as the Quantum Approximate Optimization Algorithm (QAOA) on representative problems can serve as a standard for evaluating different hardware platforms. In order for quantum optimization to rival classical approaches for practical problems, it is imperative to transcend beyond artificial problems with low circuit depth.}

\subsection{Quantum effects and optical machines}
\label{Quantum effects and optical machines}

% Quantum effects
As we can see, various optical hardware uses different mechanisms for its operation. It can have the primary mechanism's pure classical, quantum, or hybrid nature. Even in the case of operating near the classical limit, the quantum effects can be essential and greatly influence the actual operation regime.

%CIM example
For example, it was shown that the nonlinear stochastic dynamics of the CIM in the presence of quantum noise could be efficiently exploited to sample degenerate ground and low-energy spin configurations of the Ising model on the example of Max-Cut problems \cite{ng2022efficient}. Both quantum noise and optical nonlinearities play an essential role in system dynamics. Removing these essential elements will result in the degradation of sampling performance. The supplementary numerical results beyond the classical simulation complement the description of the quantum mechanism's role in the CIM operation. Another work \cite{sankar2021benchmark} studies the performance scaling of three quantum algorithms for combinatorial optimization, such as CIM performance, discrete adiabatic quantum computation, and the Dürr-Høyer algorithm for quantum minimum finding that
is based on Grover's search. Authors claim that the CIM performance is dramatically better for solving Max-Cut problems. Moreover, the CIM is competitive against various heuristic solvers implemented on CPUs, GPUs, and FPGAs.

% EP system
Many optical devices fall under the category of open quantum systems. Such a formalism is necessary to account for many complicated effects. For this purpose, a Markovian open quantum systems framework has been developed \cite{lindblad1976generators,gorini1976completely}. Such effective dynamics for the reduced density matrix of the system give rise to the Lindblad-form master equation, which allows one to trace such effects as equilibration with the pump and decay processes, thermalisation of the system and different aspects of interaction with the environment. Although the numerical methods for such processes are quite complicated, one usually develops approximated schemes that account for the omitted effects. There are many more systems where this approach could be beneficial for describing subtle but essential features—another example, except for the CIM, is the exciton-polariton system frequently mentioned before. Furthermore, one should pay attention to other microscopic processes in the EP system since such consideration gives more degrees of freedom to inspect compared to the simple mean-field theory \cite{levinsen2019microscopic,bastarrachea2021attractive,efimkin2021electron}.

\section{Final remarks}
\label{Final remarks}

\subsection{Benchmarking optical machines}
\label{Benchmarking optical machines}

% Problems in benchmarking
So far, the research and development of optical hardware are experiencing significant growth. The main problem is to compare the capabilities of optical machines as they are often tested on different problems of variable sizes and difficulty. Thus it is hard to figure out the scaling properties of the particular mechanism from either experiment or numerical emulation procedure. Another problem is lying in the biased results, which can be cherry-picked for better demonstrative purposes. In general, it is hard to find extensive, complete, up-to-date and unbiased results comparing different types of optical hardware.

% References
Although the majority of the NN optical architectures can be compared using standard metrics concerning the accuracy of the particular datasets and the required workload, we outline what is known and how some of the optical machines can be compared. For example, in \cite{cai2020power} comparisons between memristors, GPU, D-Wave and the CIMs were made using the same set of dense $60$-node Max-Cut graphs. The time to solution (with $99$ \% probability to reach optimal solution) was $600 \mu$ s for the CIM and $1000$ s for the D-Wave. \nb{It was shown that the Ising machine, which is based on optoelectronic feedback systems, resolved Max-Cut optimization problems on both regular and frustrated graphs with 100 spins, exhibiting comparable or superior performance to CIMs based on DOPOs \cite{CIM6}. Since OEO-based CIMs can be realized as integrated photonic circuits, the flexible spin coupling can be achieved optically using programmable silicon photonic circuits. This will fully exploit the high bandwidth of the optical system and result in a significant acceleration over existing CIM concepts.}

%In \cite{CIM6}, it was reported that the Ising machine based on optoelectronic feedback systems solved Max-Cut optimization problems on regular and frustrated graphs with $100$ spins showing similar or better performance compared to CIMs based on DOPOs. Since OEO-based CIMs can be implemented as integrated photonic circuits, the flexible spin coupling can be made optically with programmable silicon photonic circuits. This will take full advantage of the high bandwidth of the optical system and bring a significant speedup over existing CIM concepts. 

% Similar research directions
Establishing universal benchmarks will attract more people since understanding the hardware's successes and failures on particular problems allows one to maximize utility. Moreover, such a research direction shares similar issues with the ongoing studies on the NN architectures and phase transitions in the statistical approaches to the computational problems \cite{zdeborova2016statistical}, which is cross-beneficial for all of the domains.

\subsection{The most promising applications for optical computing}
\label{The most promising applications for optical computing}

Our subjective perspective is that modern optical computing has the potential to give a significant computational advantage in three major applied areas: {\bf Neural networks, Nonlinear optimization, and Statistical sampling.}

Optical hardware is a promising platform to get accceleration for these applications, with many computational advantages coming from the hybrid-quantum/classical mode of operation. The optics naturally supports these tasks but also benefit from many more factors, such as specific architectures and their interplay with the natural properties of light systems.

% Optimization benefits
For example, a mode selection mechanism is one of the beneficial regimes of operation for a quantum system spanning a high-dimensional space of possible solutions and finding an optimal one while settling to the first possible coherent state with a large occupation. Another component is classical dynamical system behaviour that can mimic the NN dynamics, following the classical gradient dynamics on a changing energy landscape while tunnelling through barriers to the nearest energy minimum. The task is achieved if this minimum corresponds to the optimum solution to the problem. Finally, a similar mechanism is responsible for sampling the landscape's low-energy subspace.

\subsection{Future perspectives}
\label{Future perspectives}

% NFLT
The 'no free lunch' (NFL) theorem in optimisation states that any two optimisation algorithms have the same performance averaged across all possible problem instances. This theorem applies to the hardware instead of the algorithms with many more implications. One of the consequences of the NFL theorem is the correspondence between the solver/hardware structure and the hardness of the problem with the best-case and worst-case scenarios.

% Taking advantage of the NFL
To use optical spin machines to speed up the solutions to specific industry real-life problems, one needs to think hard about the range of  application that may go far from the QUBO. These application has to closely match the optical machine's operational principle to take advantage of all the potential advantages. Many questions need to be addressed before the optical spin machines become useful for the real-life applications. 
  Which platforms should we use for comparison between different machines? What is the importance of optical quantum vs classical, classical vs classical, hybrid vs classical, optical vs other physics-based hardware advantages? How  does hardware performance compare to the  best algorithms run on traditional systems? To answer these questions, we need to introduce a standard for fair comparisons between the machines and approaches. 
   Which section of the workflow is more advantageous to optimize? Should sections that are closer to hardware or closer to a user  be more important?   How to properly optimise pre-processing and post-processing?
  How do we evaluate results and which metric should be used? The proximity between the found solutions of QUBO can be evaluated using, for instance, the Hamming distance, the distance in the  energy space, the ratio of the energies, the accuracy of the neural architecture, or some other the generalisation of the error metrics can be used.
   How do we evaluate optimisation performance in several important dimensions, e.g. the computation time, the solution quality, the energy efficiency, the input scope, etc. all can be used for evaluation.
   How do we account for overheads concerning the given architecture and the task-specific constraints? How to optimize the implementation, translating, embedding, tuning, post-processing?
   How do we find inputs that are hard and relevant? How do we avoid using trivial ones? Drawing the possible phase diagrams should be built for the parametrised tasks.
   How do we maximise the generality of the conclusions, and to what extent if we can only test some combinations of inputs and hardware.

% Results
 The advantage will come when we (i) develop purpose-built solutions tuned to specific applications, (ii) develop hybrid algorithms and approaches (e.g. including ML as a part of the hybrid solutions) and (iii) leverage programmable accelerators for core tasks. More research is needed to bring the potential of optical (or any other unconventional) computing systems to real-life applications. Answering the critical questions will bring us closer to a better understanding of the underlying principles of unconventional optical machines, improve their performance and hence achieve a significant practical impact.

\acknowledgements
N.G.B. thanks the Julian Schwinger Foundation grant JSF-19-02-0005 for  the financial support. 

%\bibliography{references}{}
%\bibliography{references,refs,Refs-pr2}{}
\bibliographystyle{ieeetr}
\bibliography{references}{}

\end{document}